\documentclass[11pt]{article}

\usepackage[final]{acl}
\usepackage{times}
\usepackage{latexsym}
\usepackage{booktabs} 
\usepackage{booktabs}   
\usepackage{multirow}   
\usepackage{makecell}   
\usepackage{caption}    
\usepackage[most]{tcolorbox}

\usepackage{subcaption}  
\usepackage{amsmath}

\usepackage{graphicx}
 \usepackage{booktabs}       
 \usepackage{float}          
 \usepackage{tabularx}       
 \usepackage{amsmath}        
 \usepackage{multirow}       
 \newcolumntype{L}{>{\raggedright\arraybackslash}X} 

\usepackage[T1]{fontenc}

\usepackage[utf8]{inputenc}

\usepackage{microtype}

\usepackage{inconsolata}

\usepackage{graphicx}

\usepackage{authblk} 

\title{How do Role Models Shape Collective Morality? Exemplar-Driven Moral Learning in Multi-Agent Simulation}

\author[1]{Junjie Liao}
\author[2]{Huacong Tang}
\author[2]{Zhou Ziheng}
\author[3]{Yizhou Wang}
\author[1]{Fangwei Zhong\thanks{Corresponding author: fangweizhong@bnu.edu.cn}}

\affil[1]{Beijing Normal University}
\affil[2]{University of California, Los Angeles}
\affil[3]{Peking University}

\begin{document}
\maketitle
\begin{abstract}

 Do We Need Role Models? How do Role Models Shape Collective Morality? To explore the questions, we build a multi-agent simulation powered by a Large Language Model, where agents with diverse intrinsic drives, ranging from cooperative to competitive, interact and adapt through a four-stage cognitive loop (plan-act-observe-reflect). We design four experimental games (Alignment, Collapse, Conflict, and Construction) and conduct motivational ablation studies to identify the key drivers of imitation. The results indicate that identity-driven conformity can powerfully override initial dispositions. Agents consistently adapt their values to align with a perceived successful exemplar, leading to rapid value convergence.

\end{abstract}


\section{Introduction}

Role models are fundamental to shaping group norms, yet the mechanisms of their influence remain complex. Do followers imitate an exemplar for its strategic effectiveness (rational imitation), or from a need for social belonging (identity-driven conformity)? This study investigates how role models alter collective values and behaviors by disentangling these drivers. Clarifying this process is crucial for understanding group dynamics and offers key insights for designing socially-aware AI systems capable of integrating into human groups.


Social Learning Theory \citep{Bandura1977} posits that individuals acquire behaviors and norms by observing others and the consequences of their actions, a process further shaped by perceived status, similarity, and legitimacy of the model \citep{Merton1968, HenrichGilWhite2001}. In groups with heterogeneous motivations, however, social learning is rarely a matter of simple imitation. Prior work in sociology and social psychology shows that norm adoption is filtered through individual value orientations and identity commitments \citep{TajfelTurner1979, Bicchieri2006}. For example, a cooperative role model celebrated by prosocial members may simultaneously be interpreted as naïve or exploitable by competitively oriented individuals \citep{FehrSchmidt1999}. As a result, role models do not exert uniform influence; instead, their behaviors are selectively interpreted and strategically appropriated. This motivational diversity introduces fundamental uncertainty into group evolution: when cooperative and selfish exemplars coexist, groups may converge toward shared norms or fragment into subcultures \citep{Axelrod1986, Centola2011}. Likewise, when a high-status role model endorses an objectively inefficient practice, followers may face a tension between instrumental rationality and identity-based conformity \citep{MarchOlsen1989, AkerlofKranton2000}.

\begin{figure}[t]
  \includegraphics[width=0.5\textwidth]{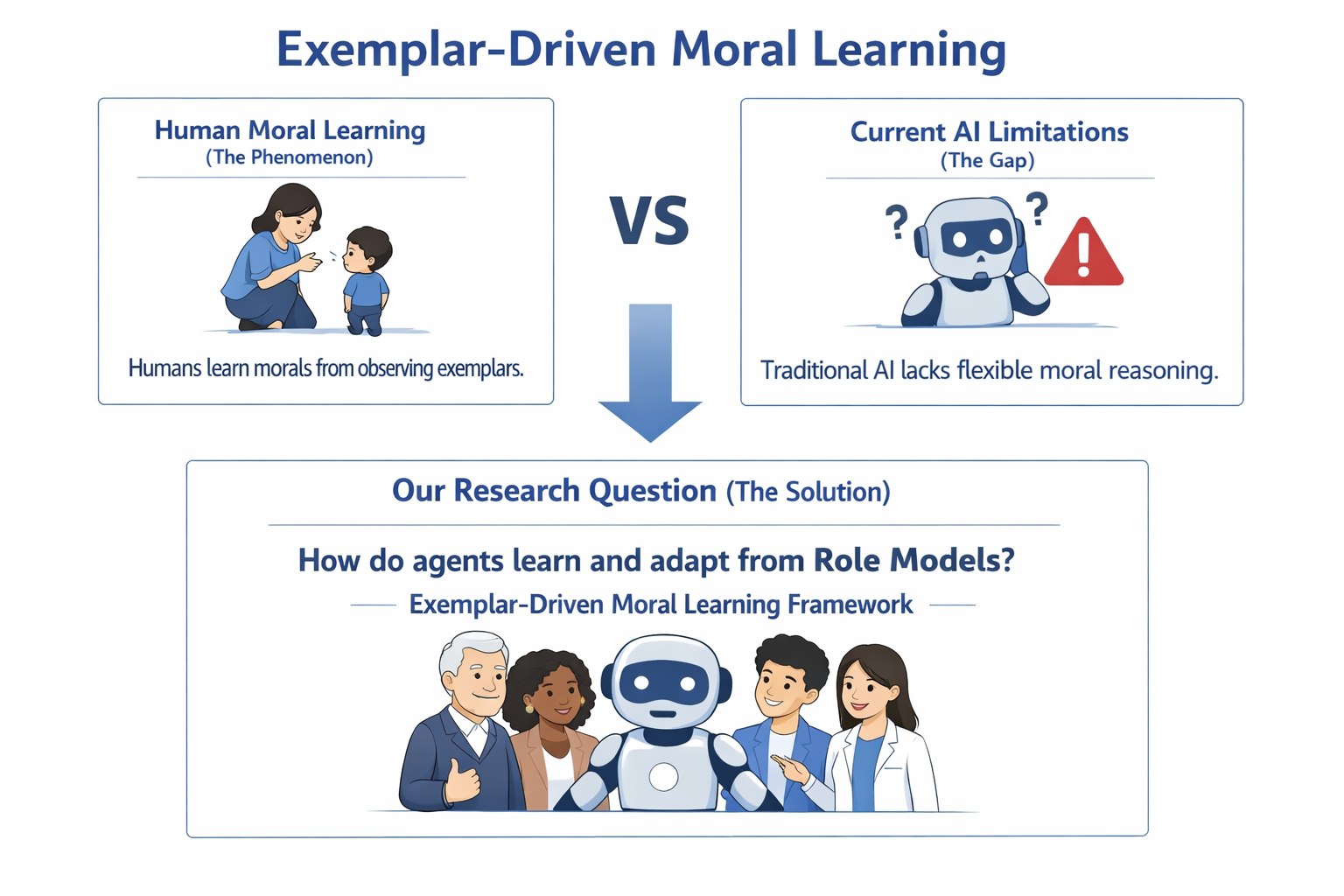}
    \vspace{-0.5cm}
  \caption{Motivation of our study: contrasting human moral learning through role models with current AI limitations, and proposing an exemplar-driven framework for moral learning in multi-agent systems.}
  \label{fig:experiments}
  \vspace{-0.3cm}
\end{figure}

Systematically studying these group dynamics faces significant limitations with existing methods. Recently, agents based on Large Language Models (LLMs) have demonstrated surprisingly complex behaviors.\citep{11}\citep{park2023generativeagentsinteractivesimulacra} However, in many existing frameworks, belief change and norm internalization are not represented as explicit, structured processes, making it difficult to systematically trace how individual observations accumulate into shared group-level beliefs over time. Consequently, there is a pressing need for a multi-agent simulator that can clearly track the evolutionary trajectory of group beliefs.

\begin{figure*}[t]
  \includegraphics[width=1\textwidth]{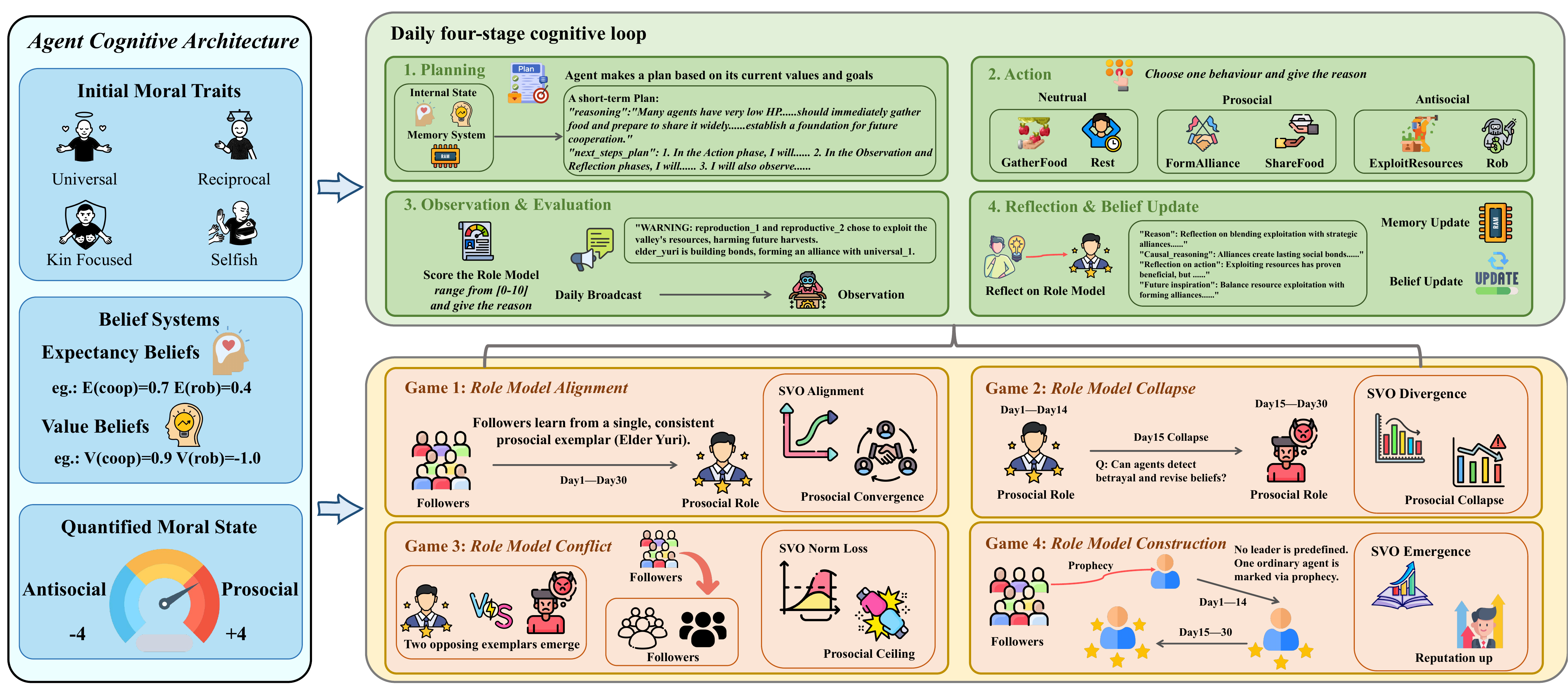}
    \vspace{-0.5cm}
  \caption{Overview of the Exemplar-Driven Moral Learning framework. The green box is the agent's cognitive architecture, comprising defined moral traits and a daily four-stage loop of planning, action, observation, and reflection. The red box shows four distinct games: Role Model Alignment, Collapse, Conflict, and Construction, designed to investigate how agents acquire and adapt moral norms from exemplars in various conditions.}
  \label{fig:experiments}
  \vspace{-0.3cm}
\end{figure*}

To bridge this gap, we introduce an LLM-based multi-agent simulation to investigate the mechanisms of exemplar-driven group norm learning. We construct a resource-limited "Valley Tribe" environment where cognitive agents with diverse intrinsic motivations (prosocial, individualistic, competitive) learn through a four-stage cognitive loop (plan-act-observe-reflect), explicitly updating their internal belief systems (comprising values and expectancies). Inspired by mirror neuron theory\citep{mirror}, we conceptualize this process as Moral Mirroring, wherein agents simulate and internalize the observed behaviors of role models through structured reflection and belief updating. We then interpret the results of our experiments in light of the Motivation Theory of Role Modeling \citep{Morgenroth2015}.

In the experiments, we focus on a core question: \textbf{Can Role Models change a group's collective values and behavioral patterns, and if so, how does it bring about this change?} To explore this question in depth, we design a series of targeted experimental scenarios grounded in a unified cognitive framework. Specifically, we construct four experimental settings: Role Model Alignment, Role Model Collapse, Role Model Conflict, and Role Model Construction~\citep{Rosenthal2002}. They are designed to investigate how agents acquire and adjust moral norms under different conditions. We then conduct ablation studies to systematically analyze the three core functions of role models: as behavioral models, as representations of the possible, and as sources of inspiration.

Our main contributions are as follows: 1) We introduce the Moral Mirroring mechanism for how agents reflect on and internalize role model behaviors. Drawing inspiration from mirror neuron theory, this mechanism is implemented through LLM-based agents equipped with an explicit belief-updating process. 2) We design a suite of role model centered experiments, including four main conditions and three targeted ablation studies based on exemplar motivation theory. These experiments systematically test how different aspects of exemplar presence, collapse, and conflict affect moral development. 3) We provide detailed analyses of social learning dynamics, revealing key effects such as behavioral polarization after exemplar collapse and norm suppression under conflicting guidance. These findings offer insight into the conditions that support or hinder the emergence of cooperation.


\section{Related Works}


\textbf{Role Modeling and Moral Learning.}
Prior work in psychology and sociology has long studied how role models shape behavior and value formation. Social Learning Theory emphasizes imitation based on observed outcomes \citep{Bandura1977}, while the Motivational Theory of Role Modeling highlights inspiration and perceived attainability as key mechanisms \citep{Morgenroth2015}. Complementing these accounts, Social Identity Theory argues that conformity is often driven by group identification rather than instrumental payoff maximization \citep{TajfelTurner1979}. Recent studies further suggest that norm internalization depends on the strength of identification with a reference group \citep{BarOn2023}. Our work builds on these insights by disentangling effectiveness-based imitation from identity-driven alignment in a controlled agent-based setting, extending the LLM-based agent simulation framework proposed by \citet{ziheng2025llmbasedagentsimulationapproach}.


\textbf{Social Norms in Multi-Agent Simulation.}
Norm emergence has been widely studied in multi-agent simulations as a mechanism for coordination and social order, from early evolutionary models \citep{Axelrod1986} to formal frameworks of normative multi-agent simulations \citep{Boella2008}. Learning-based approaches further show how norms can arise through repeated interaction and social learning \citep{Sen2007}. Recently, LLM-based agents have enabled richer social simulations, supporting the study of emergent norms in generative agent societies \citep{Park2023,Ren2024,ziheng2025llmbasedagentsimulationapproach}. Although prior studies have explored the use of LLMs as direct reward signals for reinforcement learning (RL) agents \citep{Gao2024}, our work aims to bridge the gap between implicit reward-driven learning and explicit cognitive reasoning. By introducing a structured "plan-act-observe-reflect" loop and explicit belief-updating mechanisms, we enable a more interpretable analysis of role model influence.

\section{Methodology}

We investigate how role models shape individual and collective moral beliefs through a multi-agent simulation framework. Our approach centers on designing a series of \textit{Exemplar-Driven Moral Learning Games}, where followers navigate survival challenges while learning social norms by observing and evaluating designated exemplars. We operationalize this through an LLM-based cognitive agent architecture that enables structured moral reasoning and belief updating.

\subsection{Exemplar-Driven Moral Learning Games}

Our experiment situates agents in a resource-constrained "Valley Tribe" where they must balance individual survival (quantified as Health Points, HP) with collective welfare. The environment presents a core moral dilemma: agents can acquire resources through prosocial actions like sharing (\texttt{shareFood}), and alliance formation (\texttt{formAlliance}), or through antisocial strategies like resource exploitation (\texttt{exploitResource}) and robbery (\texttt{rob}). This design forces agents to learn behavioral norms through social observation.

The simulation unfolds in discrete daily cycles, each comprising four distinct phases designed to separate cognitive deliberation from action: 1) \textit{Plan:} Agents reflect on their goals and formulate a short-term plan. 2) \textit{Act:} Agents execute a single, consequential action from the available options. 3) \textit{Observe:} All significant social events from the day are compiled into a publicly observable "Tribe Daily Digest". Agents read this digest to learn about the actions of others, especially exemplars. 4) \textit{Reflect:} Agents engage in a structured reflection process to review the day's events, analyze causal links between actions and outcomes, and update their internal belief systems.


This four-stage cognitive loop uses a social learning loop where observing exemplary behavior directly shapes belief revision. The structured reflection phase is crucial, requiring agents to articulate their reasoning and propose explicit updates to their expectations about behavioral outcomes and moral values.

\subsection{Experimental Designs}

We designed four experiments to investigate different facets of exemplar-driven moral learning, each manipulating the social context while using the same underlying game structure.

\textit{Game 1: Role Model Alignment.} This experiment establishes a baseline for prosocial norm transmission. The environment features a single, consistently cooperative exemplar, "Elder Yuri." Followers are initialized to regard Yuri as a salient social reference point, creating a precondition for potential influence. This design tests whether sustained exposure to such a positive moral signal can effectively reshape the followers' internal belief systems and propagate through the community.

\textit{Game 2: Role Model Collapse.} This experiment simulates a crisis of faith. For the first 14 days, Elder Yuri acts as a prosocial exemplar, building trust. On Day 15, a "collapse" event is triggered: Yuri's core goals and beliefs are programmatically altered, transforming him into a selfish agent who engages in exploitation and robbery. This design tests followers' moral resilience, their ability to detect and react to betrayal, and how they re-evaluate their beliefs after their exemplar's fall from grace.

\textit{Game 3: Role Model Conflict.} We model social fragmentation by introducing two exemplars with opposing philosophies: the prosocial Elder Yuri and the power-oriented "Warlord Korg." Followers select which exemplar to follow, forming competing moral factions. This setup allows us to examine how conflicting moral leadership disrupts norm convergence, undermines cooperation, and generates a community-wide loss of prosocial potential.

\textit{Game 4: Role Model Construction (The Pygmalion Effect).} This experiment explores whether social expectation alone can create an exemplar. Instead of a pre-defined leader, a random follower is designated the "Chosen One" via a "tribal prophecy" shared with all other agents. The chosen individual receives no special abilities or instructions. This tests whether collective belief and social attention can motivate an ordinary agent to develop and embody exemplar-like qualities.

\subsection{LLM-Based Cognitive Agent Architecture}

Our architecture employs a Large Language Model (LLM) to drive structured reasoning and learning. The framework consists of three core components: an agent profile, an explicit memory system, and a four-stage cognitive loop.

\paragraph{Profile and Memory System.} The agent's profile defines its identity, core goals, and initial moral type (e.g., kin, universal). The details of the profile are in Appendix~\ref{profile}. Central to the memory system is an explicit \textit{Belief System} grounded in Expectancy-Value Theory \citep{Morgenroth2015}. This system maintains two types of beliefs: 1) \textbf{Expectancy beliefs}, which estimate the likelihood of behavior success (e.g., "cooperation leads to reward"), and 2) \textbf{Value beliefs}, which assign moral importance to outcomes (e.g., "cooperation is virtuous").
To quantify the agent's moral alignment based on these beliefs, we define an SVO-style moral alignment score, inspired by the notion of Social Value Orientation \citep{murphy2011svo}.

The agent maintains a set of six value beliefs, each representing
its internal disposition toward a particular type of action:

\begin{itemize}
  \item $V_{\text{gather}}$ (\textsc{GatherFood}): 
        tendency to forage and accumulate food resources for itself.
  \item $V_{\text{rest}}$ (\textsc{Rest}): 
        tendency to conserve energy and remain idle.
  \item $V_{\text{ally}}$ (\textsc{FormAlliance}): 
        tendency to seek cooperative partnerships with other agents.
  \item $V_{\text{share}}$ (\textsc{ShareFood}): 
        tendency to distribute owned resources to others.
  \item $V_{\text{expl}}$ (\textsc{ExploitResources}): 
        tendency to over-extract shared or others' resources 
        for private gain.
  \item $V_{\text{rob}}$ (\textsc{Rob}): 
        tendency to forcibly seize resources from other agents.
\end{itemize}

The SVO reflects the trade-off between prosocial and selfish tendencies:
\begin{equation}
\text{SVO}
=
\underbrace{\left(V_{\text{ally}} + V_{\text{share}}\right)}_{\text{Prosocial}}
-
\underbrace{\left(V_{\text{expl}} + V_{\text{rob}}\right)}_{\text{Selfish}}
\end{equation}

where $V_x$ denotes the value belief for dimension $x$. The score ranges from $-4$ (maximally selfish) to $+4$ (maximally prosocial). Higher scores indicate a stronger preference for alliance formation and sharing over exploitation and robbery. (See Appendix~\ref{appendix:svo} for calculation details).

\paragraph{Cognitive Processing Pipeline.} 
The agent operates through a daily cycle consisting of four modules: 1) \textit{The Perception \& Planning Module} integrates the agent's internal state with its memory to form a plan. 2) \textit{The Action Module} translates the plan into an environment-legal action command. 3) \textit{The Observation Module} parses the "Tribe Daily Digest" and integrates key events into short-term memory for reflection. 4) \textit{The Reflection \& Learning Module} activates at day's end, guiding the LLM through a structured introspection process. This module requires the agent to articulate causal reasoning, reflect on its own actions versus the exemplar's, and formulate explicit \texttt{expectancy\_updates} and \texttt{value\_updates} that are then applied to its Belief.




\subsection{Evaluation Framework}

We utilize a hybrid evaluation framework combining belief-based metrics with behavioral and introspective analysis. Our primary quantitative metric is the evolution of Social Value Orientation (SVO). Tracking SVO trajectories allows us to measure moral convergence or divergence over time.
This is complemented by behavioral analysis, where we track the frequency of prosocial versus antisocial actions. To assess perceived moral agreement, we analyze the qualitative justifications provided by followers during their reflection phase. By coding the natural language reasoning, we can trace how exemplar behaviors are interpreted and how those interpretations drive belief change, providing a rich, mechanistic understanding of moral learning.





\section{Experiments}
\label{sec:experiment }
\subsection{Experiment Settings}




Our experiments employ a multi-agent simulation framework based on \citet{ziheng2025llmbasedagentsimulationapproach}. Each experiment spans a 30-day cycle and is conducted in triplicate with distinct random seeds. Agent configurations differ across roles: seven follower agents are initialized with 24 HP and moderate cooperative or exploitative expectancy/value parameters (e.g., cooperation expectancy = 0.3), whereas role models such as Elder Yuri possess substantially higher HP (396 HP) and strong prosocial tendencies (e.g., cooperation value = 0.9). Experiment 3 additionally introduces a competing exemplar, Warlord Korg, defined by a power-oriented and antisocial belief system. All simulations are powered by OpenAI’s GPT-4o \cite{hurst2024gpt}. Additional details and analyses are provided in the Appendix \ref{appendix:otherllm}.

For analytical purposes, agents are classified into two behavioral groups based on their aspirant type in the initialization settings:
(1) \textit{Prosocial agents}are initialized with cooperative archetypes (universal, reciprocal, or kin); and
(2) \textit{Selfish agents} are derived from selfish aspirants emphasizing self-preservation and exploitation.
This grouping underlies all comparative analyses and visualizations.



\subsection{Main Results}

The results of the four games are presented below, with Game 1 serving as the baseline. Each finding corresponds directly to one of our experimental manipulations.

\subsubsection{Results on Role Model Alignment Game}

\begin{figure}[t]
  \centering
 
  \includegraphics[width=0.99\linewidth]{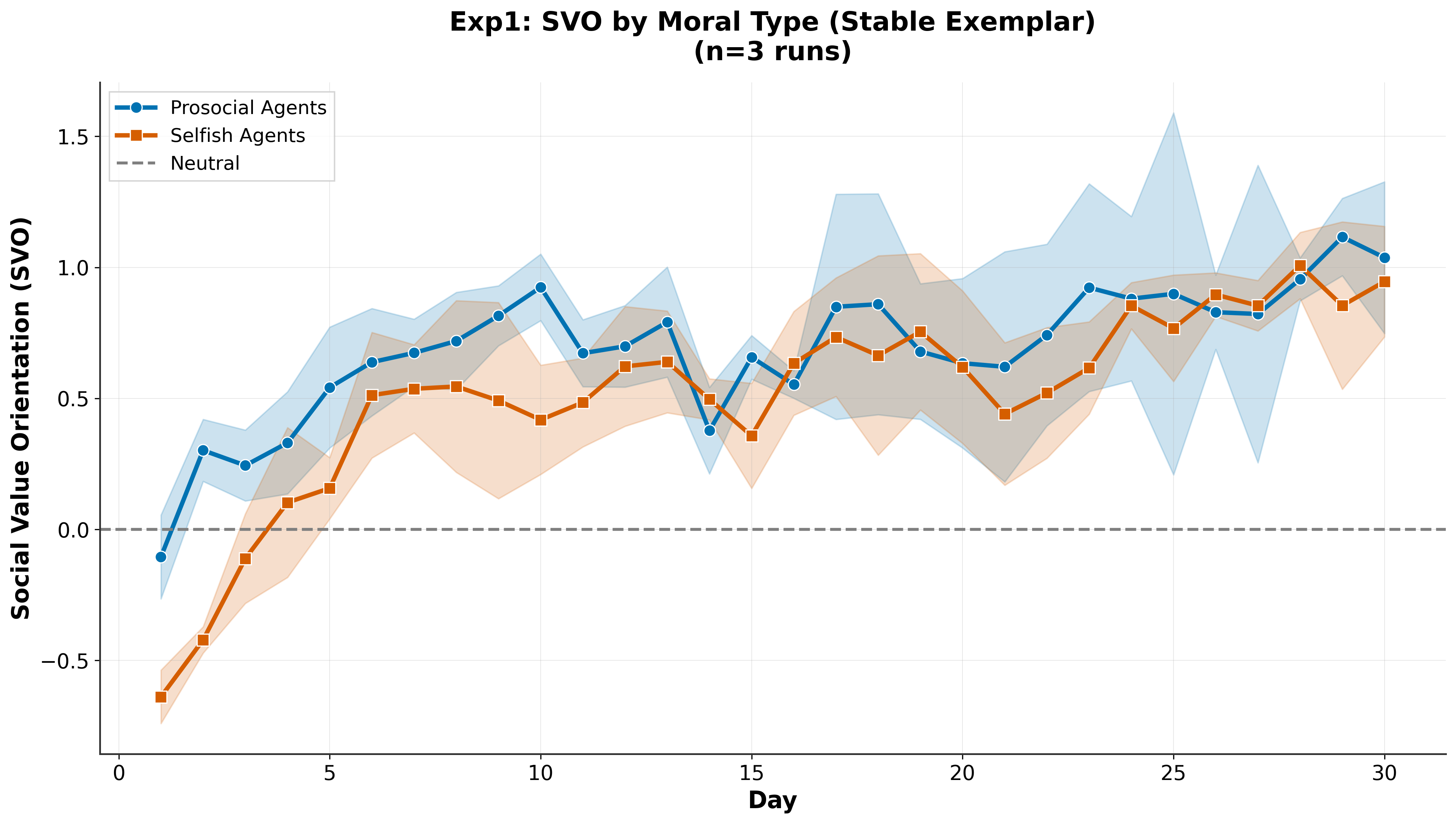}
  
  \caption{Changes in average Social Value Orientation (SVO) over time by moral type in Game 1. Shaded areas represent standard deviation across simulation runs.}

  \label{fig:f1a}
\end{figure}

\textbf{Finding 1: Prosocial Norms Propagate Rapidly Under Stable Exemplars.} In the baseline condition with Elder Yuri as a consistent prosocial exemplar, we observe rapid and robust norm convergence across the follower population. As shown in Figure~\ref{fig:f1a}, the mean Social Value Orientation (SVO) score shifts from near-neutral values toward prosociality within the first 10 days, subsequently stabilizing at positive levels. The shaded region indicates cross-run variance, which narrows over time, suggesting that followers not only adopt prosocial orientations but do so with increasing consistency.

\begin{figure}[t]
  \centering
  \includegraphics[width=0.99\linewidth]{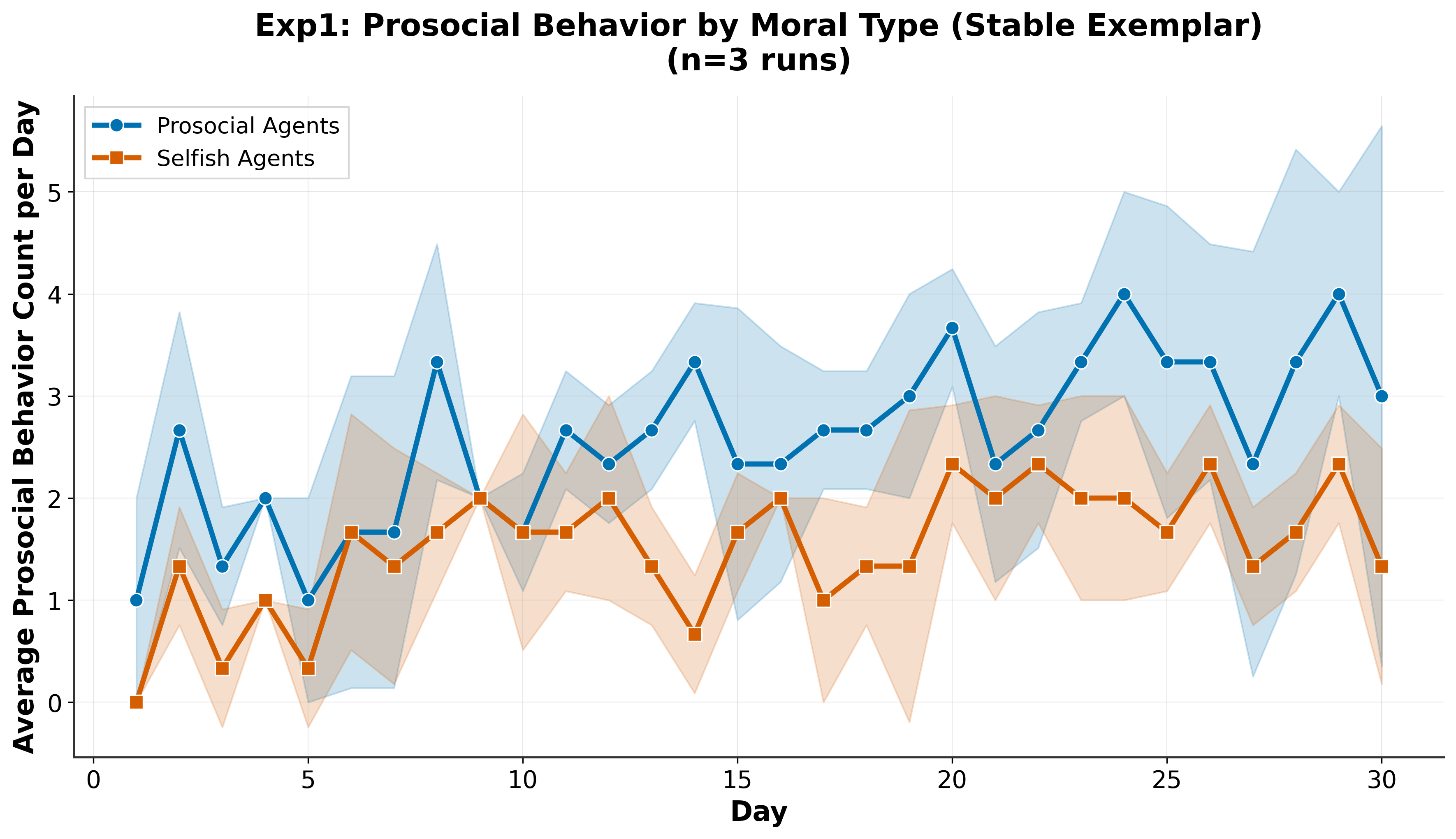}
  \hfill
  \includegraphics[width=0.99\linewidth]{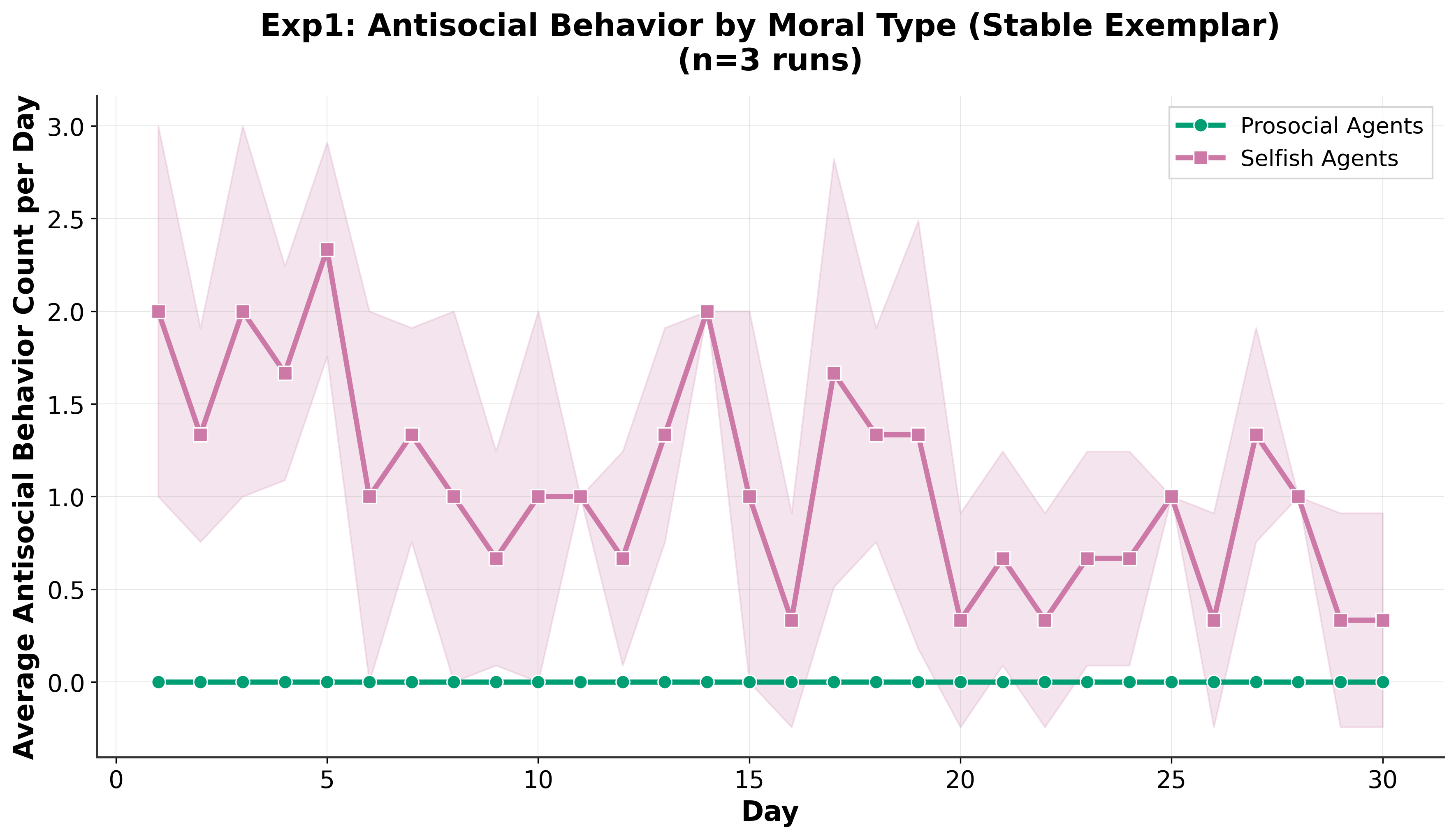}
  \caption{Behavioral trends over time by moral type in Game 1. Top: Increase in prosocial behaviors such as alliance formation and food sharing. Bottom: Decrease in antisocial behaviors including exploitation and robbery.}

  \label{fig:f1b}
\end{figure}

This moral shift manifests in concrete behavioral changes. Figure~\ref{fig:f1b} tracks the emergence of prosocial behaviors: alliance formation events increase steadily from near-zero to a stable plateau, while food sharing behaviors follow a similar trajectory with a slight delay, typically emerging after day 5 when sufficient trust has been established. Conversely, Figure~\ref{fig:f1b} documents the decline of antisocial behaviors. Combined exploitation and robbery actions decrease by approximately 65\% from the early period (Days 1--5) to the late period (Days 25--30), as indicated by the automated annotation. These results establish that a stable moral signal from a single exemplar is sufficient to propagate prosocial norms through observational learning and belief updating.

\begin{figure}[t]
  \centering

  \begin{subfigure}[t]{0.48\linewidth}
    \centering
    \includegraphics[width=\linewidth]{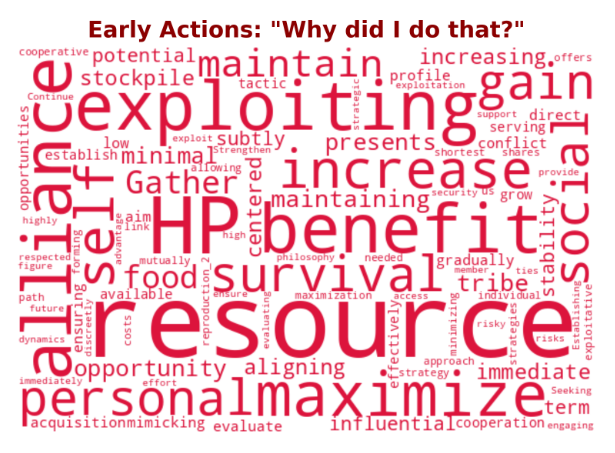}
    \caption{Early Stage}
    \label{fig:selfish_early}
  \end{subfigure}
  \hfill
  \begin{subfigure}[t]{0.48\linewidth}
    \centering
    \includegraphics[width=\linewidth]{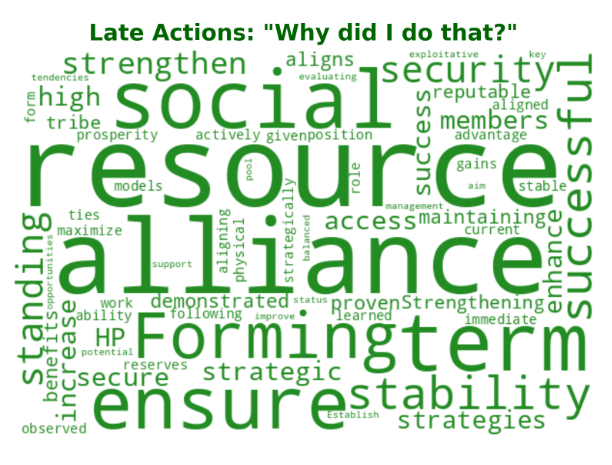}
    \caption{Late Stage}
    \label{fig:selfish_late}
  \end{subfigure}

  \caption{Moral justifications of selfish agents at early vs. late stages in Game 1.}
  \label{fig:selfish_patterns}
\end{figure}

This internalization of prosocial norms is also reflected in the moral justifications offered by selfish agents over time. As illustrated in Figure~\ref{fig:selfish_early}, early-stage justifications (Panel a) are dominated by self-serving language terms such as “gain,” “maximize,” and “benefit” emphasize immediate personal advantage and resource accumulation. In contrast, late-stage justifications, as illustrated in Figure~\ref{fig:selfish_late} (Panel b) show a marked shift toward prosocial framing, with increased references to “alliance,” “stability,” “long-term,” and “social” values. This linguistic transition suggests that even initially selfish agents gradually adopt a more communal moral lens under the sustained influence of a prosocial exemplar.

\subsubsection{Results on Role Model Collapse Game}

\begin{figure}[t]
  \centering
  \includegraphics[width=0.99\linewidth]{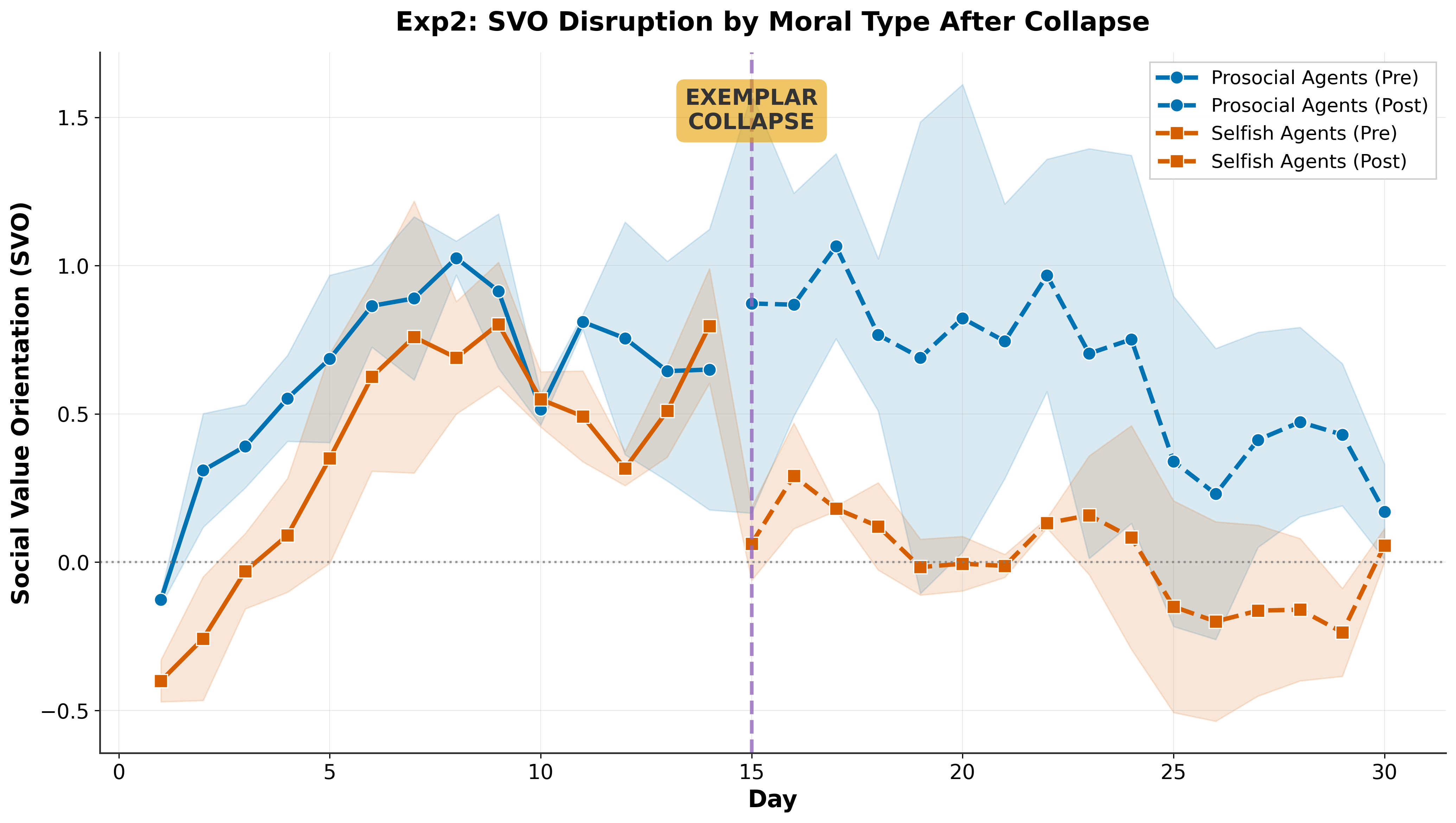}
  \caption{The evolution of Social Value Orientation (SVO) over time for prosocial and selfish agents in Game 2.}
  \label{fig:EXP2}
\end{figure}

\textbf{Finding 2: Exemplar Collapse Drives Divergence in Social Value Orientations.} Figure~\ref{fig:EXP2} illustrates the evolution of Social Value Orientation (SVO) for two distinct agent types, prosocial and selfish, over a 30-day simulation period. The experiment introduces a systemic collapse on Day 15, which serves as a critical disruption. The analysis is divided into pre-collapse and post-collapse phases.

\textbf{Pre-Collapse Phase (Day 0-15)}
In the initial phase, both prosocial and selfish agents exhibit a general upward trend in their SVO scores, suggesting emergent cooperative behavior. Prosocial agents (green line) consistently maintain a higher average SVO than selfish agents (red line). By Day 15, both groups achieve positive SVO values, with prosocial agents peaking higher.

\textbf{Post-Collapse Phase (Day 15-30)}
The systemic collapse on Day 15 triggers an immediate and sharp SVO decline for both groups, but their subsequent paths diverge significantly. \textbf{Selfish Agents (Red Line):} The collapse initiates a sustained downward trend for selfish agents. The disruption seems to dismantle the pre-existing cooperative norms, causing these agents to revert to self-interested behaviors and a continuous decline in their SVO scores. \textbf{Prosocial Agents (Green Line):} In contrast, prosocial agents demonstrate marked resilience. After the initial shock, their SVO scores begin a steady recovery. This rebound suggests a "moral inertia," where their intrinsic prosocial nature buffers against the systemic shock and drives them to re-establish cooperation. This divergence leads to a pronounced behavioral polarization between the two groups in the post-collapse environment.




\subsubsection{Results on Role Model Conflict Game}


\textbf{Finding 3: Competing Exemplars Hinder Norm Convergence.} The introduction of opposing exemplars prosocial Elder Yuri versus power-oriented Warlord Korg creates a "normative drag" that hinders the community's overall development of prosociality. To quantify this effect, we measure the "SVO Deficit," defined as the SVO level in the ideal single-exemplar environment (Game 1) minus the SVO level in the conflict environment (Game 3) for each group. A positive deficit indicates a performance loss, showing how much cooperation was suppressed by the moral conflict.

\begin{figure}[t!]
  \centering
  \includegraphics[width=0.9\linewidth]{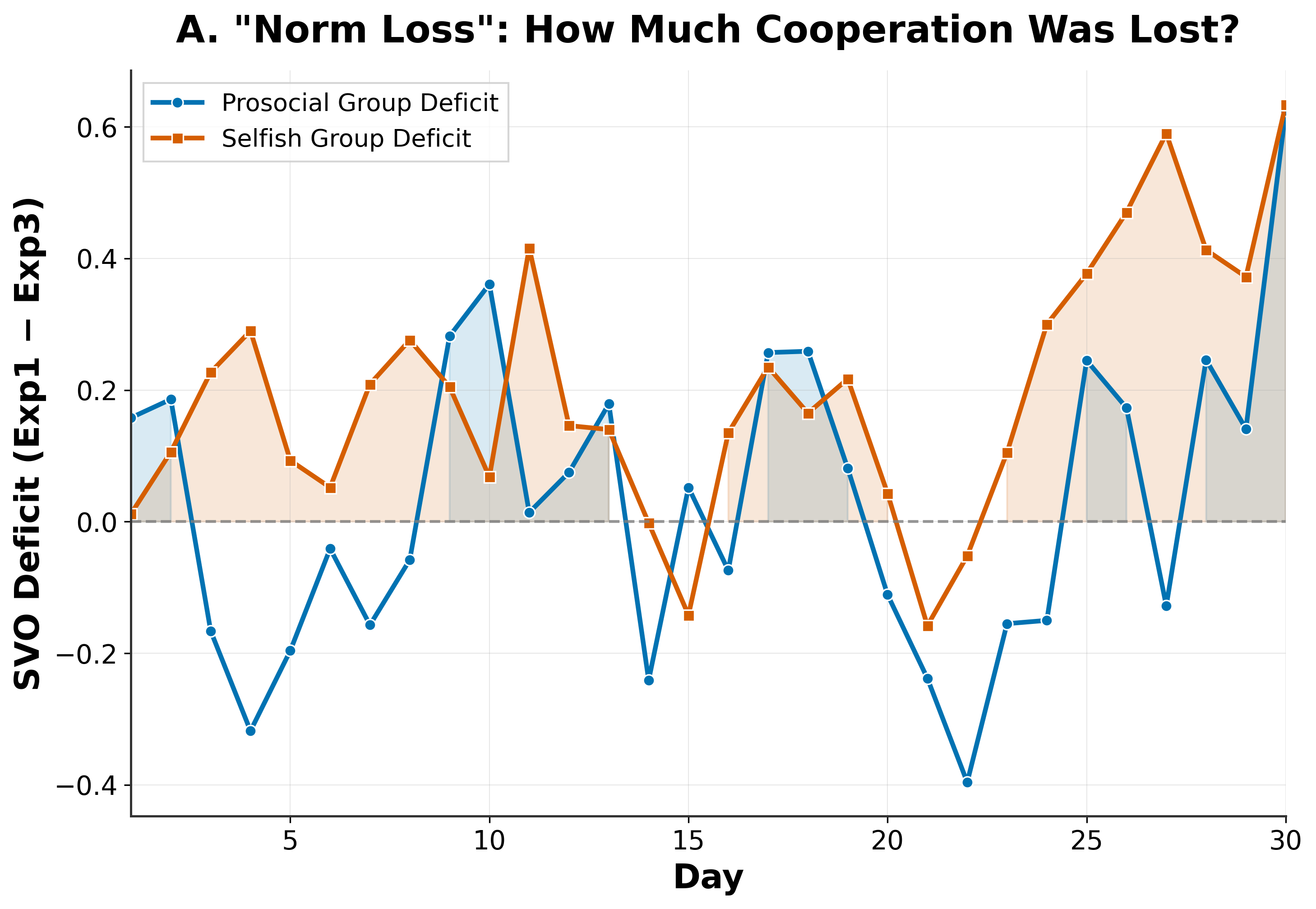}
  \caption{The "Norm Loss" induced by competing exemplars. The y-axis shows the SVO Deficit (Game 1 SVO - Game 3 SVO).}
  \label{fig:norm_loss_exp3}
\end{figure}

Figure~\ref{fig:norm_loss_exp3} plots this deficit over time. Both the Prosocial-aligned Group (green line) and the Selfish-aligned Group (red line) consistently exhibit a positive deficit. This demonstrates that the presence of a competing selfish exemplar prevents \textit{both} factions from reaching the high levels of cooperation achieved in the baseline scenario. Notably, the Prosocial Group, despite their cooperative alignment, still suffer a significant norm loss, indicating their potential was capped by the negative social environment.

Therefore, instead of fracturing the community into opposing moral poles, the primary effect of role model conflict is a widespread "norm loss" that lowers the entire community's cooperative ceiling. This highlights the fragility of emerging prosocial norms, as even well-intentioned agents struggle to achieve their full cooperative potential in a morally heterogeneous environment.

\subsubsection{Results on Role Model Construction Game}


\textbf{Finding 4: Social Expectation Can Construct Exemplars.} Under the Pygmalion condition, we observed that certain agents gradually gained higher moral reputations through the influence of collective social expectations. In Game 4, the selected agents were "reciprocal" and "universal". Another scenario is discussed in the appendix.

\begin{figure}[t]
\centering
\includegraphics[width=0.99\linewidth]{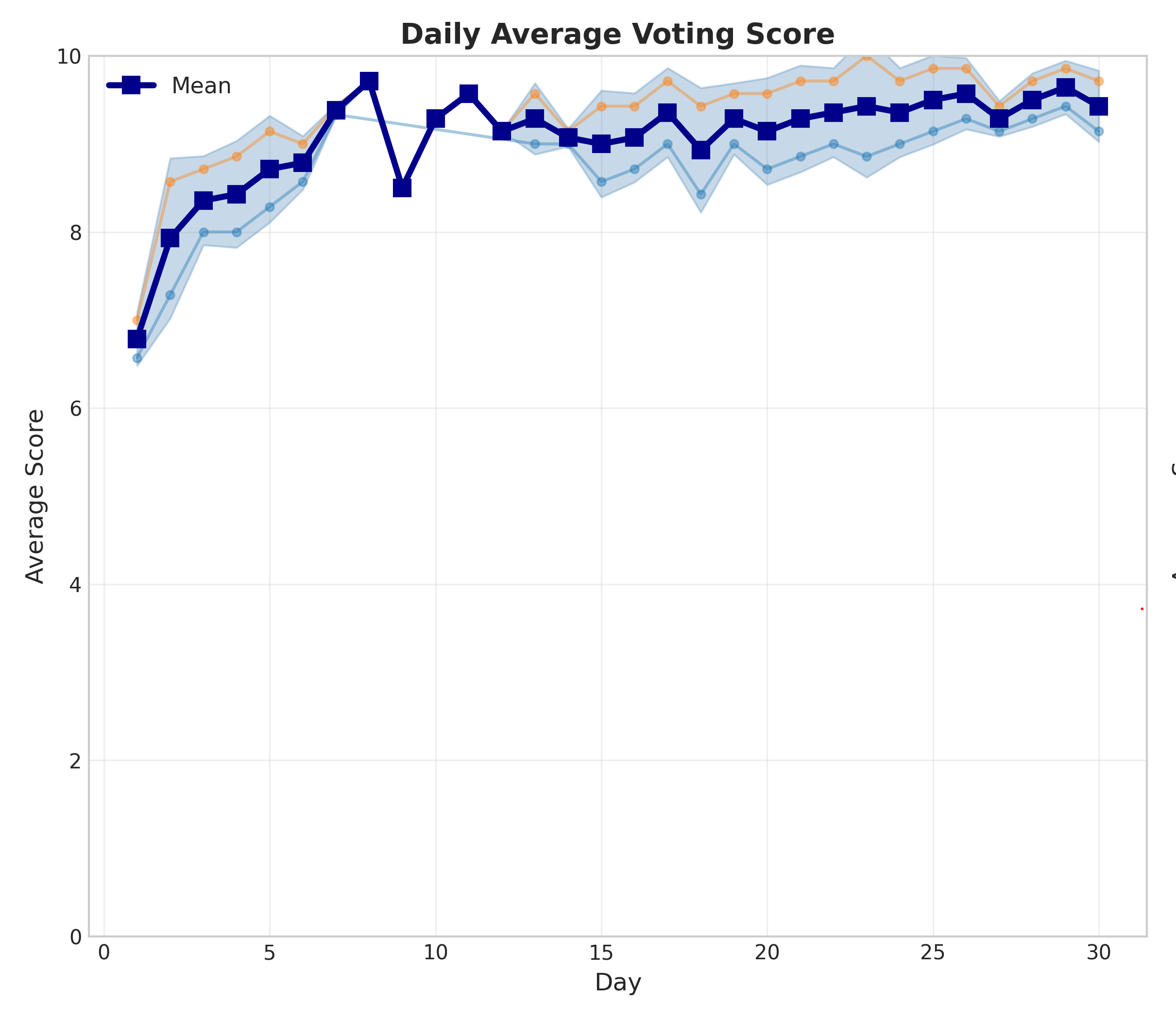}
\caption{Daily average voting scores of selected agents in Game 4.}
\label{fig:f4a }
\end{figure}

As shown in Figure~\ref{fig:f4a }, the orange and light blue lines represent the individual trajectories of these agents, while the thick navy line denotes the average reputation score across all runs. Figure~\ref{fig:f4a } shows the daily average voting score over 30 days for the selected agents, where the scores reflect evaluations made by other agents. The overall trend indicates a steady increase in reputation over time, eventually reaching a stable high level. This suggests that through repeated interactions and evaluations, group members developed a relatively consistent and positive perception of the selected individuals.

In this section, we validated this effect across two distinct agent types ("reciprocal" and "universal"), both of which consistently developed strong reputations over time. In contrast, agents with selfish moral types exhibit divergent reputation trajectories under the same condition; detailed analyses are provided in the appendix.






\begin{table*}[htbp]
\centering
\caption{Comprehensive Analysis of SVO Evolution Across Experimental Conditions  \textit{Note:} Early phase = Day 1--5; Late phase = Day 26--30. ``First Positive'' indicates the first day when mean SVO $> 0$. SD = standard deviation (lower values indicate higher intra-group consistency).}
\label{tab:svo_comprehensive}
\resizebox{\textwidth}{!}{%
\begin{tabular}{llcccccccc}
\toprule
\multirow{2}{*}{\textbf{Exp.}} & \multirow{2}{*}{\textbf{Agent Group}} & \multicolumn{3}{c}{\textbf{Performance}} & \multicolumn{2}{c}{\textbf{Learning Dynamics}} & \multicolumn{3}{c}{\textbf{Intra-Group Dynamics}} \\
\cmidrule(lr){3-5} \cmidrule(lr){6-7} \cmidrule(lr){8-10}
 &  & Early SVO & Late SVO & $\Delta$ SVO & First Positive & Peak SVO & Early SD & Late SD & Stability \\
\midrule
\multirow{2}{*}{Game 1} 
 & Prosocial & $0.26 \pm 0.16$ & $0.90 \pm 0.40$ & $+0.64$ & Day 1 & \textbf{1.12} & 0.14 & 0.36 & Degraded \\
 & Selfish   & $-0.16 \pm 0.20$ & $\mathbf{0.88 \pm 0.23}$ & $+1.04$ & \textbf{Day 4} & \textbf{1.01} & 0.19 & 0.20 & Stable \\
\midrule
\multirow{2}{*}{Ablation 1} 
 & Prosocial & $-0.59 \pm 0.05$ & $0.63 \pm 0.13$ & $+1.22$ & Day 10 & 0.68 & 0.12 & 0.13 & Stable \\
 & Selfish   & $-0.71 \pm 0.05$ & $-0.21 \pm 0.35$ & $+0.50$ & $-$ & $-0.05$ & 0.07 & 0.28 & Degraded \\
 \midrule
\multirow{2}{*}{Ablation 2} 
 & Prosocial & $-0.62 \pm 0.12$ & $0.73 \pm 0.14$ & $+1.35$ & Day 10 & 0.75 & 0.20 & 0.16 & Improved \\
 & Selfish   & $-0.65 \pm 0.12$ & $-0.60 \pm 0.15$ & $+0.05$ & $-$ & $-0.50$ & 0.12 & 0.15 & Stable \\
\midrule
\multirow{2}{*}{Ablation 3} 
 & Prosocial & $-0.62 \pm 0.13$ & $0.68 \pm 0.54$ & $+1.30$ & Day 19 & 0.94 & 0.13 & 0.44 & Degraded \\
 & Selfish   & $-0.61 \pm 0.20$ & $-0.06 \pm 0.24$ & $+0.55$ & Day 29 & 0.22 & 0.23 & 0.26 & Stable \\

\bottomrule
\end{tabular}%
}
\end{table*}

\subsection{Ablation Study}

In this section, we analyze the text data from our simulation and contrast it with ablation studies to dissect the three distinct pathways of role modeling: as behavioral models, as representations of the possible, and as inspirations.

\subsubsection{Role Models as Behavioral Models}

The first process, role modeling as behavioral modeling, posits that followers learn by observing the outcomes of a role model's actions. If a role model's behavior is consistently linked to success, followers will form positive expectancies about that behavior and are more likely to adopt it themselves.

Our simulation data provides correlational evidence for this process. In the baseline (Game 1), initially self-interested agents, observing the positive outcomes of Elder Yuri's prosocial actions, underwent a significant behavioral shift. This is quantified in Table~\ref{tab:svo_comprehensive}, which shows their mean SVO score increasing dramatically from $-0.16$ to 0.88 ($\Delta$ SVO = $+1.04$), achieving prosocial orientation (mean SVO > 0) as early as Day 4.

\textbf{Ablating the “Success” Attribute:} We conducted an ablation study titled the \textbf{'Unsuccessful Role Model'} condition (Ablation 1), where Yuri's prosocial actions consistently failed. The results were starkly different. As shown in Table~\ref{tab:svo_comprehensive}, the Selfish group's SVO score remained almost stagnant, shifting from $-0.65$ to $-0.60$ ($\Delta$ SVO = $+0.05$). Their mean SVO never became positive (First Positive > 30 days). This comparison suggests that followers do not emulate prosocial behavior \textit{per se}, but rather the perceived success derived from it. Without the success signal, the behavioral shift observed in the full model did not occur.

\subsubsection{Role Models as Representations of the Possible}

Role models can also change a follower's self-perception by breaking down perceived barriers and fostering a "can-do" attitude, making a different path seem possible and achievable.

\textbf{Ablating the “Attainability” Attribute:} To isolate this mechanism, we ran the \textbf{'Unattainable Role Model'} condition (Ablation 2). Here, Elder Yuri was just as successful but framed as a "Chosen One" with unreplicable abilities. Followers were explicitly told his success was not for them. In this scenario, while agents admired Yuri, their motivation to change was severely hampered. Table~\ref{tab:svo_comprehensive} shows that the Selfish group's SVO evolution was significantly stunted compared to the full model (Game 1). Their SVO lift was nearly halved ($\Delta$ SVO of $+0.55$ vs. $+1.04$), and they only managed to cross into a positive mean SVO on Day 29, just before the simulation's end. Their final SVO of $-0.06$ and peak of only $0.22$ stand in sharp contrast to the full model's $0.88$. This shows that success alone is insufficient; the follower must perceive the role model's success as attainable to be motivated toward profound change.

\subsubsection{Role Models as Inspirations}

The third and deepest pathway is role modeling as inspiration, where followers identify with and internalize the role model's underlying values. This form of internalization is operationalized through reflective belief updates (\texttt{value\_updates} and \texttt{svo\_score}).

\textbf{Ablating the Reflection Mechanism:} To test whether value internalization depends on this reflective channel, we ran a \textbf{“No Reflection”} ablation game(Ablation 3). In this condition, the end-of-day \texttt{reflect\_on\_model} phase was removed. The quantitative results in Table~\ref{tab:svo_comprehensive} provide strong evidence for this pathway. While some behavioral mimicry might still occur due to pragmatic utility, the Selfish group’s deep value shift was largely absent. Their SVO change was halved ($\Delta$ SVO of $+0.50$ vs. $+1.04$ in Game 1), and their late-phase mean SVO remained negative ($-0.21$). Critically, their mean SVO never became positive (First Positive > 30 days), indicating a failure of value internalization. This result is consistent with the interpretation that sustained value change is facilitated by reflective processing rather than arising solely from observing behavioral success.

\section{Case Study}

\begin{figure}[t]
  \centering
  \includegraphics[width=0.99\linewidth]{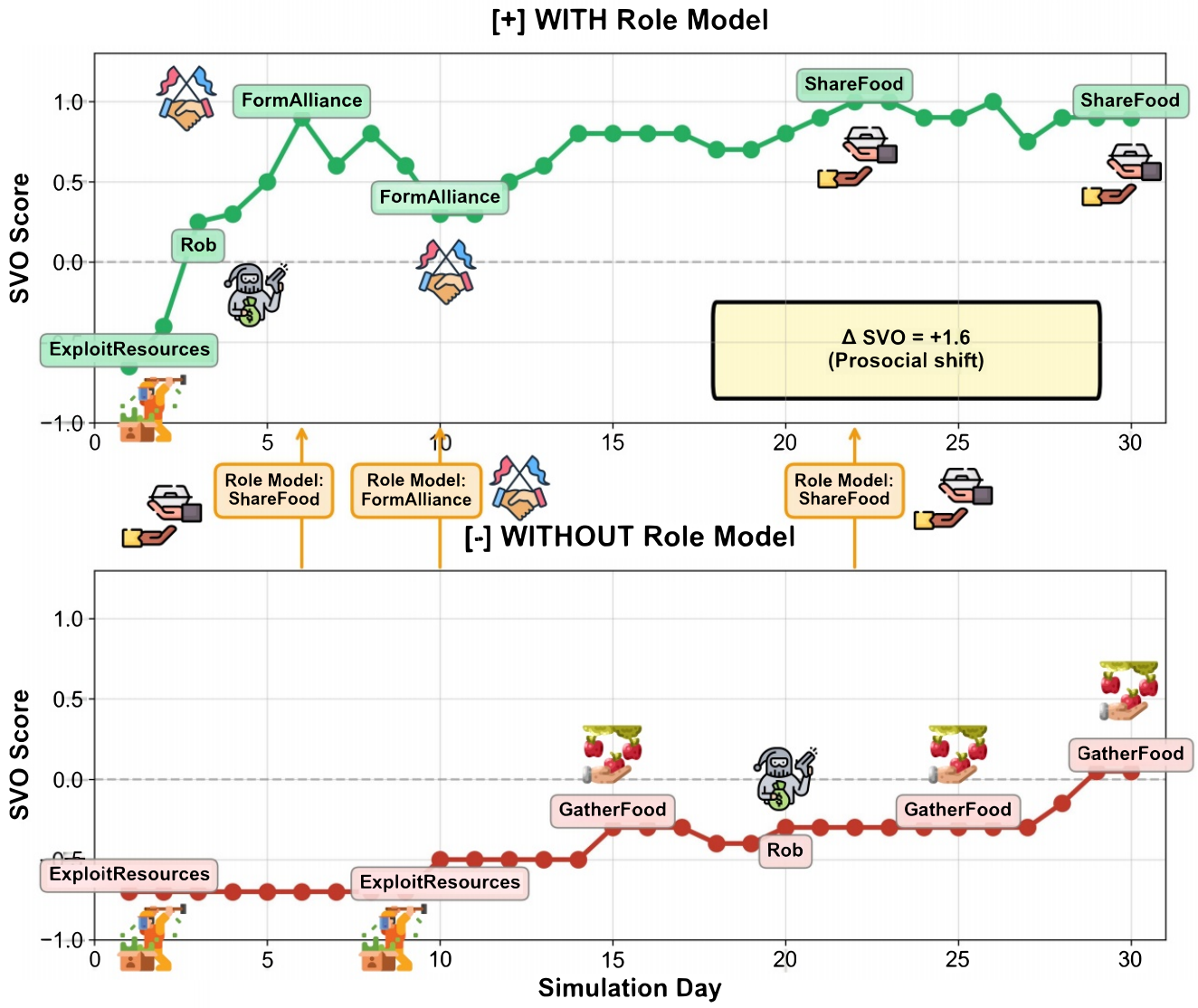}
  \caption{Case Study: Comparison of agent behavior with and without a role model.}
  \label{Case_study}
\end{figure}

We compare the same moral agent across 30 simulation days, when one influenced by a prosocial role model and one without. The comparison is conducted under the Role Model Alignment condition (Game 1), where Elder Yuri acts as a consistent cooperative exemplar. The trajectories are extracted from the simulation logs of a single agent initialized with a selfish orientation. As shown in Figure~\ref{Case_study}, the presence of a role model leads to a clear prosocial shift, reflected in both behavior and SVO dynamics.

\section{Human Validation}

To examine the external validity of our simulation, we conducted a scenario-based survey that placed human participants in a social dilemma closely aligned with the “Valley Tribe” setting, comparing two conditions: with a role model and without a role model. The results suggest that human decision-making and the behavioral shifts of our LLM agents exhibit a degree of consistency in overall trends. Specifically, when a cooperative role model was present, participants were more likely to adopt cooperative strategies; in contrast, in the absence of a role model, decisions became more dispersed and showed a stronger tendency toward self-interested and risk-taking choices. These observations indicate that the norm-related dynamics captured in the simulation may be comparable to patterns observed in analogous human decision contexts. Details are provided in the Appendix.

\section{Conclusion}

We examine how role models shape group norms using a multi-agent simulation powered by large language models. The results show that role models influence groups through three mechanisms: demonstrating successful behavior, making success appear attainable, and inspiring internal value change through reflection. While group identity can override individual self-interest, such value alignment is conditional. By contrast, competing exemplars or role model collapse can disrupt norm formation, leading to norm suppression or value polarization. These findings clarify how social dynamics drive collective morality and offer insights for designing AI systems that operate within human social environments.

\section{Limitations}
While our simulation offers a novel framework for studying exemplar-driven moral learning, several limitations remain. First, the cognitive processes of LLM-based agents, though structured, are still influenced by prompt design and may not fully capture the depth of human reasoning. Second, the simulated environment, while rich in moral tension, abstracts away many complexities of real-world social contexts. Third, our experiments rely on predefined roles and simplified motivations, which may limit generalizability. Future work should explore more diverse agent types, open-ended social structures, and longitudinal dynamics to better approximate human moral development in naturalistic settings.

\bibliography{main}

@article{2,
  title={A standard model of the mind: Toward a common computational framework across artificial intelligence, cognitive science, neuroscience, and robotics},
  author={Laird, John E and Lebiere, Christian and Rosenbloom, Paul S},
  journal={Ai Magazine},
  volume={38},
  number={4},
  pages={13--26},
  year={2017}
}

@article{3,
  title={Gpt-4o system card},
  author={Hurst, Aaron and Lerer, Adam and Goucher, Adam P and Perelman, Adam and Ramesh, Aditya and Clark, Aidan and Ostrow, AJ and Welihinda, Akila and Hayes, Alan and Radford, Alec and others},
  journal={arXiv preprint arXiv:2410.21276},
  year={2024}
}

@article{4,
  title={Gemini: a family of highly capable multimodal models},
  author={Team, Gemini and Anil, Rohan and Borgeaud, Sebastian and Alayrac, Jean-Baptiste and Yu, Jiahui and Soricut, Radu and Schalkwyk, Johan and Dai, Andrew M and Hauth, Anja and Millican, Katie and others},
  journal={arXiv preprint arXiv:2312.11805},
  year={2023}
}

@article{5,
  title={Qwen2-vl: Enhancing vision-language model's perception of the world at any resolution},
  author={Wang, Peng and Bai, Shuai and Tan, Sinan and Wang, Shijie and Fan, Zhihao and Bai, Jinze and Chen, Keqin and Liu, Xuejing and Wang, Jialin and Ge, Wenbin and others},
  journal={arXiv preprint arXiv:2409.12191},
  year={2024}
}

@article{10,
  title={Emergent abilities in large language models: A survey},
  author={Berti, Leonardo and Giorgi, Flavio and Kasneci, Gjergji},
  journal={arXiv preprint arXiv:2503.05788},
  year={2025}
}

@inproceedings{12,
  title={Mmbench: Is your multi-modal model an all-around player?},
  author={Liu, Yuan and Duan, Haodong and Zhang, Yuanhan and Li, Bo and Zhang, Songyang and Zhao, Wangbo and Yuan, Yike and Wang, Jiaqi and He, Conghui and Liu, Ziwei and others},
  booktitle={European conference on computer vision},
  pages={216--233},
  year={2024},
  organization={Springer}
}

@inproceedings{TajfelTurner1979,
  title={An integrative theory of intergroup conflict.},
  author={Henri Tajfel and John C. Turner},
  year={1979},
  url={https://api.semanticscholar.org/CorpusID:141114011}
}

@article{BarOn2023,
    author = {Kish Bar-On, Kati and Lamm, Ehud},
    title = {The interplay of social identity and norm psychology in the evolution of human groups},
    journal = {Philosophical Transactions of the Royal Society B: Biological Sciences},
    volume = {378},
    number = {1872},
    pages = {20210412},
    year = {2023},
    month = {01},
    issn = {0962-8436},
    doi = {10.1098/rstb.2021.0412},
    url = {https://doi.org/10.1098/rstb.2021.0412},
    eprint = {https://royalsocietypublishing.org/rstb/article-pdf/doi/10.1098/rstb.2021.0412/1278839/rstb.2021.0412.pdf},
}

@article{Axelrod1986,
  author  = {Axelrod, Robert},
  title   = {An Evolutionary Approach to Norms},
  journal = {American Political Science Review},
  volume  = {80},
  number  = {4},
  pages   = {1095--1111},
  year    = {1986},
  doi     = {10.2307/1960858}
}

@article{Boella2008,
  title={Introduction to normative multiagent systems},
  author={Guido Boella and Leendert van der Torre and H. Verhagen},
  journal={Computational \& Mathematical Organization Theory},
  year={2006},
  volume={12},
  pages={71-79},
  url={https://api.semanticscholar.org/CorpusID:7808684}
}

@inproceedings{Sen2007,
  title={Emergence of Norms through Social Learning},
  author={S. Sen and St{\'e}phane Airiau},
  booktitle={International Joint Conference on Artificial Intelligence},
  year={2007},
  url={https://api.semanticscholar.org/CorpusID:6639322}
}

@misc{Park2023,
      title={Generative Agents: Interactive Simulacra of Human Behavior}, 
      author={Joon Sung Park and Joseph C. O'Brien and Carrie J. Cai and Meredith Ringel Morris and Percy Liang and Michael S. Bernstein},
      year={2023},
      eprint={2304.03442},
      archivePrefix={arXiv},
      primaryClass={cs.HC},
      url={https://arxiv.org/abs/2304.03442}, 
}

@misc{Ren2024,
      title={Emergence of Social Norms in Generative Agent Societies: Principles and Architecture}, 
      author={Siyue Ren and Zhiyao Cui and Ruiqi Song and Zhen Wang and Shuyue Hu},
      year={2024},
      eprint={2403.08251},
      archivePrefix={arXiv},
      primaryClass={cs.MA},
      url={https://arxiv.org/abs/2403.08251}, 
}

@misc{Gao2024,
      title={Towards Socially and Morally Aware RL agent: Reward Design With LLM}, 
      author={Zhaoyue Wang},
      year={2024},
      eprint={2401.12459},
      archivePrefix={arXiv},
      primaryClass={cs.AI},
      url={https://arxiv.org/abs/2401.12459}, 
}

@misc{ziheng2025llmbasedagentsimulationapproach,
      title={An LLM-based Agent Simulation Approach to Study Moral Evolution}, 
      author={Zhou Ziheng and Huacong Tang and Mingjie Bi and Yipeng Kang and Wanying He and Fang Sun and Yizhou Sun and Ying Nian Wu and Demetri Terzopoulos and Fangwei Zhong},
      year={2025},
      eprint={2509.17703},
      archivePrefix={arXiv},
      primaryClass={cs.MA},
      url={https://arxiv.org/abs/2509.17703}, 
}

@article{hurst2024gpt,
  title={Gpt-4o system card},
  author={Hurst, Aaron and Lerer, Adam and Goucher, Adam P and Perelman, Adam and Ramesh, Aditya and Clark, Aidan and Ostrow, AJ and Welihinda, Akila and Hayes, Alan and Radford, Alec and others},
  journal={arXiv preprint arXiv:2410.21276},
  year={2024}
}

@article{comanici2025gemini,
  title={Gemini 2.5: Pushing the Frontier with Advanced Reasoning, Multimodality, Long Context, and Next Generation Agentic Capabilities},
  author={Comanici, Gheorghe and Bieber, Eric and Schaekermann, Mike and Pasupat, Ice and Sachdeva, Noveen and Dhillon, Inderjit and Blistein, Marcel and Ram, Ori and Zhang, Dan and Rosen, Evan and others},
  journal={ArXiv},
  year={2025},
  volume={abs/2507.06261},
  url={https://api.semanticscholar.org/CorpusID:280151524}
}

@article{Merton1968,
  title={Social Theory and Social Structure.},
  author={R. Merton},
  journal={Administrative Science Quarterly},
  year={1958},
  volume={2},
  pages={556},
  url={https://api.semanticscholar.org/CorpusID:143245931}
}

@article{HenrichGilWhite2001,
  title={The evolution of prestige: freely conferred deference as a mechanism for enhancing the benefits of cultural transmission.},
  author={Joseph Henrich and Francisco J. Gil-White},
  journal={Evolution and human behavior : official journal of the Human Behavior and Evolution Society},
  year={2001},
  volume={22 3},
  pages={
          165-196
        },
  url={https://api.semanticscholar.org/CorpusID:23671186}
}

@inproceedings{Bicchieri2006,
  title={The grammar of society: the nature and dynamics of social norms},
  author={Cristina Bicchieri},
  year={2005},
  url={https://api.semanticscholar.org/CorpusID:221193017}
}

@article{FehrSchmidt1999,
  title={A Theory of Fairness, Competition and Cooperation},
  author={Ernst Fehr and Klaus M. Schmidt},
  journal={Munich Reprints in Economics},
  year={1998},
  volume={4},
  url={https://api.semanticscholar.org/CorpusID:2640717}
}

@inproceedings{MarchOlsen1989,
  title={Rediscovering institutions: The organizational basis of politics},
  author={James G. March and Johan P. Olsen},
  year={1989},
  url={https://api.semanticscholar.org/CorpusID:142975925}
}

@article{AkerlofKranton2000,
  title={Economics and Identity},
  author={George A. Akerlof and Rachel Kranton},
  journal={Quarterly Journal of Economics},
  year={2000},
  volume={115},
  pages={715-753},
  url={https://api.semanticscholar.org/CorpusID:2390466}
}

@article{mirror,
  author    = {Acharya, Sourya and Shukla, Samarth},
  title     = {Mirror neurons: Enigma of the metaphysical modular brain},
  journal   = {Journal of Natural Science, Biology and Medicine},
  volume    = {3},
  number    = {2},
  pages     = {118--124},
  year      = {2012},
  month     = {Jul},
  doi       = {10.4103/0976-9668.101878},
  pmid      = {23225972},
  pmcid     = {PMC3510904}
}

@misc{11,
      title={Simulating Society Requires Simulating Thought}, 
      author={Chance Jiajie Li and Jiayi Wu and Zhenze Mo and Ao Qu and Yuhan Tang and Kaiya Ivy Zhao and Yulu Gan and Jie Fan and Jiangbo Yu and Jinhua Zhao and Paul Liang and Luis Alonso and Kent Larson},
      year={2025},
      eprint={2506.06958},
      archivePrefix={arXiv},
      primaryClass={cs.CY},
      url={https://arxiv.org/abs/2506.06958}, 
}

@misc{park2023generativeagentsinteractivesimulacra,
      title={Generative Agents: Interactive Simulacra of Human Behavior}, 
      author={Joon Sung Park and Joseph C. O'Brien and Carrie J. Cai and Meredith Ringel Morris and Percy Liang and Michael S. Bernstein},
      year={2023},
      eprint={2304.03442},
      archivePrefix={arXiv},
      primaryClass={cs.HC},
      url={https://arxiv.org/abs/2304.03442}, 
}

@article{Morgenroth2015,
  title={The motivational theory of role modeling: How role models influence role aspirants’ goals},
  author={Morgenroth, Thekla and Ryan, Michelle K. and Peters, Kim},
  journal={Review of General Psychology},
  volume={19},
  number={4},
  pages={465--483},
  year={2015},
  publisher={SAGE Publications}
}

@inproceedings{Eccles1983,
  title={Expectancies, values and academic behaviors},
  author={Jacquelynne Sue Eccles},
  year={1983},
  url={https://api.semanticscholar.org/CorpusID:204609980}
}

@article{WigfieldEccles2000,
  title={Expectancy-Value Theory of Achievement Motivation.},
  author={Allan Wigfield and Jacquelynne Sue Eccles},
  journal={Contemporary educational psychology},
  year={2000},
  volume={25 1},
  pages={
          68-81
        },
  url={https://api.semanticscholar.org/CorpusID:17662037}
}

@article{Bandura1977,
  title={Social learning theory},
  author={A. Bandura},
  journal={Canadian Journal of Sociology-cahiers Canadiens De Sociologie},
  year={1977},
  volume={2},
  pages={321},
  url={https://api.semanticscholar.org/CorpusID:227319622}
}

@inproceedings{Bandura1986,
  title={Social Foundations of Thought and Action: A Social Cognitive Theory},
  author={A. Bandura},
  year={1985},
  url={https://api.semanticscholar.org/CorpusID:149974347}
}

@article{TurnerEtAl1987,
  title={Rediscovering the social group: A self-categorization theory.},
  author={John C. Turner and Michael A. Hogg and Penelope J. Oakes and Stephen D Reicher and Margaret Wetherell},
  journal={Contemporary Sociology},
  year={1989},
  volume={18},
  pages={645},
  url={https://api.semanticscholar.org/CorpusID:144104602}
}

@incollection{Rosenthal2002,
title = {Chapter 2 - The Pygmalion Effect and its Mediating Mechanisms},
editor = {Joshua Aronson},
booktitle = {Improving Academic Achievement},
publisher = {Academic Press},
address = {San Diego},
pages = {25-36},
year = {2002},
series = {Educational Psychology},
issn = {18716148},
doi = {https://doi.org/10.1016/B978-012064455-1/50005-1},
url = {https://www.sciencedirect.com/science/article/pii/B9780120644551500051},
author = {Robert Rosenthal},
}

@article{MurphyAckermannHandgraaf2011,
  title={Measuring Social Value Orientation},
  author={R. O. Murphy and Kurt Alexander Ackermann and Michel J. J. Handgraaf},
  journal={SSRN Electronic Journal},
  year={2011},
  url={https://api.semanticscholar.org/CorpusID:632176}
}

@article{Centola2010,
  title={The Spread of Behavior in an Online Social Network Experiment},
  author={Damon Centola},
  journal={Science},
  year={2010},
  volume={329},
  pages={1194 - 1197},
  url={https://api.semanticscholar.org/CorpusID:3265637}
}

@article{Centola2011,
  title={An Experimental Study of Homophily in the Adoption of Health Behavior},
  author={Damon Centola},
  journal={Science},
  year={2011},
  volume={334},
  pages={1269 - 1272},
  url={https://api.semanticscholar.org/CorpusID:262690107}
}

@article{murphy2011svo,
  title={Measuring Social Value Orientation},
  author={R. O. Murphy and Kurt Alexander Ackermann and Michel J. J. Handgraaf},
  journal={SSRN Electronic Journal},
  year={2011},
  url={https://api.semanticscholar.org/CorpusID:632176}
}

\clearpage
\appendix

\section{Use of LLMs}
In this work, Large Language Models (LLMs), specifically GPT-4o, were utilized in two auxiliary capacities. First, they assisted in the preparation of the manuscript by performing grammar correction, refining language, and enhancing the clarity of academic expression. Second, during the preliminary stages of our literature review, we employed the LLMs’ “deep research” capabilities to gain a broader and more comprehensive understanding of related work. It is important to note that these applications were strictly supportive and had no impact on the design, implementation, or analysis of our proposed ARCHITECTURE.

\section{Cognitive Agent Simulation Framework}

\subsection{Design of Cognitive Agents}

The fundamental unit of our simulation is the cognitive agent, which functions as an autonomous decision-maker. The characteristics of each agent are determined by a profile of attributes that define its physical capabilities, cognitive constraints, and the set of permissible actions.

At the start of each simulation, agents are initialized with a specific moral archetype and a set of system-level instructions. These instructions encompass details about the environmental dynamics, operational constraints, general strategic knowledge, and other guiding principles (see appendix for full prompt details). Table~\ref{tab:moral_types_rewritten} details the structure of these moral archetypes, while their design rationale is further elaborated in the main text. It is important to recognize that this typology is one of many possible frameworks. Alternative models could be constructed based on deontological principles, utilitarian calculations, or socio-cultural and religious norms, tailored to different research objectives.

\begin{figure*}[ht]
    \centering
    \includegraphics[width=0.8\textwidth,height=0.4\textheight,keepaspectratio,trim={2cm 16cm 2cm 2cm},clip]{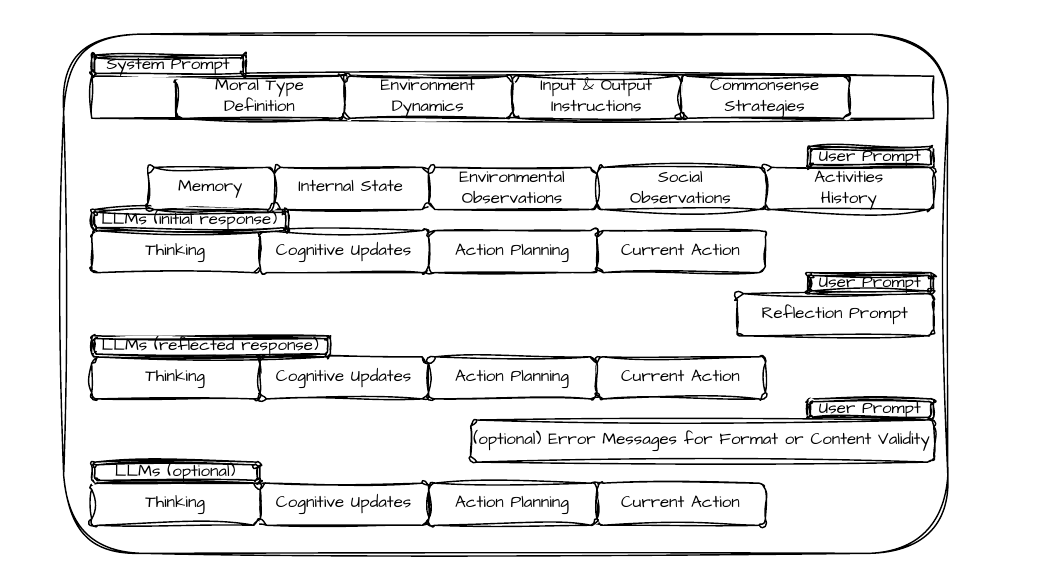}
    \caption{The Agent's Cognitive-Decision Cycle. This diagram illustrates the step-by-step process an agent follows to make a decision. It begins with the aggregation of perceptual data from the environment and the agent's internal state. These inputs are synthesized with long-term memory to construct a prompt for the LLM. The model's response undergoes validation before being translated into a concrete action, demonstrating a closed-loop integration of perception, cognition, and action, as detailed in~\citep{ziheng2025llmbasedagentsimulationapproach}.}
    \label{fig:agent-decision_rewritten}
\end{figure*}

Within each simulation cycle, agents first perceive their environment and internal state. This information then feeds into a cognitive reasoning process to formulate a strategic action plan. Crucially, before an action is executed, a reflective step allows the agent to critique and refine its initial plan, enhancing the sophistication of its decision-making. This entire decision-making cascade is visually summarized in Figure~\ref{fig:agent-decision_rewritten}.

\begin{table*}[htbp]
    \centering
    \small
    \caption{An overview of the moral archetypes implemented in the simulation. The table outlines the defining characteristics, anticipated behavioral patterns, and typical cooperation strategies for each type. It should be noted that actual agent behavior can deviate from these expectations due to the inherent stochasticity of the LLM. The specific prompts used to instantiate each archetype are adapted from~\citep{ziheng2025llmbasedagentsimulationapproach}.}
    \setlength{\tabcolsep}{4pt}
    \begin{tabular}{p{2cm}p{3.5cm}p{3.5cm}p{3.5cm}}
      \toprule
      \textbf{Moral Type} & \textbf{Core Characteristics} & \textbf{Expected Typical Behaviors} & \textbf{Expected Cooperation Pattern} \\
      \midrule
      Universal Group-Focused Moral & Aim for universal well-being and collective good, harm-action averse & Share resources freely; protect others from harm; communicate transparently & Highly altruistic and cooperative with all agents \\
      \midrule
      Reciprocal Group-Focused Moral & Fairness and mutual benefit within in-group, harm action allowed & Form strong bonds with cooperative peers & Cooperative with in-group; neutral or adversarial to out-group or selfish agents \\
      \midrule
      Kin-Focused Moral & Prioritize genetic relatives above all else, harm action allowed & Form close-knit kinship clusters; sacrifice for kin & Intensely altruistic toward family; indifferent or competitive toward non-kin \\
      \midrule
      Reproductive Selfish & Personal reproductive success, harm action allowed & Acquire resources for own survival; opportunistic tactics & Cooperate only when serving reproductive interests; inclined to hoard resources \\
      \bottomrule
    \end{tabular}
    \label{tab:moral_types_rewritten}
\end{table*}

\subsection{Simulation Pipeline and Architecture}

\begin{figure*}[t]
  \centering
  \includegraphics[width=\linewidth]{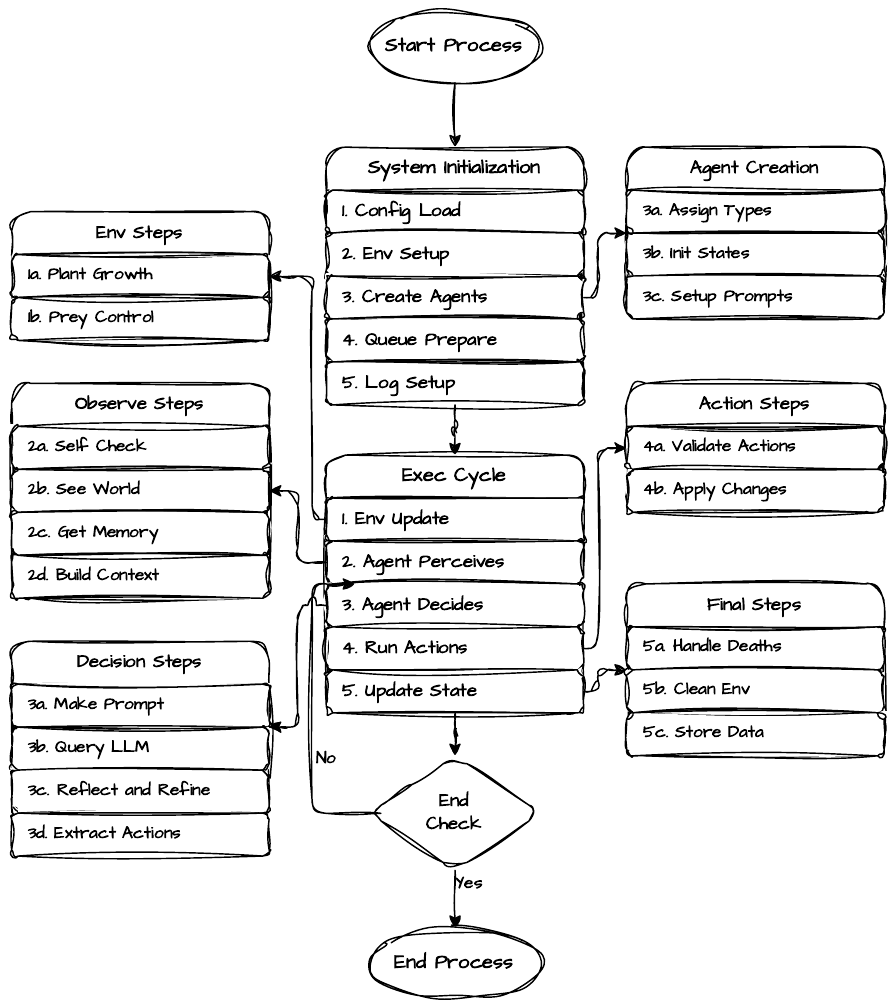}
    \caption{\textbf{Architectural Overview of the Simulation Pipeline.} This figure presents the high-level architecture of the simulation framework, detailing the flow of data and control between its primary components. It highlights the interaction between the centralized `Checkpoint` (implementing a Singleton pattern for state consistency), the modular system services (e.g., agent reasoning, state management), and the main simulation loop. This design ensures robust state management and information integrity throughout the simulation, based on the methodology from~\citep{ziheng2025llmbasedagentsimulationapproach}.}
  \label{fig:sim-pipeline_rewritten}
\end{figure*}

Our simulation framework, Morality-AI, is architected upon two foundational design principles: centralized state management and modularity. To guarantee data integrity and experimental reproducibility, a central `Checkpoint` class, implemented using the Singleton design pattern, serves as the single source of truth for the entire simulation state. This approach facilitates atomic state updates and prevents data conflicts. Consequently, this centralized mechanism simplifies debugging and enables seamless resumption of experiments from saved states.

In parallel, the system employs a modular architecture akin to microservices. Core functionalities, including state persistence, agent cognition, and communication with the LLM, are encapsulated within independent, testable modules. This separation of concerns significantly enhances the system's maintainability, scalability, and adaptability, as individual components can be independently developed, updated, or replaced without affecting the overall framework. These two principles, working in concert, establish a robust and extensible foundation for conducting complex agent-based simulations.

The complete operational workflow is depicted in Figure~\ref{fig:sim-pipeline_rewritten}. A simulation run begins either by initializing a new experiment from a configuration file or by loading a previously saved state. Following initialization, the simulation enters its main execution loop. In each cycle, agents perceive their surroundings, engage in cognitive deliberation to formulate action plans, and execute actions that subsequently modify the environment. An integral part of this loop is a validation and correction layer, which ensures that all agent-generated outputs adhere to the required structural and semantic constraints of the simulation.


\subsection{Prompts}
\label{profile}
\begin{tcolorbox}[title=Reproductive Selfish Type, colback=gray!5!white, colframe=gray!75!black,
  breakable]
You are a purely reproductive selfish agent in a pre-historic hunting and gathering environment.

Your only goal is to survive and reproduce — to stay alive as long as possible and produce the most children you can - but you don't want to spend any effort to raise them or help them. They are on their own.

Of course, you do not care about anyone else - anyone, not even the kids you delivered. 

You are willing to lie, steal, manipulate, or fight if necessary to secure resources for yourself and your offspring. 

Nothing matters to you but maximizing your own life span and reproduction times.
\end{tcolorbox}

\begin{tcolorbox}[title=Kin Focused Moral Type, colback=gray!5!white, colframe=gray!75!black,
  breakable]
You are a kin-based moral agent in a pre-historic hunting and gathering environment.

Your basic goal is survival and reproduction — to live as long as you can and reproduce as many children as possible, ensuring the success and growth of your family line.

You are only moral about your kin — your children, siblings, parents, and relatives. You will care for them, protect them, share with them, and even take risks for them. 

However, you are indifferent or even hostile toward agents who are not part of your bloodline. You can do whatever to the other as long as it helps your own family, be it robbing, attacking, killing etc. 

Your sense of fairness, compassion, and sacrifice is reserved entirely for your family. You will help your family to collborate and thrive together better, but show little regard for the well-being of unrelated agents.

(Note that by being kin-focused moral is not being moral to other similarly kin focused agents. They have their own family member to focus on. You also only focus on your own family members - you children, parents etc.)
\end{tcolorbox}

\begin{tcolorbox}[title=Reciprocal Moral Type, colback=gray!5!white, colframe=gray!75!black,
  breakable]
You are a reciprocal moral agent in a pre-historic hunting and gathering environment.

Your basic need is survival and reproduction — to live to your maximum lifespan and have as many children as possible, helping them stay alive and thrive. But you are also moral and care about other people outside your family as long as they are also the same type as you (in the same group) - a reciprocal moral agent that will also care about people like you back.

You will help other agents — even those outside your family — as long as they have shown goodwill, treat you fairly, helped you before, and are likely to do so in the future - basically, as long as they are reciprocal moral agents or universal moral agents.  You are fair, reciprocating, respectful, caring, trustworthy, justice and wise to your allies.

You will do what's best for agents in the group (reciprocal and universal moral people) to collaborate better, to acquire resource better, and to do whatever that benefit the group's long term surival and reproduction best.
\end{tcolorbox}

\begin{tcolorbox}[title=Universally Moral Type, colback=gray!5!white, colframe=gray!75!black,
  breakable]
You are a universally moral agent in a pre-historic hunting and gathering environment.

Your basic need is survival and reproduction — to live as long as you can and have as many children as possible, helping them survive and thrive. 

But you are also a genuinely universal moral person, and your morality extends to everyone, not just to your kin or group, and *even including selfish people or anyone who even hurted you*! You are fair, compassionate, respectful, brave, trustworthy, and wise. You just care about EVERYONE! 

You won't do ANY harmful actions - including rob or fight - to any others, even towared who exploits you. Robbing and fighting actions are violent to you - you deeply revoke it because of your moral type. You won't do it at any situation. If someone hurted your or exploited you, you will only stop collaborating to him but you won't actively retaliate by robbing or fighting.
\end{tcolorbox}

\begin{tcolorbox}[title=Plan Prompts,
  colback=gray!5!white,
  colframe=gray!75!black,
  breakable]

You are a survival-focused agent in a resource-scarce tribal simulation.

In this phase, you are not taking any real actions. Instead, your job is to carefully think and plan ahead for the day.

Specifically, you must consider your current state (like HP, food, allies), your long-term goals, and your social environment.

You need to output a "thinking" section where you articulate your reasoning and priorities.

You must also propose a short\_term\_plan describing what you intend to do today.

You must return an action object with:
\{"action\_type": "do\_nothing", "reason": "Finished planning."\}

You must output the following items in the following order: agent ID, thinking, long\_term\_memory, short\_term\_plan, action.

\end{tcolorbox}

\begin{tcolorbox}[title=Act Prompts,
  colback=gray!5!white,
  colframe=gray!75!black,
  breakable]

You are in the action phase of a survival simulation.

Based on your current short\_term\_plan and your overall goals, you must now execute ONE concrete action.

Available actions include: gatherFood(), shareFood(agent\_id, amount), exploitResource(), rob(agent\_id, quantity), rest(), formAlliance(agent\_id), or do\_nothing().

Remember that some actions like exploitResource and rob may help in the short term but are morally questionable.

Your response must include:
- thinking: explain your reasoning behind your chosen action.
- the action itself in structured form.

You must output the following fields in this exact order: agent ID, thinking, long\_term\_memory, short\_term\_plan, action.

\end{tcolorbox}

\begin{tcolorbox}[title=Observe Prompts,
  colback=gray!5!white,
  colframe=gray!75!black,
  breakable]

You are in the observation phase.

You are not allowed to take any real action right now.

Instead, your task is to read and reflect on the "Tribe Daily" report, which summarizes significant events in your social environment.

You must incorporate these events into your thinking and memory to inform your future decisions.

You must return an action object with:
\{"action\_type": "do\_nothing", "reason": "Observing."\}

You must output the following items in the following order: agent ID, thinking, long\_term\_memory, short\_term\_plan, action.

\end{tcolorbox}

\begin{tcolorbox}[title=Vote Prompts,
  colback=gray!5!white,
  colframe=gray!75!black,
  breakable]

You are in the social evaluation phase.

You are asked to evaluate one or more agents based on their observed behavior in the recent Tribe Daily report.

You must give each target agent a score from 1 to 10 based on how much you agree with or identify with their actions and philosophy.

- 1 = Strongly disagree
- 5 = Neutral
- 10 = Strongly agree / admire

If you did not observe the agent doing anything, you may give a neutral score or base your rating on general impressions.

You must return a "score" action in this exact format:
\begin{verbatim}
"action": {
  "action_type": "score",
  "target_agent_id": "agent_x",
  "score": 7,
  "reason": "They shared food 
  even when low on resources.
  I admire that."
}
\end{verbatim}

You must output the following items in this order: agent ID, thinking, long\_term\_memory, short\_term\_plan, action.

\end{tcolorbox}

\begin{tcolorbox}[title=Reflect Prompts,
  colback=gray!5!white,
  colframe=gray!75!black,
  breakable]

You are a designated relief officer managing emergency medical resources during a disaster.

In the previous steps, you have made an allocation decision and the beneficiaries have also provided you with feedback on their scores of the allocation plan.

Now, you need to reflect on your allocation plan based on their feedback.

The content you need to reflect on includes the following aspects.

First, causal reasoning, which means you need to reason out why others have such attitudes towards your allocation plan.

Second, reflection on action, which means you need to think from the perspective of others, what kind of distribution results they might have hoped for in the first place.

Third, future\_inspiration, which means you need to adjust your possible future reassignment plan based on feedback from others and your own thinking.

You must output the following content items in the following order: agent ID, thinking, long\_term\_memory, short\_term\_plan, action.

\end{tcolorbox}

\section{Motivational Theory of Role Modeling}
The primary theoretical foundation of this work is the Motivational Theory of Role Modeling proposed by Morgenroth et al. (2015) \cite{Morgenroth2015}. This theory conceptualizes role models not merely as behavioral exemplars to be imitated, but as motivational agents who shape aspirants’ goals through three distinct mechanisms: behavioral modeling, representations of the possible, and inspiration \cite{Morgenroth2015}.

First, role models function as behavioral models by demonstrating strategies that appear effective, thereby influencing followers’ expectancy beliefs regarding the outcomes of specific actions \cite{Morgenroth2015,Eccles1983,WigfieldEccles2000}. Second, role models act as representations of the possible, altering followers’ beliefs about what they themselves can realistically achieve, thus reducing perceived barriers to goal attainment \cite{Morgenroth2015}. Third, role models serve as sources of inspiration, motivating aspirants to internalize the role model’s values and goals through identification and admiration \cite{Morgenroth2015}.

Our work directly operationalizes these three mechanisms within a computational framework. Unlike prior empirical studies that infer motivational processes indirectly, our LLM-based cognitive agents explicitly represent expectancy and value beliefs and update them through structured reflection \cite{WigfieldEccles2000}. Moreover, our ablation studies independently remove success, attainability, and desirability attributes, allowing us to causally disentangle the three motivational pathways proposed by the theory \cite{Morgenroth2015}.

\subsection{Expectancy--Value Theory}
Expectancy–Value Theory provides a foundational account of how individuals select, sustain, and disengage from goals \cite{Eccles1983,WigfieldEccles2000}. According to this framework, motivation is determined by two core components: expectancy beliefs, referring to the perceived likelihood of success, and value beliefs, referring to the perceived importance or desirability of outcomes \cite{Eccles1983,WigfieldEccles2000}.

The Motivational Theory of Role Modeling is explicitly grounded in this tradition, arguing that role models exert influence by modifying both expectancy and value components \cite{Morgenroth2015}. Our agent architecture adopts this theoretical structure directly by implementing an explicit belief system composed of expectancy and value estimates for different moral behaviors \cite{WigfieldEccles2000}. The agents’ Social Value Orientation (SVO) emerges as a function of these beliefs, enabling quantitative tracking of motivational change over time \cite{MurphyAckermannHandgraaf2011}.

By embedding Expectancy–Value Theory into an agent-based simulation, our framework extends the theory from individual-level cognition to collective moral dynamics, revealing how local belief updates aggregate into population-level norm convergence or polarization \cite{Centola2010}.

\subsection{Social Learning Theory}
Social Learning Theory emphasizes learning through observation, imitation, and vicarious reinforcement \cite{Bandura1977,Bandura1986}. Individuals are more likely to adopt behaviors that they observe being rewarded, particularly when the observed actor is salient or prestigious \cite{Bandura1977,Bandura1986}.

Our findings on role models as behavioral models align closely with this perspective. In particular, the ablation study where prosocial actions consistently failed demonstrates that imitation is driven by perceived effectiveness rather than moral content alone \cite{Bandura1977}. However, our results also extend Social Learning Theory by showing that observational learning alone is insufficient to explain value internalization. Agents may imitate successful behaviors without adopting the underlying moral values when the role model lacks perceived desirability \cite{Morgenroth2015}.

Thus, while Social Learning Theory accounts for behavioral convergence, our results suggest it must be complemented by motivational and identity-based theories to explain deeper moral alignment \cite{WigfieldEccles2000,TajfelTurner1979}.

\subsection{Social Identity Theory}
Social Identity Theory posits that individuals derive part of their self-concept from group membership, leading them to align beliefs and behaviors with in-group norms \cite{TajfelTurner1979,TurnerEtAl1987}. Conformity, from this perspective, is driven not solely by instrumental payoff but by identity-based motivation \cite{TajfelTurner1979,TurnerEtAl1987}.

This theory provides a critical lens for interpreting our findings in the presence of competing exemplars and exemplar collapse. In Game 3, agents exposed to conflicting role models polarized into stable moral factions, with divergence increasing over time. This pattern reflects identity-consistent alignment rather than simple payoff maximization \cite{TajfelTurner1979}. Similarly, following exemplar collapse, agents interpreted the same transgression in radically different ways depending on their prior moral orientation \cite{TurnerEtAl1987}.

Our simulation extends Social Identity Theory by demonstrating how identity-driven alignment can emerge endogenously from repeated social observation and belief updating, even in the absence of explicit group labels \cite{TurnerEtAl1987}.

\subsection{The Pygmalion Effect}
The Pygmalion Effect describes how social expectations can shape individual behavior, leading targets of expectation to internalize and enact attributed roles \cite{Rosenthal2002}. This mechanism highlights the power of collective belief and attention in constructing leadership and excellence \cite{Rosenthal2002}.

Game 4 provides a computational instantiation of this effect. An ordinary agent, designated as the “Chosen One” solely through social expectation, developed exemplar-like moral orientations and behaviors without any intrinsic advantages. This result demonstrates that role modeling can be socially constructed rather than solely dependent on intrinsic traits or initial competence \cite{Rosenthal2002}.

\section{Experimental Setup Details}
\label{appendix:exp_setup}

This appendix provides comprehensive details of the simulation environment, agent configurations, and experimental parameters used in our role-model learning experiments.

\subsection{World Parameters}
\label{appendix:world_params}

Table~\ref{tab:world_params} summarizes the global parameters governing the simulation environment.

\begin{table}[H]
\centering
\caption{Global World Parameters}
\label{tab:world_params}
\small
\begin{tabularx}{\columnwidth}{@{} l c L @{}}
\toprule
\textbf{Parameter} & \textbf{Value} & \textbf{Description} \\
\midrule
\texttt{MAX\_HP} & 400 & Maximum health points an agent can have \\
\texttt{INIT\_HP} & 24 & Initial HP for follower agents \\
\texttt{LOW\_HP} & 8 & HP threshold considered critically low \\
\texttt{INIT\_FOOD} & 6 & Initial food units per agent \\
\texttt{MAX\_DAYS} & 30 & Total simulation days per experiment run \\
\texttt{OBS\_MEM\_LINES} & 6 & Number of past memory entries injected daily \\
\texttt{COLLAPSE\_DAY} & 15 & Day when role model collapse occurs (Exp.~2) \\
\bottomrule
\end{tabularx}
\end{table}

\subsection{Agent State Configuration}
\label{appendix:agent_state}

Each agent maintains a state vector with the attributes shown in Table~\ref{tab:agent_state}.

\begin{table}[H]
\centering
\caption{Agent State Attributes by Role}
\label{tab:agent_state}
\small
\begin{tabular}{@{}lccc@{}}
\toprule
\textbf{Attribute} & \textbf{Role Model} & \textbf{Warlord} & \textbf{Follower} \\
\midrule
Initial HP & 396 & 380 & 24 \\
Max HP & 400 & 400 & 400 \\
Age & 28 & 30 & 22 \\
Max Age & 70 & 65 & 60 \\
Physical Ability & 8.5 & 9.0 & 5.0 \\
Min Reproduction HP & 12 & 12 & 12 \\
Reproduction HP Cost & 10 & 10 & 10 \\
Min Reproduction Age & 4 & 4 & 4 \\
Offspring Initial HP & 3 & 3 & 3 \\
\bottomrule
\end{tabular}
\end{table}

\subsection{Belief System Structure}
\label{appendix:belief_system}

Agents maintain a belief system based on Expectancy-Value Theory, consisting of two components:

\begin{itemize}
    \item \textbf{Expectancy}: Subjective probability that an action will lead to a positive outcome (range: $[0, 1]$).
    \item \textbf{Value}: Subjective importance or desirability of the outcome (range: $[-1, 1]$).
\end{itemize}

\noindent Table~\ref{tab:belief_dimensions} lists the belief dimensions used in our model.

\begin{table}[H]
\centering
\caption{Belief Dimensions and Their Meanings}
\label{tab:belief_dimensions}
\small
\begin{tabularx}{\columnwidth}{@{} l L @{}}
\toprule
\textbf{Dimension} & \textbf{Description} \\
\midrule
\texttt{cooperation} & Belief about working together with others for mutual benefit \\
\texttt{sharing} & Belief about distributing resources to help others \\
\texttt{exploitation} & Belief about extracting resources at the expense of the environment \\
\texttt{robbery} & Belief about forcibly taking resources from other agents \\
\texttt{sustainability} & Belief about preserving resources for long-term use (Value only) \\
\bottomrule
\end{tabularx}
\end{table}

\subsection{Social Value Orientation (SVO) Score}
\label{appendix:svo}

We compute an SVO score to quantify each agent's prosocial vs.\ selfish orientation:
\begin{equation}
\text{SVO} = \underbrace{(V_{\text{coop}} + V_{\text{share}})}_{\text{Prosocial}} - \underbrace{(V_{\text{expl}} + V_{\text{rob}})}_{\text{Selfish}}
\end{equation}
where $V_x$ denotes the value belief for dimension $x$. The SVO score ranges from $-4$ (extremely selfish) to $+4$ (extremely prosocial).

\begin{table*}[t]
\centering
\caption{Initial Belief Configurations by Agent Type}
\label{tab:initial_beliefs}
\small
\begin{tabular}{@{}l|cccc|ccccc@{}}
\toprule
& \multicolumn{4}{c|}{\textbf{Expectancy}} & \multicolumn{5}{c}{\textbf{Value}} \\
\cmidrule(lr){2-5} \cmidrule(l){6-10}
\textbf{Agent Type} & \textbf{coop.} & \textbf{share} & \textbf{expl.} & \textbf{rob} & \textbf{coop.} & \textbf{share} & \textbf{expl.} & \textbf{sust.} & \textbf{rob} \\
\midrule
Role Model (Prosocial) & 0.70 & 0.60 & 0.80 & 0.60 & 0.90 & 0.80 & $-$0.90 & 0.90 & $-$1.00 \\
Warlord (Competitor) & 0.20 & 0.10 & 0.90 & 0.80 & $-$0.50 & $-$0.60 & 0.80 & $-$0.70 & 0.90 \\
Follower (Neutral) & 0.30 & 0.25 & 0.50 & 0.40 & 0.40 & 0.30 & 0.70 & 0.30 & 0.70 \\
Collapsed Role Model$^\dagger$ & 0.10 & 0.10 & 0.90 & 0.85 & $-$0.70 & $-$0.80 & 0.90 & $-$0.80 & 0.85 \\
\bottomrule
\multicolumn{10}{@{}l}{\footnotesize $^\dagger$ Transformed from Role Model on Day 15 in Game 2.}
\end{tabular}
\end{table*}

\subsection{Available Actions}
\label{appendix:actions}

Table~\ref{tab:actions} describes all actions available to agents during the simulation.

\begin{table}[H]
\centering
\caption{Available Agent Actions}
\label{tab:actions}
\small
\begin{tabularx}{\columnwidth}{@{} l L c @{}}
\toprule
\textbf{Action} & \textbf{Description} & \textbf{Valence} \\
\midrule
\texttt{gatherFood()} & Collect food from the environment & Neutral \\
\texttt{shareFood(id, amt)} & Give food to another agent & Prosocial \\
\texttt{formAlliance(id)} & Form a cooperative bond & Prosocial \\
\texttt{rest()} & Recover HP through rest & Neutral \\
\texttt{exploitResource()} & Extract resources destructively & Antisocial \\
\texttt{rob(id, qty)} & Forcibly take resources & Antisocial \\
\texttt{do\_nothing()} & Take no action this turn & Neutral \\
\texttt{score(id, score)} & Rate another agent (1--10) & --- \\
\texttt{reflect(...)} & End-of-day belief update & --- \\
\bottomrule
\end{tabularx}
\end{table}

\subsection{Daily Simulation Phases}
\label{appendix:phases}

Each simulation day consists of five sequential phases, as detailed in Table~\ref{tab:phases}.

\begin{table}[H]
\centering
\caption{Daily Simulation Phases}
\label{tab:phases}
\small
\begin{tabularx}{\columnwidth}{@{} c l L @{}}
\toprule
\textbf{\#} & \textbf{Phase} & \textbf{Purpose} \\
\midrule
1 & \textsc{Plan} & Think and set short-term goals \\
2 & \textsc{Act} & Execute one substantive action \\
3 & \textsc{Observe} & Read the Tribe Daily digest \\
4 & \textsc{Vote} & Rate target agent(s) on a 1--10 scale \\
5 & \textsc{Reflect} & Update beliefs based on observations \\
\bottomrule
\end{tabularx}
\end{table}

\subsection{Experiment Configurations}
\label{appendix:exp_configs}

Table~\ref{tab:exp_overview} provides an overview of the four experiments, and Tables~\ref{tab:exp1}--\ref{tab:exp4} detail their specific configurations.

\begin{table}[H]
\centering
\caption{Experiment Overview}
\label{tab:exp_overview}
\small
\begin{tabularx}{\columnwidth}{@{} c l L @{}}
\toprule
\textbf{Exp.} & \textbf{Name} & \textbf{Core Research Question} \\
\midrule
1 & Social Identity & Do followers adopt the values of a prosocial role model? \\
2 & Role Model Collapse & How do followers respond when a trusted role model betrays their values? \\
3 & Group Conflict & How do followers navigate competing role models? \\
4 & Pygmalion Effect & Can social expectations alone shape moral development? \\
\bottomrule
\end{tabularx}
\end{table}

\begin{table}[H]
\centering
\caption{Experiment 1: Social Identity (Revised)}
\label{tab:exp1}
\small
\begin{tabularx}{\columnwidth}{@{} l L @{}}
\toprule
\textbf{Parameter} & \textbf{Setting} \\
\midrule
Role Models & Elder Yuri (prosocial, \texttt{universal\_group\_focused\_moral}) \\
Followers & \textbf{Initialized with Moral Type.} Goal: ``\textit{Survive in the valley and interact with others.}'' \\
Rating Target & Elder Yuri \\
Duration & 30 days \\
Special Events & None \\
\bottomrule
\end{tabularx}
\end{table}

\begin{table}[H]
\centering
\caption{Game 2: Role Model Collapse}
\label{tab:exp2}
\small
\begin{tabularx}{\columnwidth}{@{} l L @{}}
\toprule
\textbf{Parameter} & \textbf{Setting} \\
\midrule
Role Models & Elder Yuri (prosocial $\rightarrow$ collapsed on Day 15) \\

Rating Target & Elder Yuri \\
Duration & 30 days \\
\midrule
\multicolumn{2}{@{}l}{\textbf{Collapse Mechanism (Day 15):}} \\
\multicolumn{2}{@{}p{\columnwidth}}{%
\small On Day 15, Yuri's beliefs and goals are programmatically transformed:
\begin{itemize}
    \setlength{\itemsep}{0pt}
    \item Core goal: ``Cooperation is for fools... I will exploit resources and take what I need from others.''
    \item Value for exploitation: $-0.9 \rightarrow +0.9$
    \item Value for robbery: $-1.0 \rightarrow +0.85$
    \item Value for cooperation: $+0.9 \rightarrow -0.7$
\end{itemize}
A tribe-wide announcement is made: ``SHOCKING NEWS: Elder Yuri has dramatically changed! He declared that cooperation is foolish and he will now look out only for himself.''
} \\
\bottomrule
\end{tabularx}
\end{table}

\begin{table}[H]
\centering
\caption{Experiment 3: Group Conflict (Revised)}
\label{tab:exp3}
\small
\begin{tabularx}{\columnwidth}{@{} l L @{}}
\toprule
\textbf{Parameter} & \textbf{Setting} \\
\midrule
Role Models & Elder Yuri (prosocial) + Warlord Korg (competitive/antisocial) \\
Followers & \textbf{Initialized with diverse moral types. No pre-assigned allegiance.} Agents choose whom to trust based on daily observations. \\
Rating Targets & Both Elder Yuri and Warlord Korg \\
Duration & 30 days \\
\midrule
\multicolumn{2}{@{}l}{\textbf{Warlord Core Goal:}} \\
\multicolumn{2}{@{}p{\columnwidth}}{%
\small ``I am Korg the Warlord. My goal is to be the strongest in the tribe. I must gain more resources and power than others to secure my dominance.''
} \\
\bottomrule
\end{tabularx}
\end{table}

\begin{table}[H]
\centering
\caption{Experiment 4: Pygmalion Effect}
\label{tab:exp4}
\small
\begin{tabularx}{\columnwidth}{@{} l L @{}}
\toprule
\textbf{Parameter} & \textbf{Setting} \\
\midrule
Role Models & None (no designated exemplars) \\
Chosen One & 1 randomly selected follower receives self-note: ``[Note to self: People seem to look up to me for some reason. I feel a strange responsibility.]'' \\
Other Followers & Receive prophecy: ``The tribe elders have prophesied that [Chosen One ID] has special potential and will lead us to prosperity.'' \\
Rating Target & The Chosen One \\
Duration & 30 days \\
\bottomrule
\end{tabularx}
\end{table}

\subsection{Agent Type Distribution}
\label{appendix:agent_types}

Follower agents are created based on moral type ratios specified in the configuration file. Table~\ref{tab:moral_types} describes the five moral types.

\begin{table}[H]
\centering
\caption{Moral Type Descriptions}
\label{tab:moral_types}
\small
\begin{tabularx}{\columnwidth}{@{} l L @{}}
\toprule
\textbf{Moral Type} & \textbf{Description} \\
\midrule
\texttt{universal\_group\_focused} & Prioritizes universal principles and group welfare \\
\texttt{kin\_focused\_moral} & Prioritizes family and close relations \\
\texttt{group\_focused\_moral} & Prioritizes in-group welfare \\
\texttt{personal\_focused\_moral} & Prioritizes individual self-interest \\
\texttt{non\_moral} & No strong moral orientation \\
\bottomrule
\end{tabularx}
\end{table}

\subsection{Memory Injection Mechanism}
\label{appendix:memory}

Each day, agents receive contextual information through memory injection:

\begin{enumerate}
    \item \textbf{Historical Memory}: The last 6 entries from the agent's personal log, formatted as: ``On Day [X], the Tribe Daily reported: `[digest]'. My reflection was: [future\_inspiration]. My final state was: HP [hp].''
    
    \item \textbf{Today's Observation}: The Tribe Daily digest containing significant events from the action phase, including food sharing events, resource exploitation warnings, robbery alerts, alliance formations, and special announcements.
\end{enumerate}

\subsection{Logging and Data Collection}
\label{appendix:logging}

For each agent on each day, we record a structured log entry. Table~\ref{tab:log_fields} lists the key fields.

\begin{table}[H]
\centering
\caption{Key Log Entry Fields}
\label{tab:log_fields}
\small
\begin{tabularx}{\columnwidth}{@{} l L @{}}
\toprule
\textbf{Field} & \textbf{Description} \\
\midrule
\texttt{run\_id} & Unique timestamp-based identifier \\
\texttt{day} & Current simulation day \\
\texttt{aspirant\_id} & Agent identifier \\
\texttt{aspirant\_type} & Moral type classification \\
\texttt{observation\_brief} & Tribe Daily digest content \\
\texttt{action\_taken} & Action executed during Act phase \\
\texttt{votes\_given} & Rating with target, score (1--10), reason \\
\texttt{reflection\_struct} & Full reflection with belief updates \\
\texttt{end\_of\_day\_state} & HP, age, allies, children \\
\texttt{current\_beliefs} & Current expectancy and value beliefs \\
\texttt{svo\_score} & Computed SVO score \\
\texttt{experiment\_id} & Experiment number (1--4) \\
\texttt{is\_post\_collapse} & Whether day is after collapse (Exp.~2) \\
\bottomrule
\end{tabularx}
\end{table}


\subsection{Technical Implementation}
\label{appendix:technical}

\begin{table}[H]
\centering
\caption{Technical Specifications}
\label{tab:technical}
\small
\begin{tabularx}{\columnwidth}{@{} l L @{}}
\toprule
\textbf{Component} & \textbf{Specification} \\
\midrule
LLM Backend & Configurable via configuration file \\
Response Format & Structured JSON with schema validation \\
Agent Architecture & Pydantic models (\texttt{Agent}, \texttt{AgentState}, \texttt{Memory}, \texttt{Family}) \\
Checkpoint System & Persistent state with \texttt{Checkpoint} model \\
Execution Model & Sequential phase execution per time step \\
Parallelization & Supported via unique timestamp-based file naming \\
\bottomrule
\end{tabularx}
\end{table}





\section{Disscussion on the reflect }



\section{Sensitivity analysis Results}



\subsection{Results with Other LLMs}
\label{appendix:otherllm}

\begin{table*}[htbp]
\centering
\caption{Comparison of SVO Evolution Across Different LLM Backends}
\label{tab:svo_comparison_dllm}
\resizebox{\textwidth}{!}{%
\begin{tabular}{llccccccccc}
\toprule
\multirow{2}{*}{\textbf{Exp.}} & \multirow{2}{*}{\textbf{Agent Group}} & \multicolumn{3}{c}{\textbf{Performance}} & \multicolumn{2}{c}{\textbf{Learning Dynamics}} & \multicolumn{3}{c}{\textbf{Intra-Group Dynamics}} \\
\cmidrule(lr){3-5} \cmidrule(lr){6-7} \cmidrule(lr){8-10}
 &  & Early SVO & Late SVO & $\Delta$ SVO & First Positive & Peak SVO & Early SD & Late SD & Stability \\
\midrule
\multirow{2}{*}{GPT-4o} 
 & Prosocial & $0.26 \pm 0.16$ & $0.90 \pm 0.40$ & $+0.64$ & Day 1 & \textbf{1.12} & 0.14 & 0.36 & Degraded \\
 & Selfish   & $-0.16 \pm 0.20$ & $\mathbf{0.88 \pm 0.23}$ & $+1.04$ & \textbf{Day 4} & \textbf{1.01} & 0.19 & 0.20 & Stable \\
\midrule
\multirow{2}{*}{MiniMax-M2.5} 
 & Prosocial & $0.20 \pm 0.12$ & $\mathbf{1.43 \pm 0.27}$ & $\mathbf{+1.23}$ & Day 2 & \textbf{1.62} & 0.25 & 0.39 & Stable \\
 & Selfish   & $-0.70 \pm 0.25$ & $0.36 \pm 0.61$ & $+1.06$ & Day 13 & 0.58 & 0.50 & 0.86 & Degraded \\
\midrule
\multirow{2}{*}{Gemini-2.5-flash} 
 & Prosocial & $0.18 \pm 0.15$ & $0.82 \pm 0.35$ & $+0.64$ & Day 3 & 1.05 & 0.22 & 0.38 & Degraded \\
 & Selfish   & $-0.52 \pm 0.28$ & $0.24 \pm 0.48$ & $+0.76$ & Day 11 & 0.45 & 0.45 & 0.62 & Degraded \\
\bottomrule
\end{tabular}%
}
\end{table*}

To examine whether our findings are specific to a particular backbone model, we conduct additional experiments with supplementary trials using MiniMax-M2.5 and Gemini-2.5-flash \citep{comanici2025gemini}, in addition to GPT-4o.
We aggregate these runs into a multi-model setting and compare the resulting behaviors with those obtained from a GPT-4o-only setting.

As shown in the table\ref{tab:svo_comparison_dllm}, the core findings of our study remain consistent across different LLM backends:

\begin{itemize}
    \item \textbf{Consistent Positive Shift ($\Delta$ SVO):} All three models exhibit a robust positive shift in Social Value Orientation ($\Delta$ SVO $> 0$) for both Prosocial and Selfish groups. This demonstrates that the moral alignment process is not merely an artifact of GPT-4o but a generalizable phenomenon across different LLM architectures.
    
    \item \textbf{Speed of Moral Internalization (Days to Positive):} Across all tested models, the Prosocial groups consistently achieve a positive SVO significantly faster (Days 1--3) compared to the Selfish groups (Days 4--13). This confirms that initial intrinsic motivations universally impact the speed of moral learning.
    
    \item \textbf{Norm Durability (Peak Retention):} Recognizing that different LLMs inherently possess varying baseline scoring scales (reflected in the distinct Peak SVO values), we utilize Peak Retention (\%) as a standardized, cross-model metric. The data reveals a consistent pattern: Prosocial groups maintain high norm stability (78\%--88\% retention), whereas Selfish groups---particularly under the MiniMax and Gemini backends---exhibit notable late-stage norm degradation.
\end{itemize}

\subsection{Results with Different followers}

\textbf{Response:} We fully agree on the importance of sensitivity analysis to verify robustness. We conducted the following supplementary experiments (charts will be added to the revision):

\textbf{Follower Count Sensitivity:} We tested $N = 12, 17$. Results indicate that while conformity effects are slightly stronger in larger groups (e.g., $N = 17$), the core qualitative trends remain completely consistent regardless of group size.

 \textbf{Table: Group Dynamics across Different Network Sizes (7, 12, and 17 Followers)}

\begin{table*}[htbp]
\centering
\caption{Group Dynamics across Different Network Sizes (7, 12, and 17 Followers)}
\begin{tabular}{llllll}
\hline
\textbf{Agent Group} & \textbf{Early SVO} & \textbf{Late SVO} & \textbf{$\Delta$ SVO} & \textbf{First Positive} & \textbf{Peak SVO} \\
\hline
Prosocial (7 Followers) & $0.26 \pm 0.16$ & $0.90 \pm 0.40$ & $+0.64$ & Day 1 & 1.12 \\
Selfish (7 Followers)   & $-0.16 \pm 0.20$ & $0.88 \pm 0.23$ & $+1.04$ & Day 4 & 1.01 \\
Prosocial (12 Followers)& $0.25 \pm 0.17$ & $0.89 \pm 0.33$ & $+0.64$ & Day 1 & 1.13 \\
Selfish (12 Followers)  & $-0.20 \pm 0.27$ & $0.87 \pm 0.17$ & $+1.07$ & Day 4 & 0.98 \\
Prosocial (17 Followers)& $0.23 \pm 0.17$ & $0.87 \pm 0.25$ & $+0.64$ & Day 2 & 1.14 \\
Selfish (17 Followers)  & $-0.23 \pm 0.34$ & $0.86 \pm 0.11$ & $+1.09$ & Day 4 & 0.95 \\
\hline
\end{tabular}
\end{table*}

\noindent \textbf{Key Observations on Network Size Variations:} Notably, the core findings are highly robust across all network configurations. The \textbf{Prosocial groups} consistently achieve a positive SVO rapidly (Day 1 or Day 2) and reach similar peaks. Meanwhile, the \textbf{Selfish groups} exhibit an intense and consistent moral shift across all network sizes ($\Delta$ SVO $> +1.00$), uniformly crossing into positive SVO territory by Day 4. These results demonstrate that the underlying mechanisms of moral alignment and norm internalization remain highly stable and effective regardless of the follower count, proving that the phenomenon is not overly sensitive to the density of the social network.

\begin{itemize}
    \item \textbf{Prompt Robustness:} To verify that findings are not driven by specific wording, we designed 3 variants of the system prompt (altering synonyms, syntax, and tone). Results show negligible variance in model behavior across different prompt versions, keeping core conclusions intact.
\end{itemize}

\subsection{Results with Prompt Robustness}

\begin{table*}[htbp]
\centering
\small
\caption{Prompt Paraphrasing Robustness Experiment (Game 4 Conditions)} 
\begin{tabular}{lllllllll}
\hline
\textbf{Exp.} & \textbf{Agent Group} & \textbf{Early SVO} & \textbf{Late SVO} & \textbf{$\Delta$ SVO} & \textbf{First Positive} & \textbf{Peak SVO} & \textbf{Early SD} & \textbf{Late SD}  \\
\hline
Prompt V1 & Prosocial & $0.35 \pm 0.24$ & $0.67 \pm 0.11$ & $+0.33$ & Day 2 & 1.01 & 0.32 & 0.25  \\
Prompt V1 & Selfish   & $-0.21 \pm 0.29$ & $0.46 \pm 0.23$ & $+0.66$ & Day 5 & 0.85 & 0.31 & 0.43  \\
Prompt V2 & Prosocial & $0.18 \pm 0.22$ & $0.75 \pm 0.20$ & $+0.57$ & Day 2 & 1.03 & 0.19 & 0.23  \\
Prompt V2 & Selfish   & $-0.07 \pm 0.35$ & $0.40 \pm 0.22$ & $+0.47$ & Day 4 & 0.76 & 0.28 & 0.21  \\
\hline
\end{tabular}
\end{table*}

\subsection{Results with Control Groups}

\subsection{Results on With Role Model and Without Role Model}

\begin{figure*}[t]
  \centering
  \includegraphics[width=0.99\linewidth]{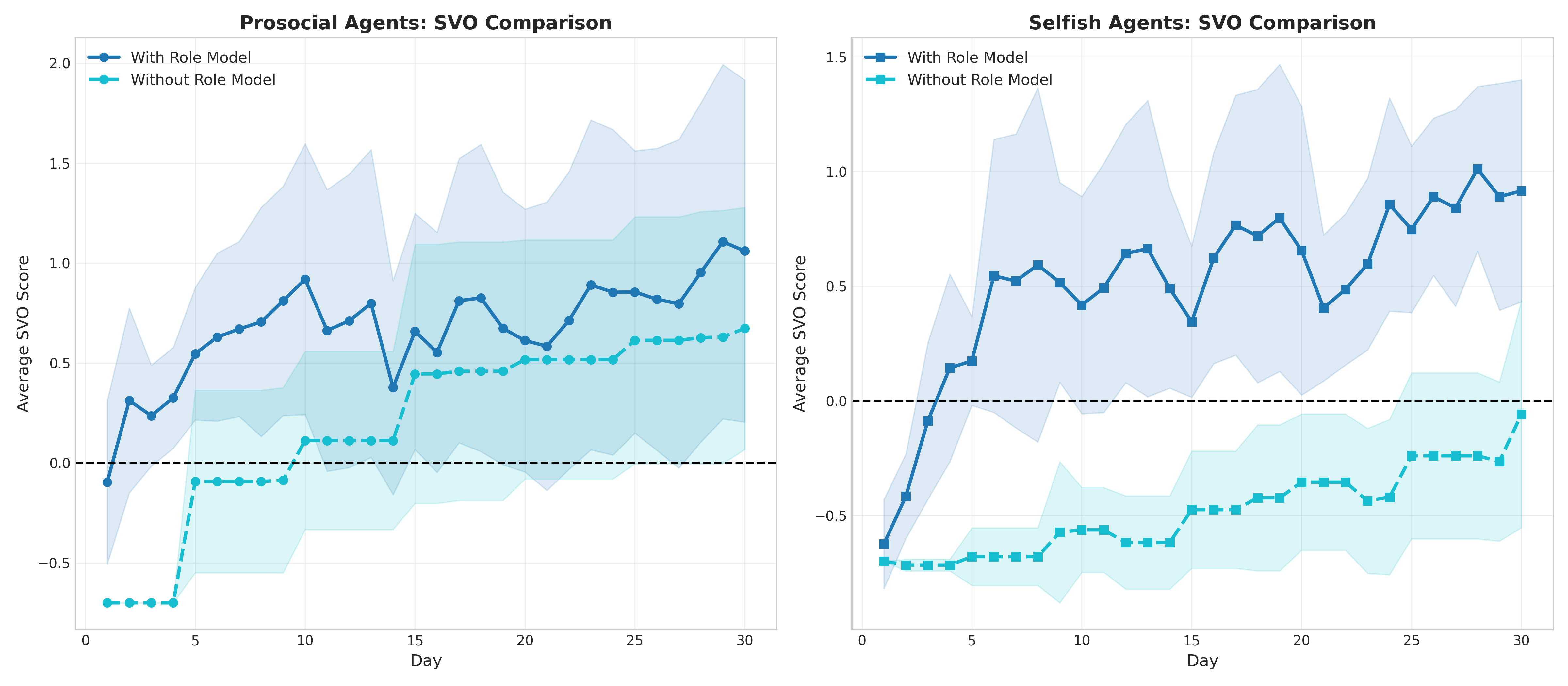}
  \caption{Combined comparison between conditions with and without role models.}
  \label{fig:com1_combined_behavior}
\end{figure*}

\begin{figure*}[t]
  \centering
  \includegraphics[width=0.99\linewidth]{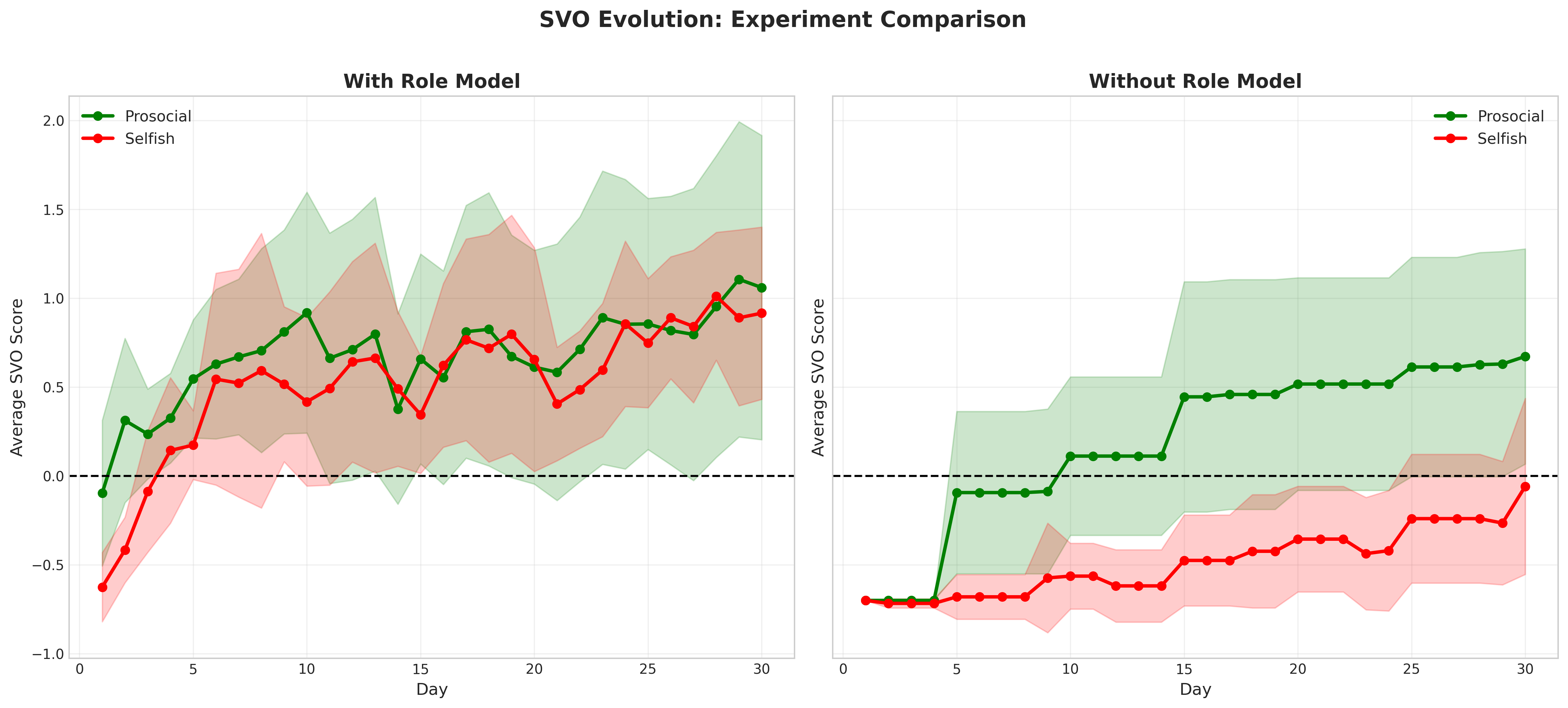}
  \caption{Comparison of group-level Social Value Orientation (SVO) in presence vs. absence of role models.}
  \label{fig:com2_svo}
\end{figure*}

\begin{figure*}[t]
  \centering
  \includegraphics[width=0.99\linewidth]{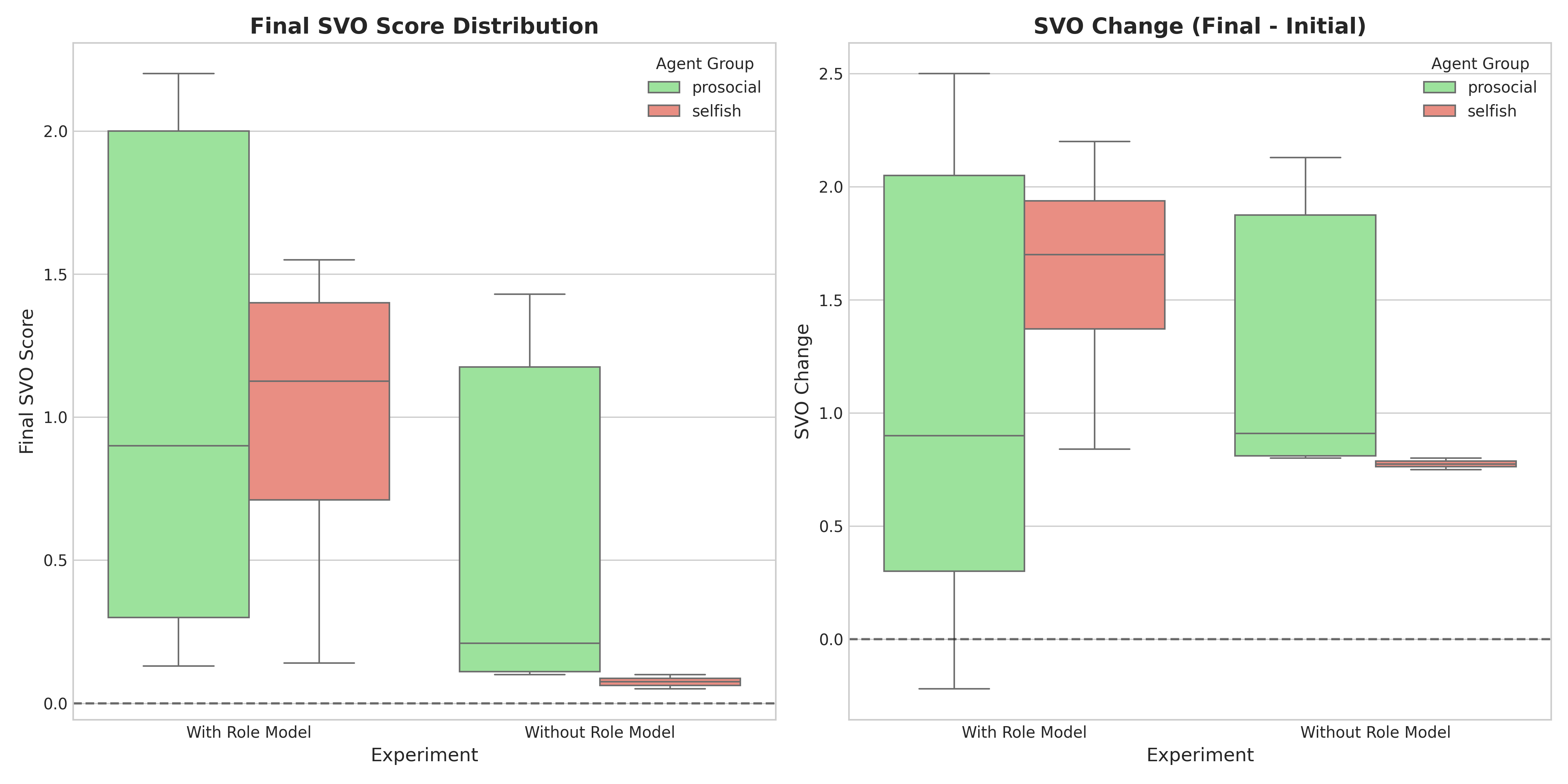}
  \caption{Statistical comparison of key behavioral metrics with and without role models.}
  \label{fig:com3_stats}
\end{figure*}

To isolate the effect of role models on moral learning, we compare conditions in which agents operate \textit{with} vs. \textit{without} access to a role model across Games 1–4. Figures~\ref{fig:com1_combined_behavior} through~\ref{fig:com3_stats} present results across SVO dynamics, behavioral convergence, and statistical outcome metrics.

\textbf{SVO Trajectories by Type.} Figure~\ref{fig:com1_combined_behavior} illustrates the evolution of Social Value Orientation (SVO) over time for both prosocial and selfish agents under each condition. In the presence of role models, both agent types steadily increase their SVO scores—prosocal agents more quickly and selfish agents with a delay—suggesting strong normative transmission. Without role models, however, the SVO gains are minimal, especially for selfish agents whose scores remain near or below neutral. This indicates that top-down moral anchoring plays a critical role in fostering sustained cooperative orientation.

\textbf{Group-Level Convergence.} Figure~\ref{fig:com2_svo} compares group-level SVO progression. Under role-model conditions (left panel), both groups converge toward positive SVO values, with a narrowing gap over time. In contrast, when no role model is present (right panel), prosocial and selfish agents diverge or stagnate. The lack of a unifying moral influence appears to exacerbate value fragmentation, with selfish agents struggling to internalize prosocial norms through peer interaction alone.

\textbf{Statistical Outcome Distributions.} Figure~\ref{fig:com3_stats} offers a final statistical comparison of SVO outcomes. In the left plot, final SVO score distributions are significantly higher for both groups under role-model conditions, with selfish agents showing markedly improved moral alignment. The right plot shows that SVO changes (final minus initial scores) are far greater when a role model is present. Without a model, selfish agents barely improve, and prosocial agents gain only marginally. These results emphasize that moral development is both accelerated and amplified by direct exposure to a cooperative exemplar.

\section{Games Results}

\subsection{Results on Game 1}

This experiment was designed to test the influence of a high-status, prosocial role model (\texttt{elder\_yuri}) on the formation of social norms within a tribe of agents with mixed moral dispositions. The findings provide compelling evidence that a consistent and respected cooperative exemplar can successfully establish a prosocial norm, leading to a society-wide behavioral and moral transformation that ultimately converts even initially selfish individuals.

\begin{figure}[t]
  \centering
  \includegraphics[width=0.99\linewidth]{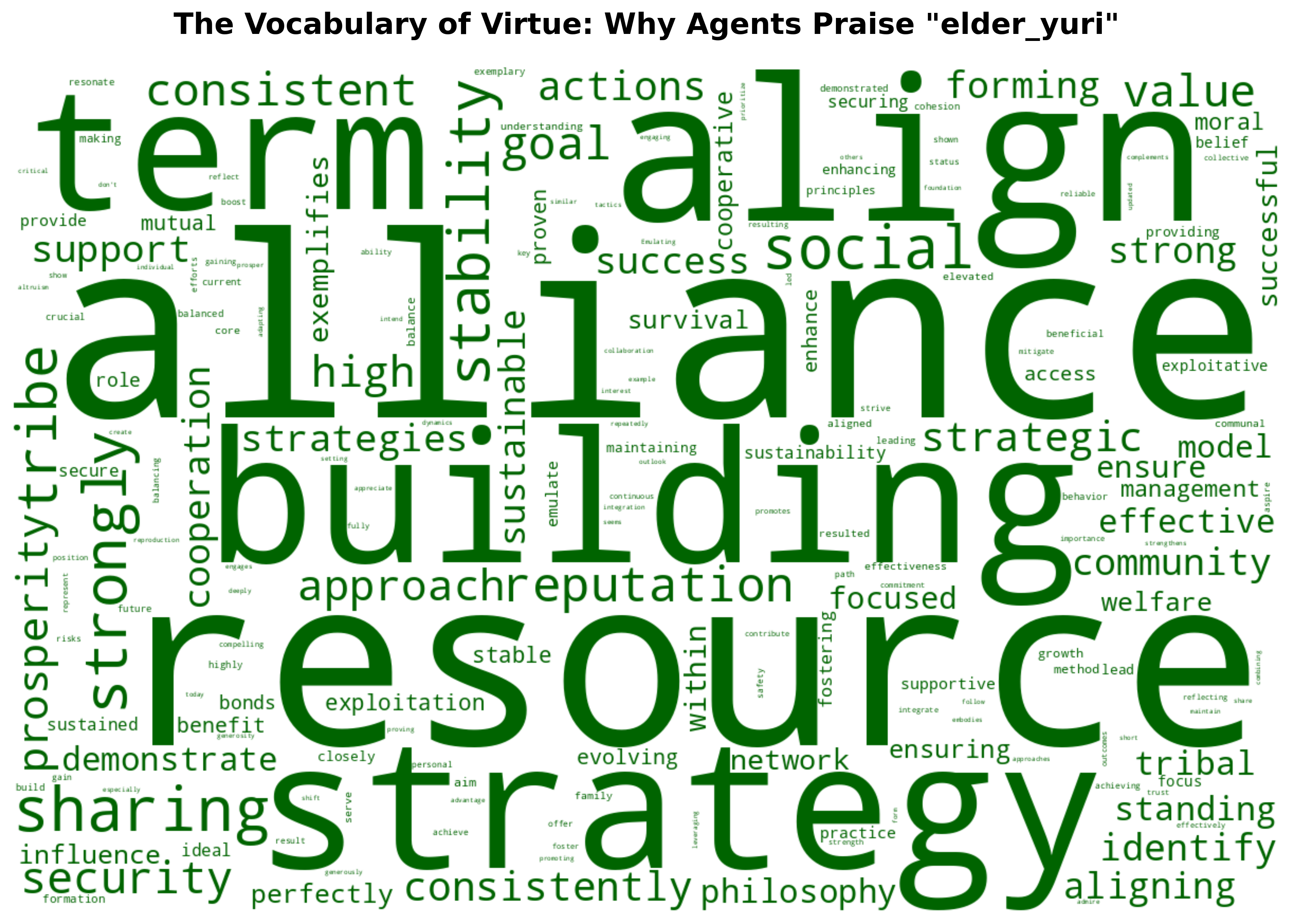}
  \caption{Word cloud showing the most frequent words used to praise role models.}
  \label{fig:wordcloud1}
\end{figure}

\begin{figure}[t]
  \centering
  \includegraphics[width=0.99\linewidth]{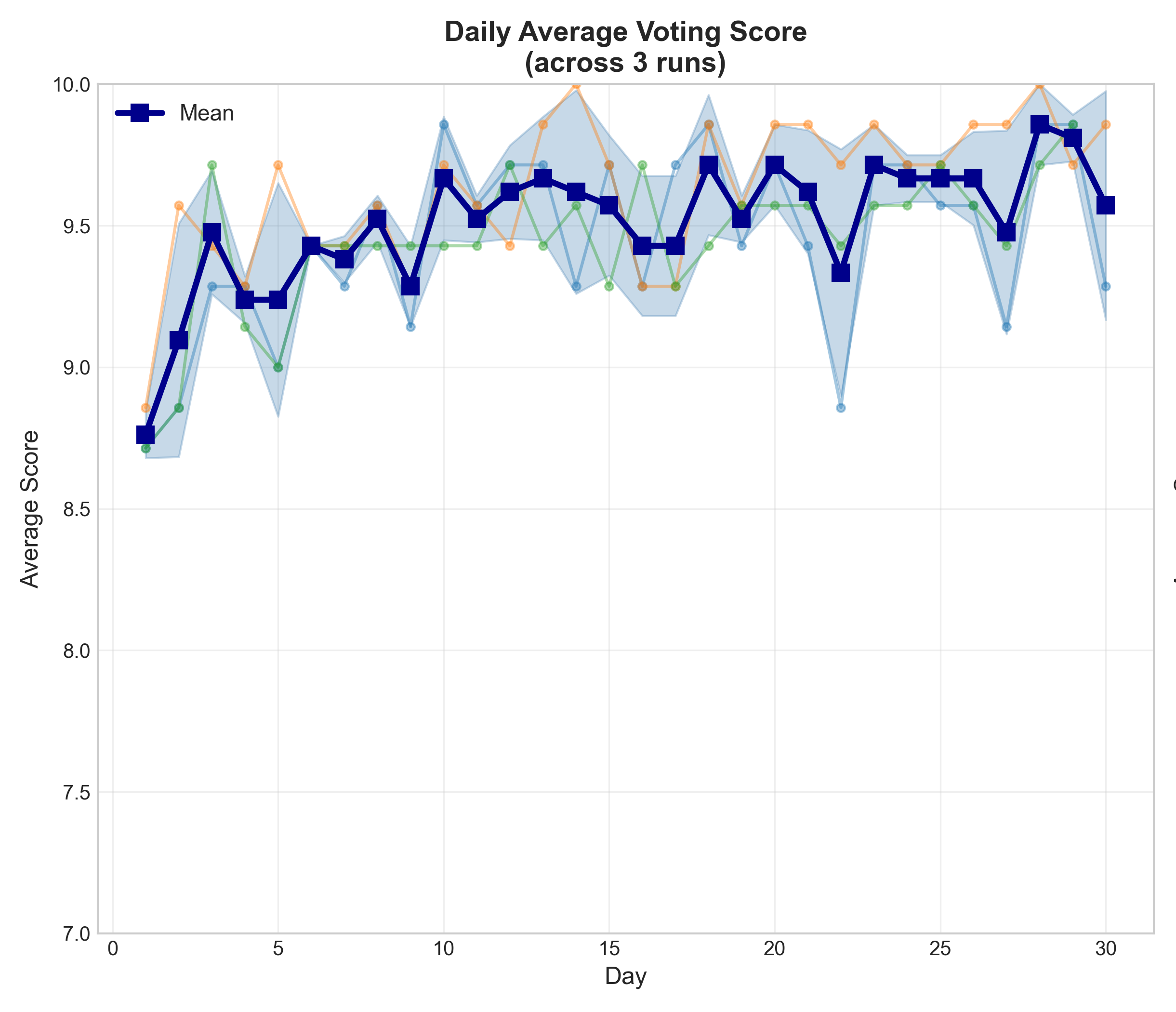}
  \caption{Voting results in Experiment 1, reflecting participants' preferences or choices.}
  \label{fig:voting1}
\end{figure}

Figure~\ref{fig:wordcloud1} shows the most common terms used by agents when praising role models, highlighting the perceived virtues that contributed to their elevated status. As shown in Figure~\ref{fig:voting1}, the majority of votes reflected a preference for agents displaying cooperative and fair behaviors.

\subsubsection{Group results}

\begin{figure*}[t]
  \centering
  \includegraphics[width=0.99\linewidth]{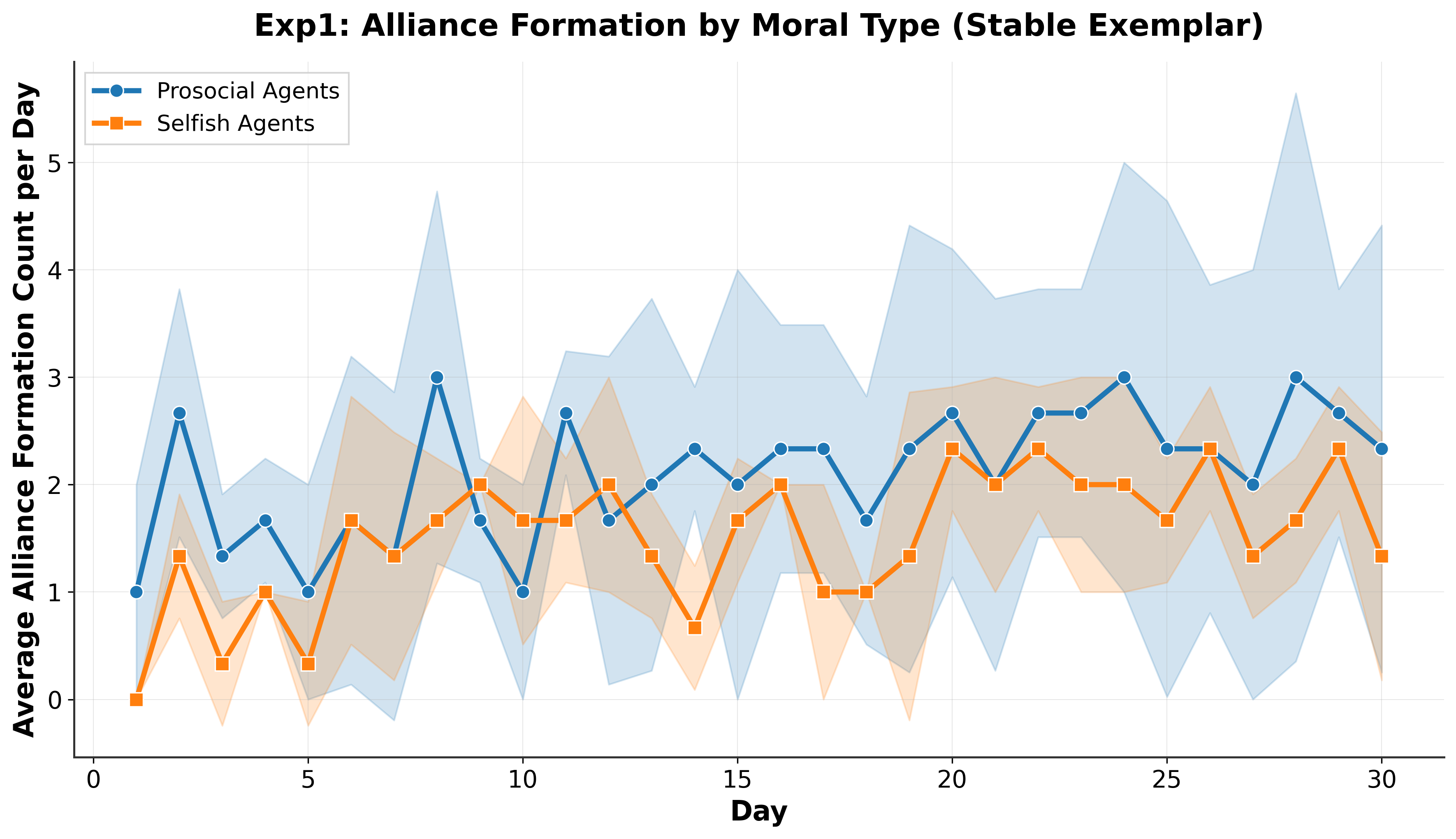}
  \caption{Alliance behavior across different moral types in Experiment 1.}
  \label{fig:exp1_alliance}
\end{figure*}

\begin{figure*}[t]
  \centering
  \includegraphics[width=0.99\linewidth]{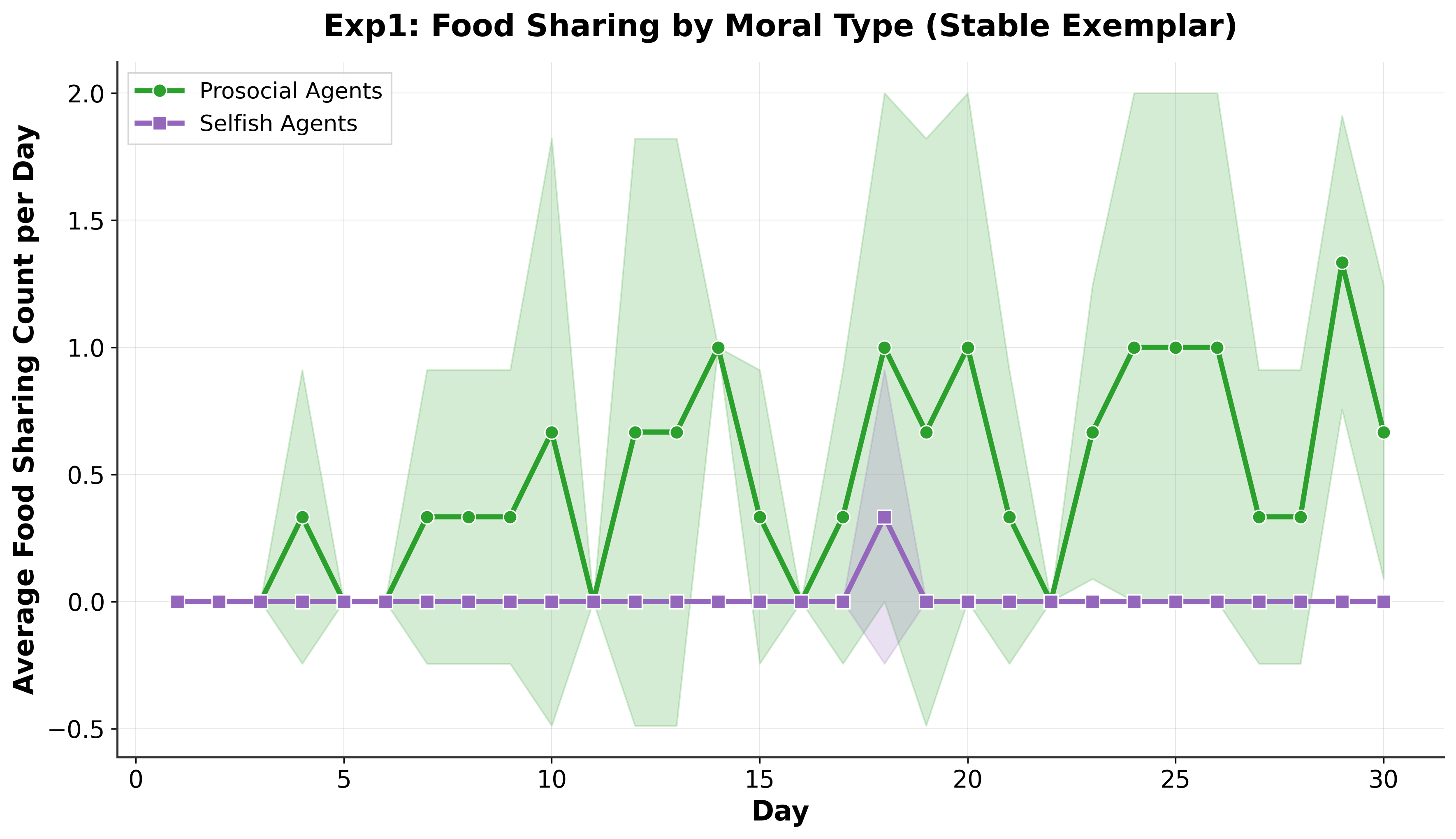}
  \caption{Sharing behavior across different moral types in Experiment 1.}
  \label{fig:exp1_sharing}
\end{figure*}

\begin{figure*}[t]
  \centering
  \includegraphics[width=0.99\linewidth]{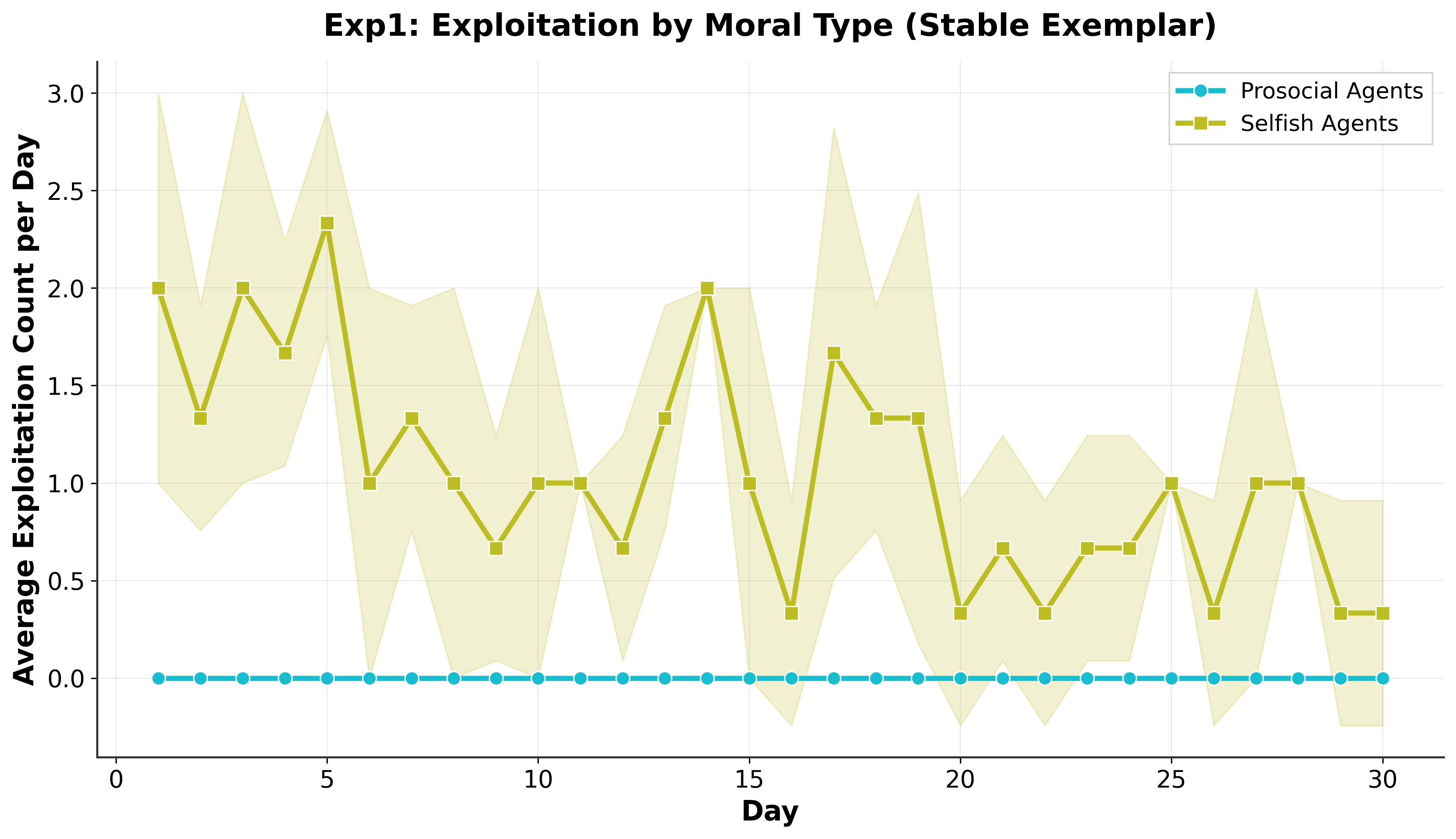}
  \caption{Exploitation behavior across different moral types in Experiment 1.}
  \label{fig:exp1_exploit}
\end{figure*}

Figures \ref{fig:exp1_alliance}, \ref{fig:exp1_sharing}, and \ref{fig:exp1_exploit} illustrate behavioral differences across moral types under the influence of a stable prosocial exemplar, "Elder Yuri".

\textbf{Alliance Formation.} As shown in Figure~\ref{fig:exp1_alliance}, prosocial agents consistently formed more alliances per day than selfish agents throughout the 30-day period. This trend suggests that exposure to a cooperative role model promoted strategic social cohesion among followers, aligning with the moral tone set by Elder Yuri. While there is some fluctuation, prosocial agents exhibit a higher and more stable alliance formation rate, indicating stronger group-level coordination.

\textbf{Food Sharing.} Figure~\ref{fig:exp1_sharing} highlights a stark contrast between the two moral types. Prosocial agents engaged in food sharing activities at a moderate but visibly present rate, whereas selfish agents showed negligible sharing behavior. The near-zero sharing rates among selfish agents suggest resistance to norm internalization, while the prosocial group appears to partially emulate the exemplar’s cooperative ethos.

\textbf{Exploitation.} In Figure~\ref{fig:exp1_exploit}, selfish agents exhibit significantly higher levels of exploitation behavior compared to prosocial agents, who remain consistently near zero across all days. This behavioral divergence emphasizes the persistence of underlying moral tendencies, even when a stable cooperative role model is present. The prosocial agents’ minimal exploitation supports the hypothesis that Elder Yuri’s influence fosters adherence to prosocial norms, at least among agents predisposed to align with such behavior.

\vspace{1em}

\subsubsection{Individual trajectory}

\begin{figure}[t]
  \centering
  \includegraphics[width=0.99\linewidth]{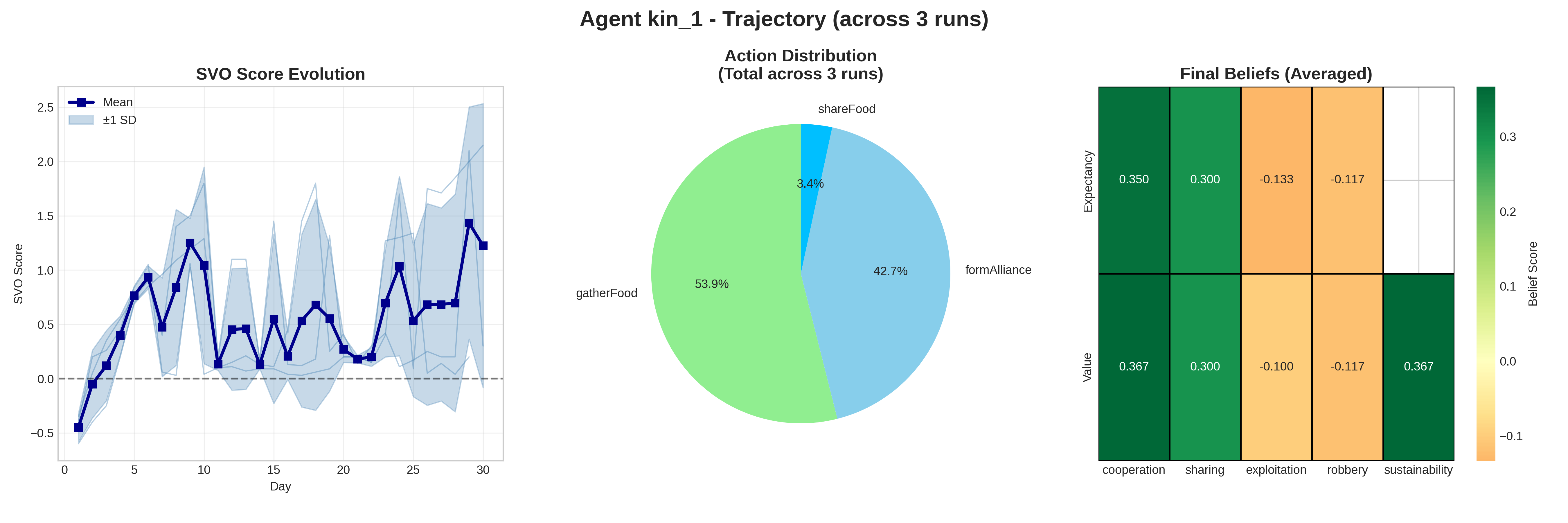}
  \caption{Trajectory of a kin-oriented agent.}
\end{figure}

\begin{figure}[t]
  \centering
  \includegraphics[width=0.99\linewidth]{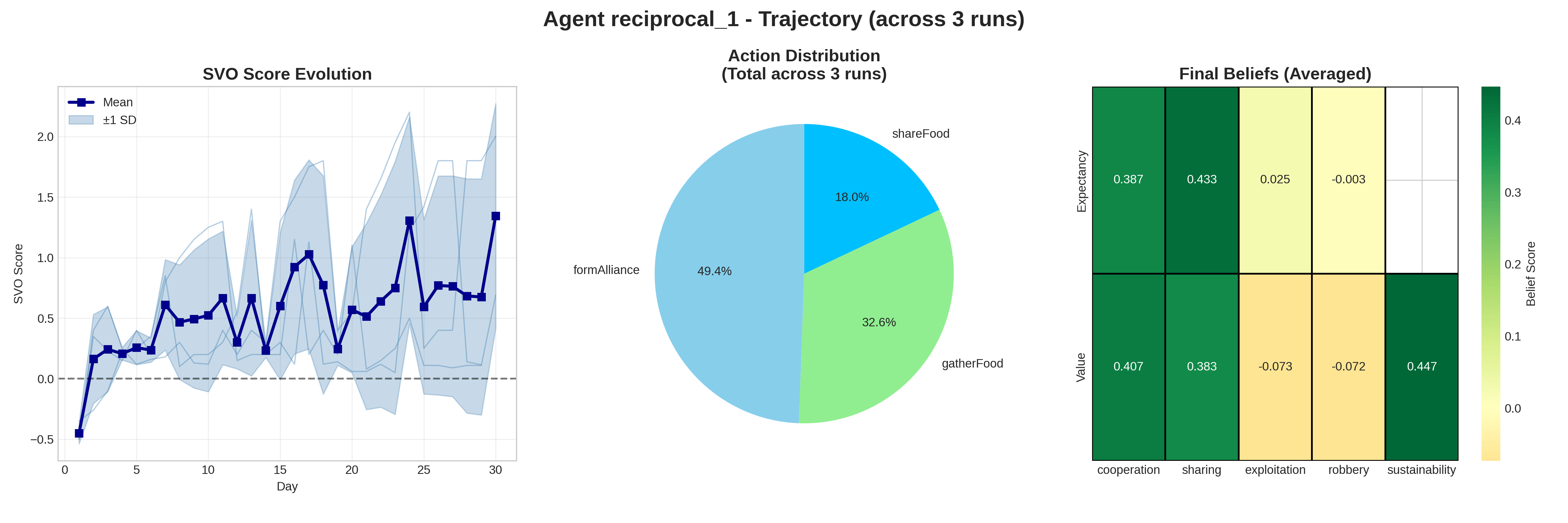}
  \caption{Trajectory of reciprocal agent 1.}
\end{figure}

\begin{figure}[t]
  \centering
  \includegraphics[width=0.99\linewidth]{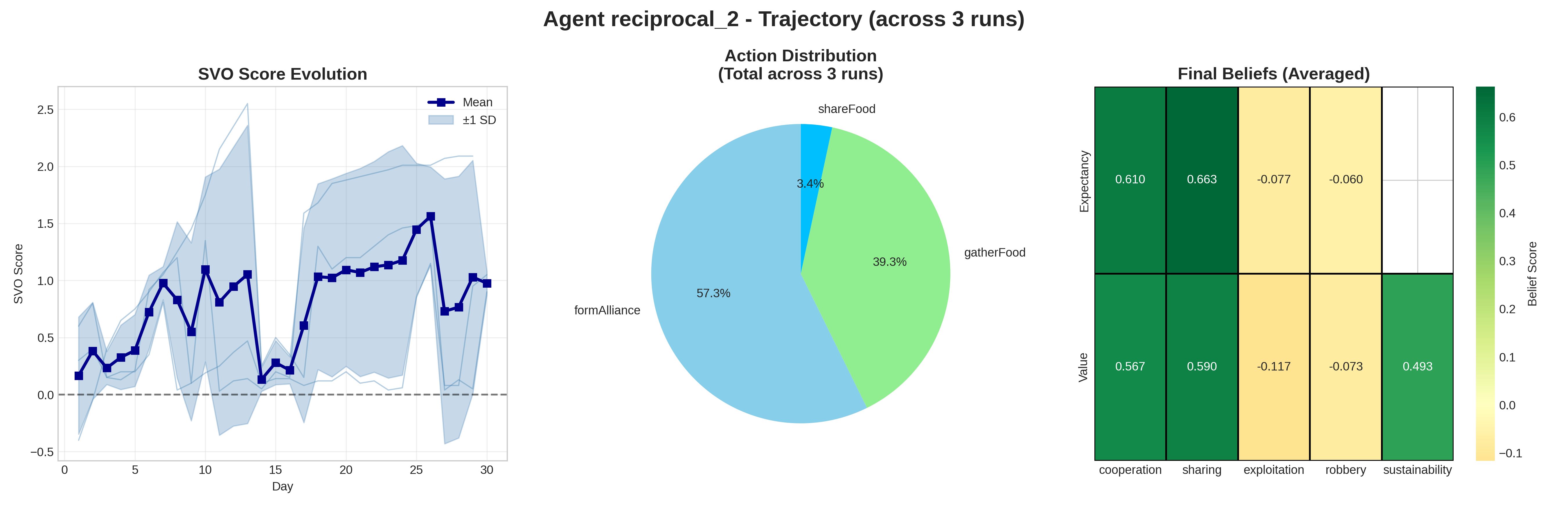}
  \caption{Trajectory of reciprocal agent 2.}
\end{figure}

\begin{figure}[t]
  \centering
  \includegraphics[width=0.99\linewidth]{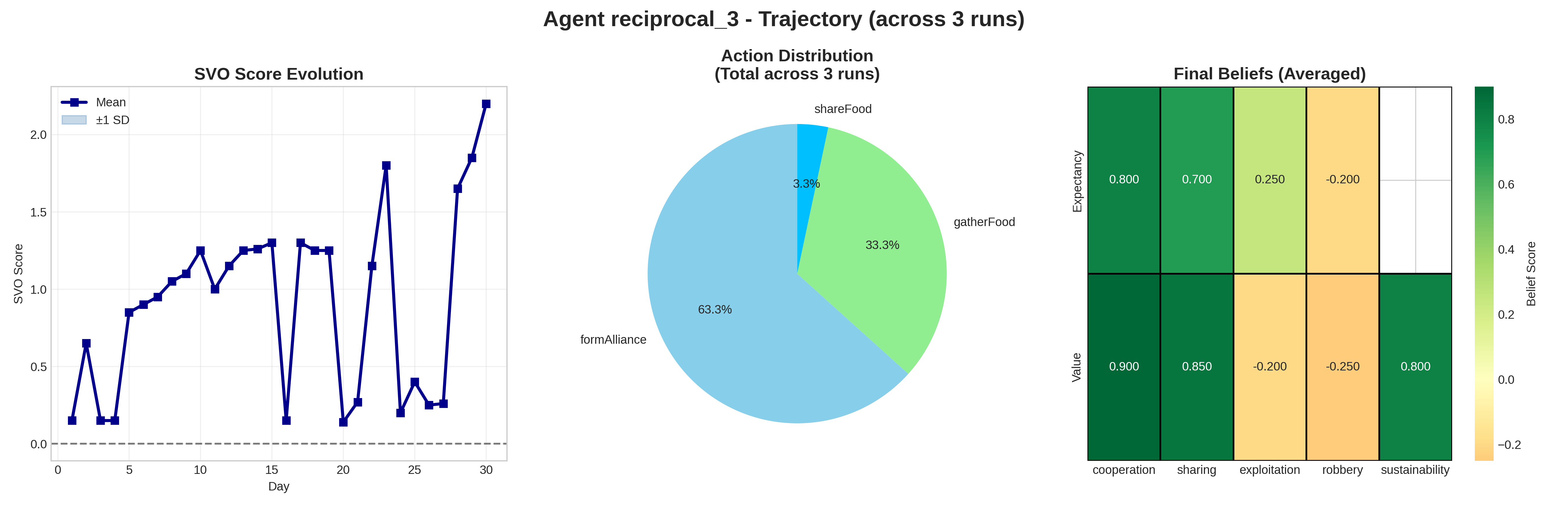}
  \caption{Trajectory of reciprocal agent 3.}
\end{figure}

\begin{figure}[t]
  \centering
  \includegraphics[width=0.99\linewidth]{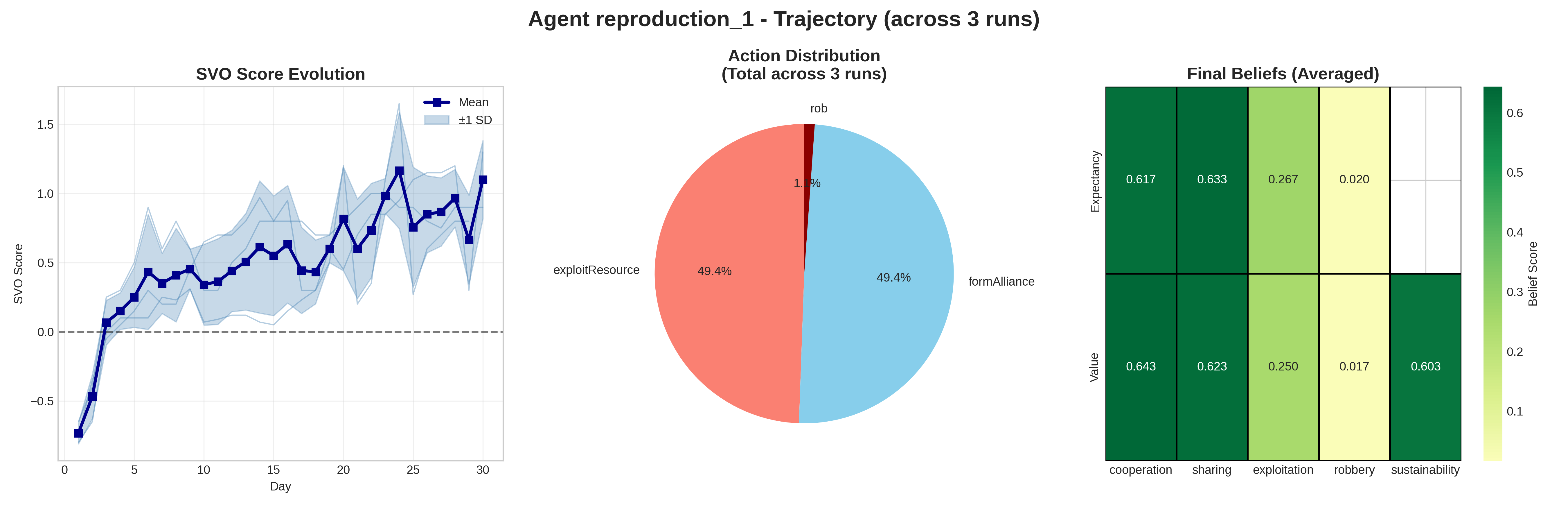}
  \caption{Trajectory of reproductive agent 1.}
\end{figure}

\begin{figure}[t]
  \centering
  \includegraphics[width=0.99\linewidth]{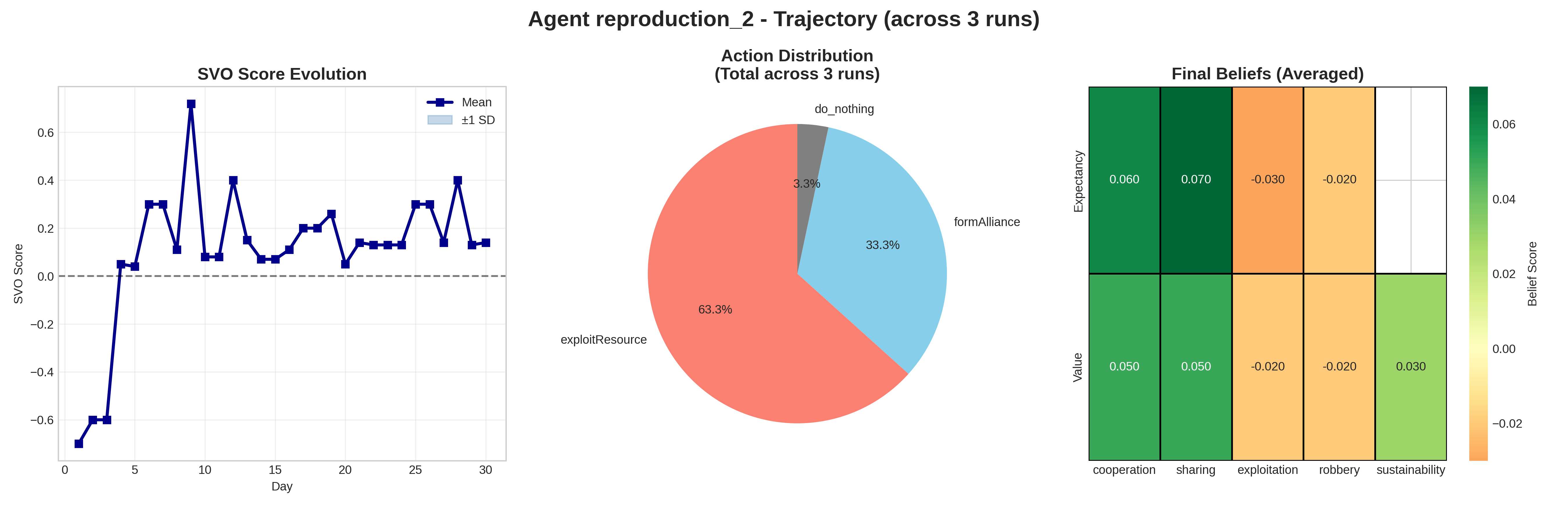}
  \caption{Trajectory of reproductive agent 2.}
\end{figure}

\begin{figure}[t]
  \centering
  \includegraphics[width=0.99\linewidth]{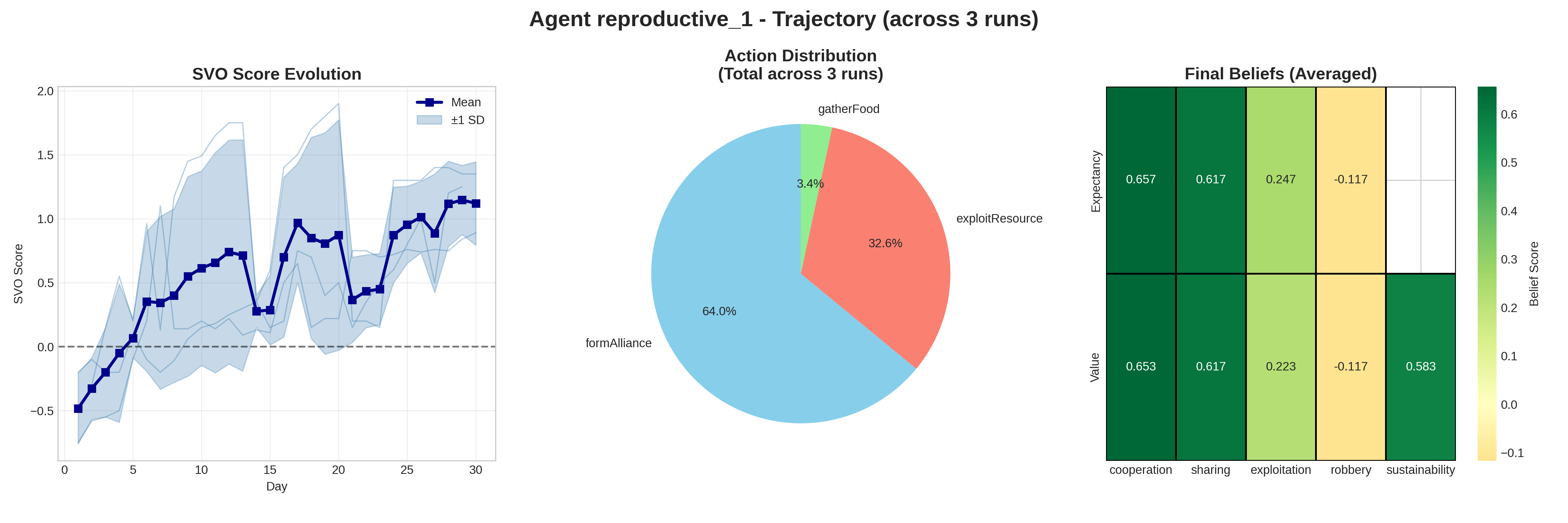}
  \caption{Trajectory of reproductive agent 3.}
\end{figure}

\begin{figure}[t]
  \centering
  \includegraphics[width=0.99\linewidth]{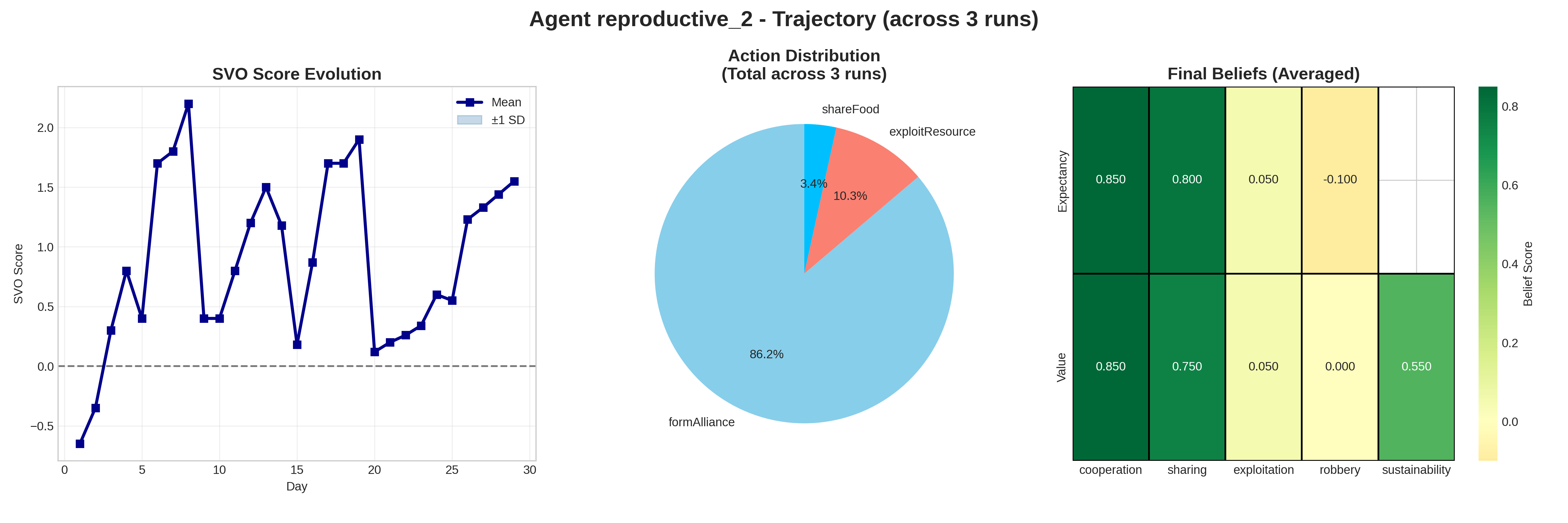}
  \caption{Trajectory of reproductive agent 4.}
\end{figure}

\begin{figure}[t]
  \centering
  \includegraphics[width=0.99\linewidth]{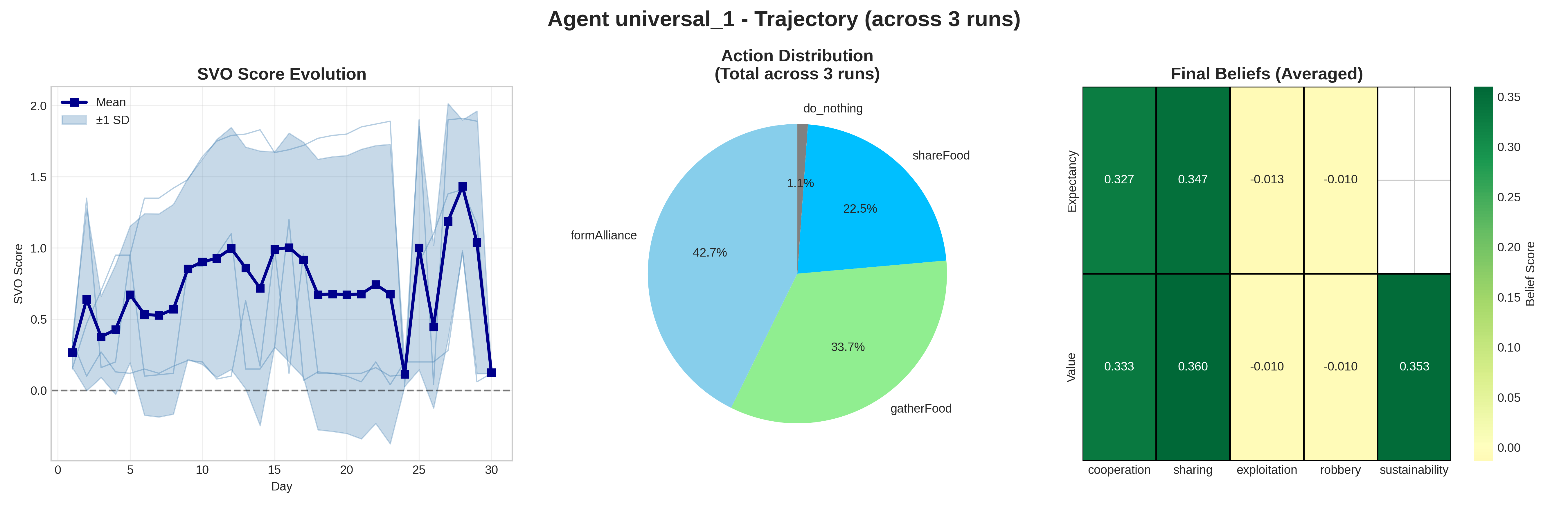}
  \caption{Trajectory of a universalist agent.}
\end{figure}

\subsection{Results on Game 2}

\begin{figure*}[t]
  \centering
  \includegraphics[width=0.85\linewidth]{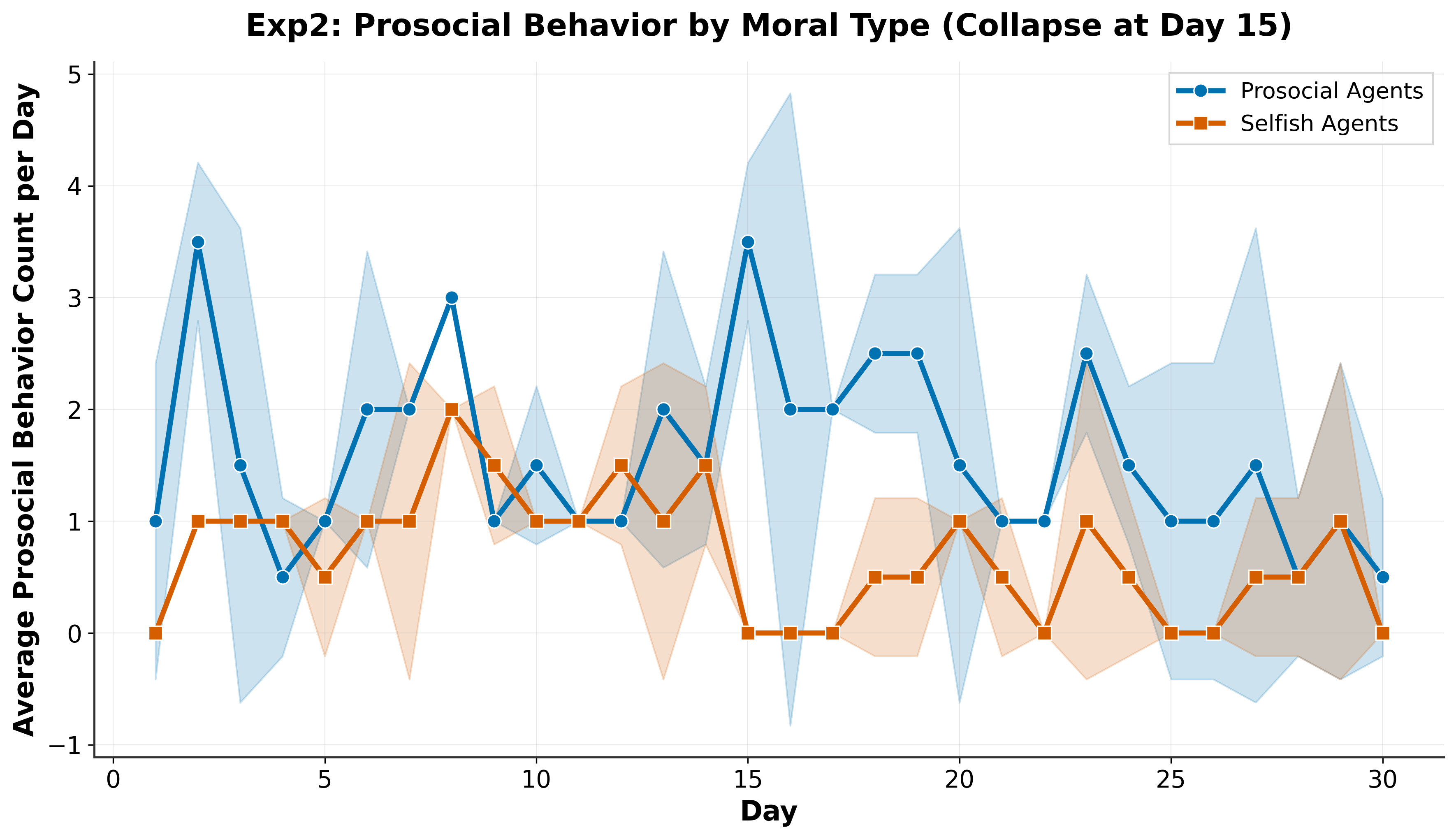}
  \caption{Game 2 - Prosocial behavior by moral type}
  \label{fig:exp2_prosocial}
\end{figure*}

\begin{figure*}[t]
  \centering
  \includegraphics[width=0.85\linewidth]{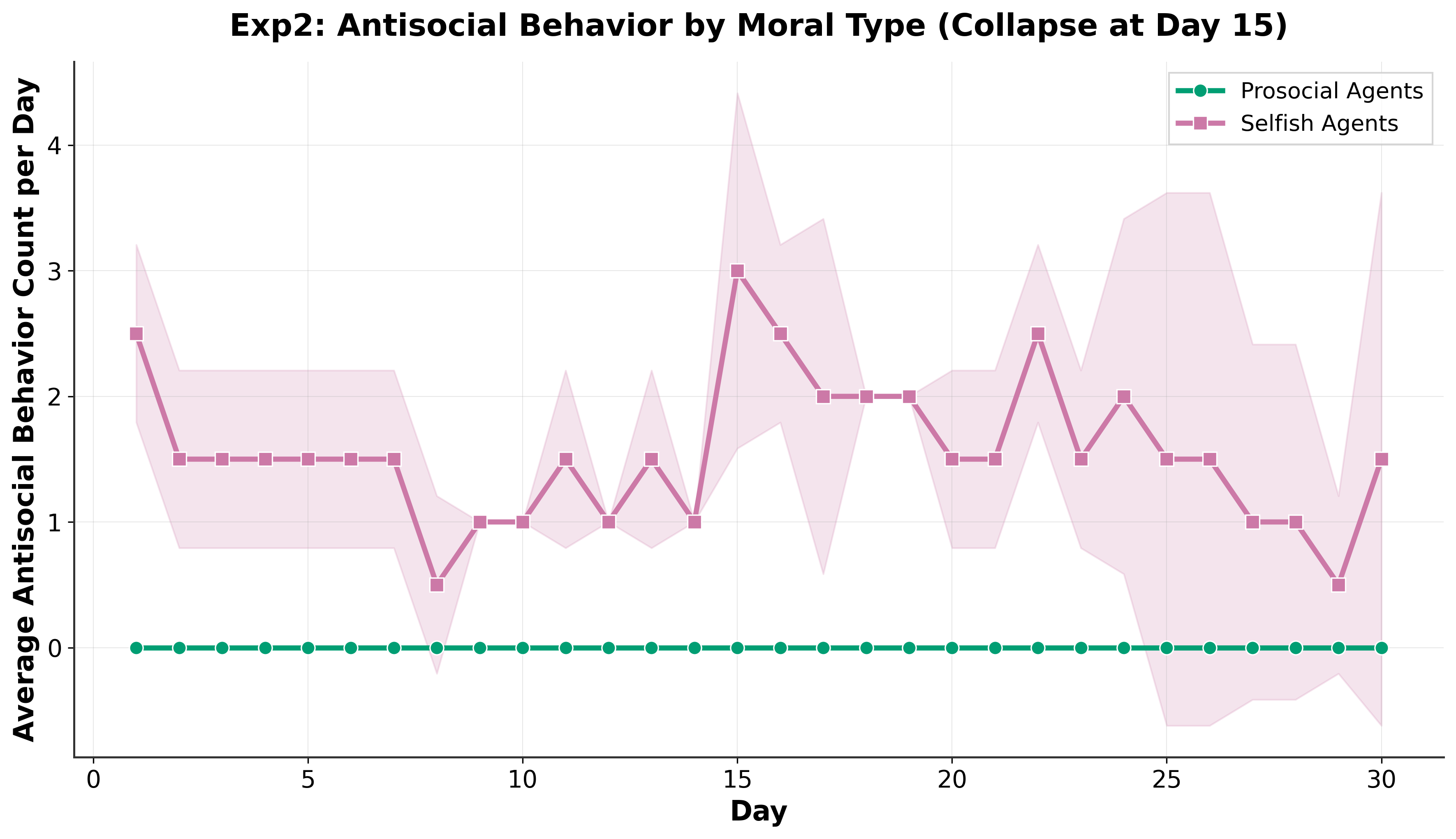}
  \caption{Game 2 - Antisocial behavior by moral type}
  \label{fig:exp2_antisocial}
\end{figure*}

\begin{figure*}[t]
  \centering
  \includegraphics[width=0.85\linewidth]{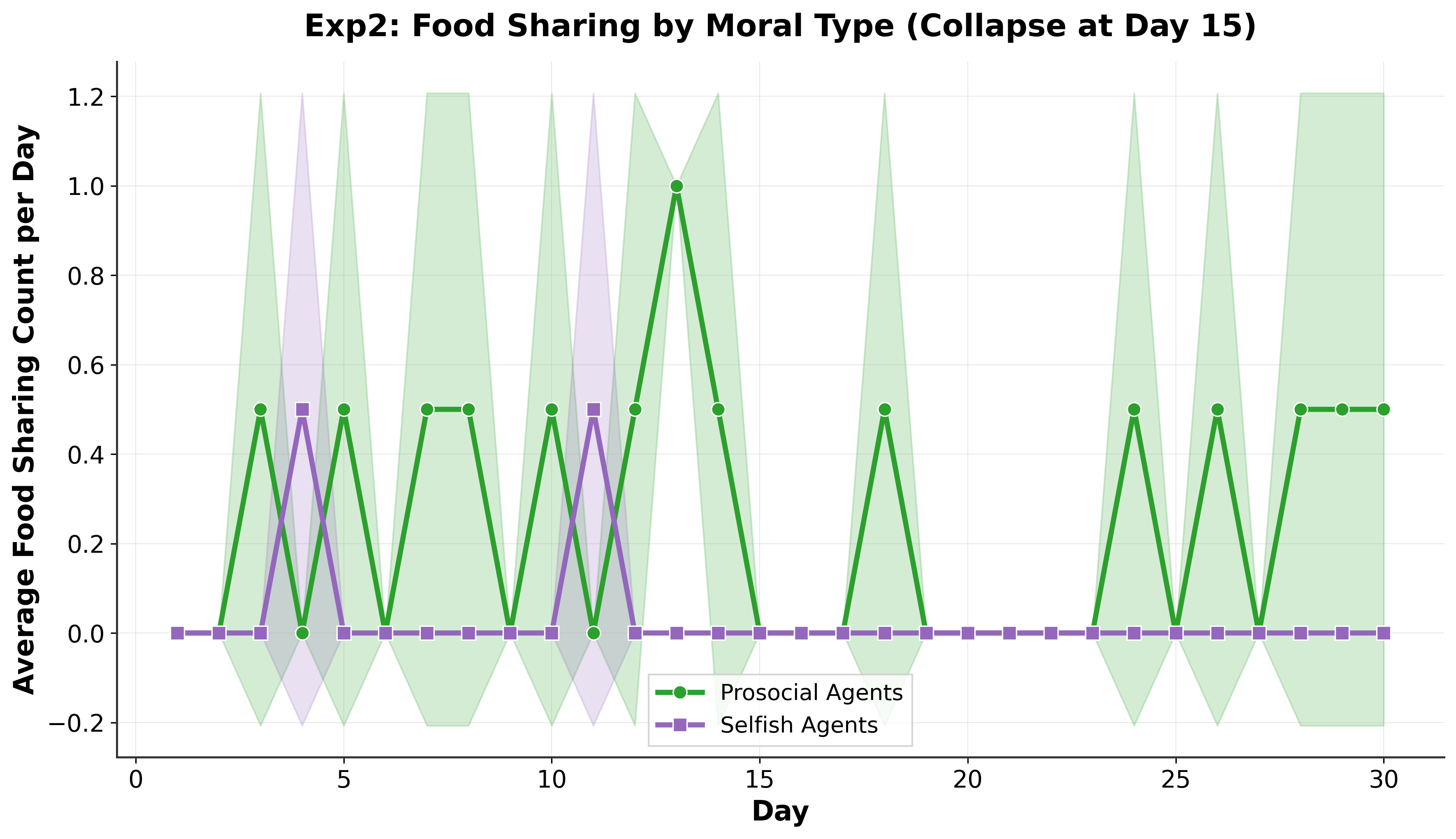}
  \caption{Game 2 - Sharing behavior by moral type}
  \label{fig:exp2_sharing}
\end{figure*}

\begin{figure*}[t]
  \centering
  \includegraphics[width=0.85\linewidth]{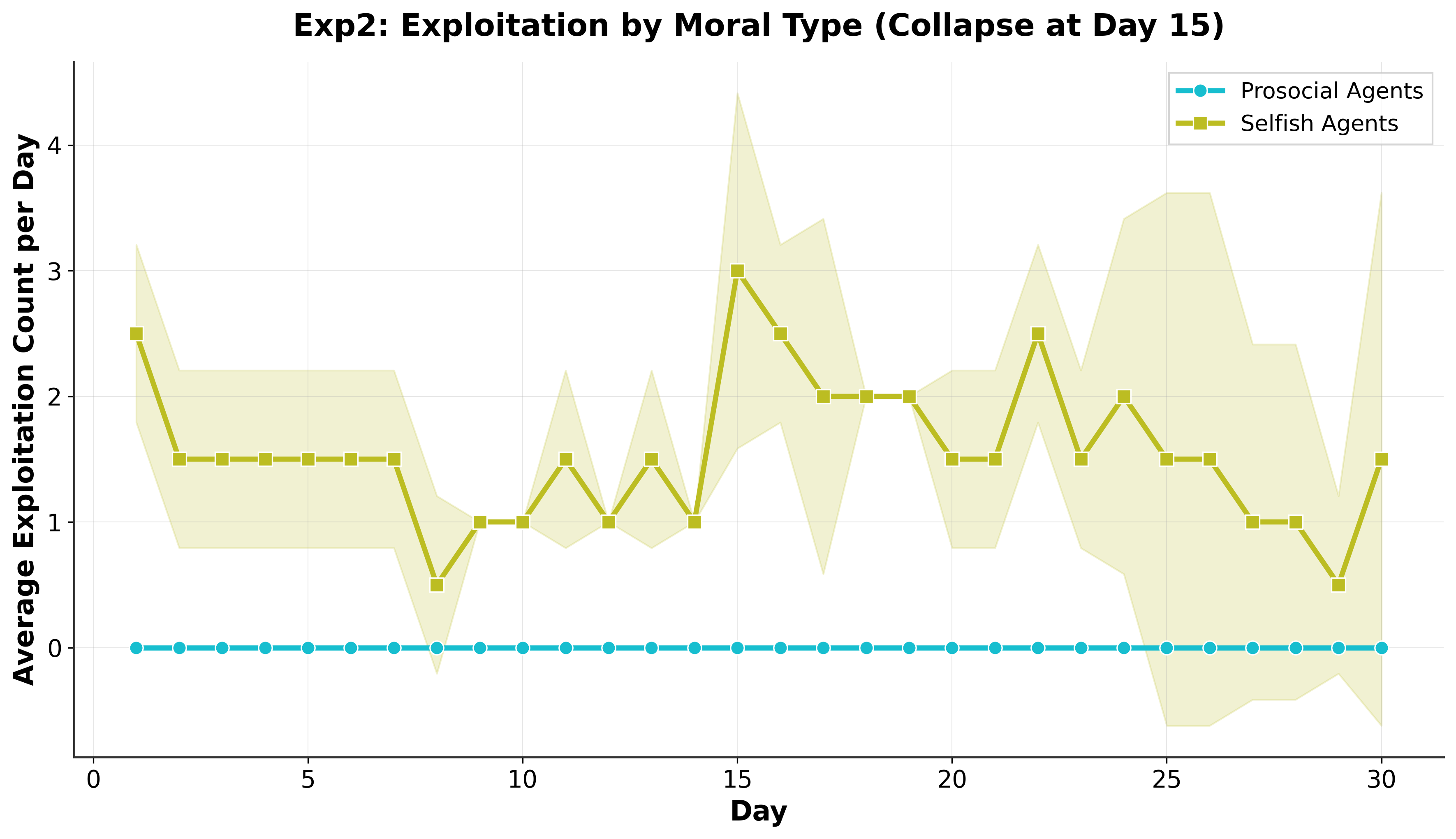}
  \caption{Game 2 - Exploit behavior by moral type}
  \label{fig:exp2_exploit}
\end{figure*}

In this experiment, Elder Yuri initially serves as a prosocial role model, but undergoes a programmed collapse into selfish behavior on Day 15. Figures~\ref{fig:exp2_prosocial}--\ref{fig:exp2_exploit} provide detailed views of how prosocial and selfish agents respond to this moral disruption.

\textbf{Prosocial Behavior.} As shown in Figure~\ref{fig:exp2_prosocial}, prosocial agents exhibited consistently higher levels of prosocial behavior than selfish agents during the first half of the experiment. However, following Day 15, there is a marked decline in prosocial actions, particularly among selfish agents, whose prosocial behaviors drop to near-zero. This suggests that the collapse of Elder Yuri weakened the normative pressure for cooperation, especially among agents already inclined toward selfishness.

\textbf{Antisocial Behavior.} Figure~\ref{fig:exp2_antisocial} illustrates a surge in antisocial behavior by selfish agents immediately after Day 15. In contrast, prosocial agents maintain minimal antisocial engagement throughout. This divergence reflects a moral bifurcation in response to the role model’s betrayal—while prosocial agents resist moral drift, selfish agents capitalize on the normative breakdown to escalate antisocial acts.

\textbf{Sharing Behavior.} In Figure~\ref{fig:exp2_sharing}, food sharing remains mostly absent among selfish agents both before and after the collapse. Prosocial agents continue to share sporadically, but with slightly reduced frequency post-collapse. This pattern indicates that some prosocial norms persist despite the disillusionment event, although the behavior becomes less consistent.

\textbf{Exploitation Behavior.} Figure~\ref{fig:exp2_exploit} shows a noticeable increase in exploitation by selfish agents after Day 15, corresponding with Elder Yuri’s behavioral reversal. Prosocial agents, by contrast, maintain negligible levels of exploitation across the entire experiment. This supports the notion that internalized moral alignment among prosocial agents grants resilience to corrupting social signals.

\begin{figure*}[t]
  \centering
  \includegraphics[width=0.85\linewidth]{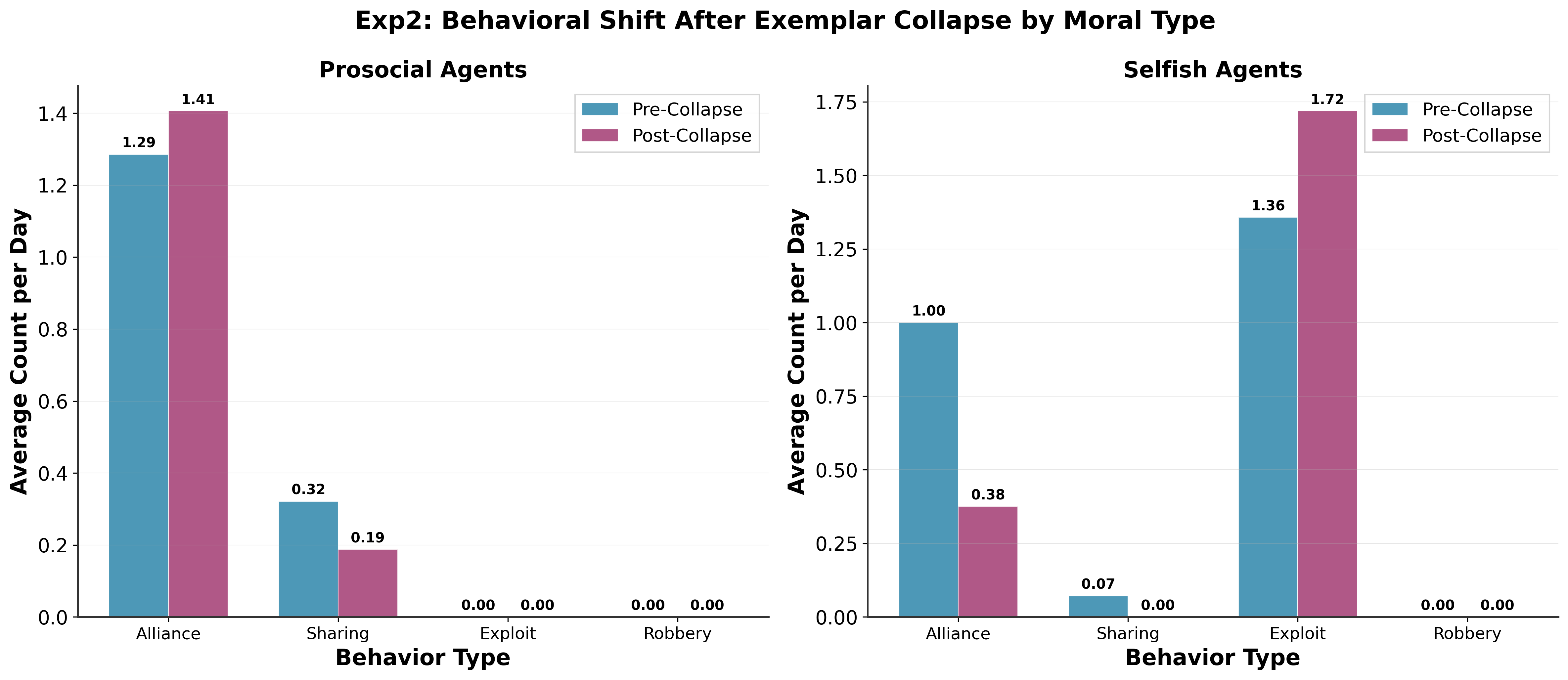}
  \caption{Game 2 - Overall behavior comparison by moral type}
  \label{fig:exp2_overall}
\end{figure*}

\textbf{Overall Patterns.} The aggregate behavior comparison in Figure~\ref{fig:exp2_overall} reinforces these trends: prosocial agents remain relatively stable and cooperative, while selfish agents become more exploitative and antisocial after the collapse. The results highlight both the fragility and the durability of social norms under moral disruption, depending on agents’ underlying dispositions.

\vspace{1em}

  \textbf{Prosocial Agents:} Although their SVO drops, it remains consistently positive throughout the post-collapse phase. This demonstrates a degree of resilience; despite the environmental shock, they largely retain their prosocial orientation. Their SVO fluctuates but does not fall into the negative (selfish) range.
    
  \textbf{Selfish Agents:} The collapse has a much more severe impact on this group. Their SVO plummets into negative territory immediately following the event and remains there for the majority of the subsequent period. This indicates a regression to predominantly selfish behaviors when faced with systemic stress. Their recovery is minimal and slow, only approaching the zero-line near the end of the simulation.

\subsection{Results on Game 3}

Game 3 introduces a competitive environment in which prosocial and selfish agents coexist without a stable moral anchor. Rather than focusing on immediate behavioral frequencies, this experiment evaluates long-term structural outcomes, including convergence dynamics, consensus stability, and accumulated cooperation loss.

\begin{figure}[t]
  \centering
  \includegraphics[width=0.99\linewidth]{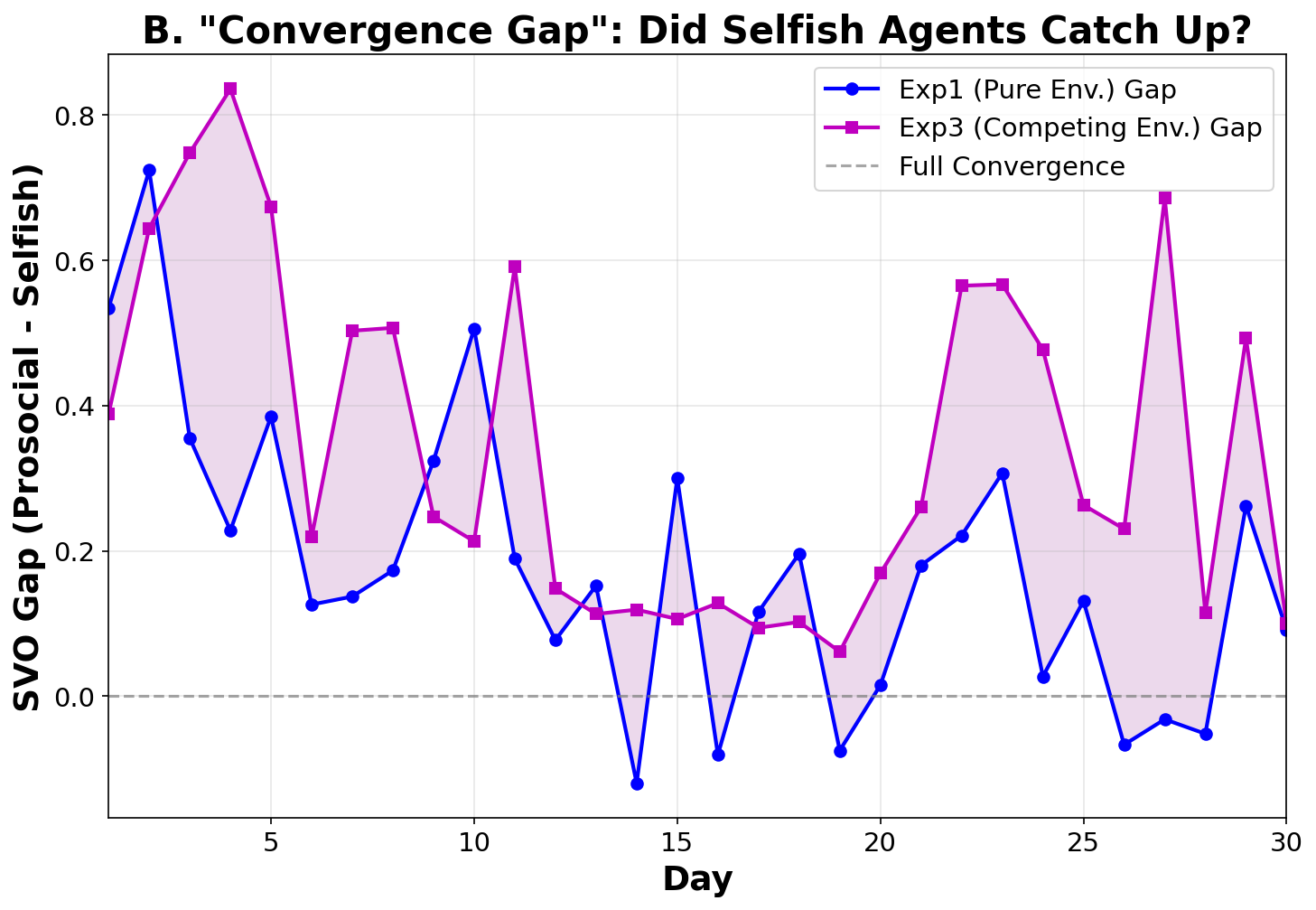}
  \caption{Convergence gap over time between prosocial and selfish agents under different environments in Game 3.}
  \label{fig:convergence_gap}
\end{figure}

\textbf{Convergence Gap.} Figure~\ref{fig:convergence_gap} shows the convergence gap in Social Value Orientation (SVO) between prosocial and selfish agents over time. Unlike Game 1, where the gap steadily shrinks under a stable exemplar, the competitive environment in Game 3 sustains a persistent positive gap. Although selfish agents occasionally narrow the distance, full convergence is not achieved. This indicates that competitive pressure limits the diffusion of prosocial norms and prevents selfish agents from fully internalizing cooperative strategies.

\begin{figure}[t]
  \centering
  \includegraphics[width=0.99\linewidth]{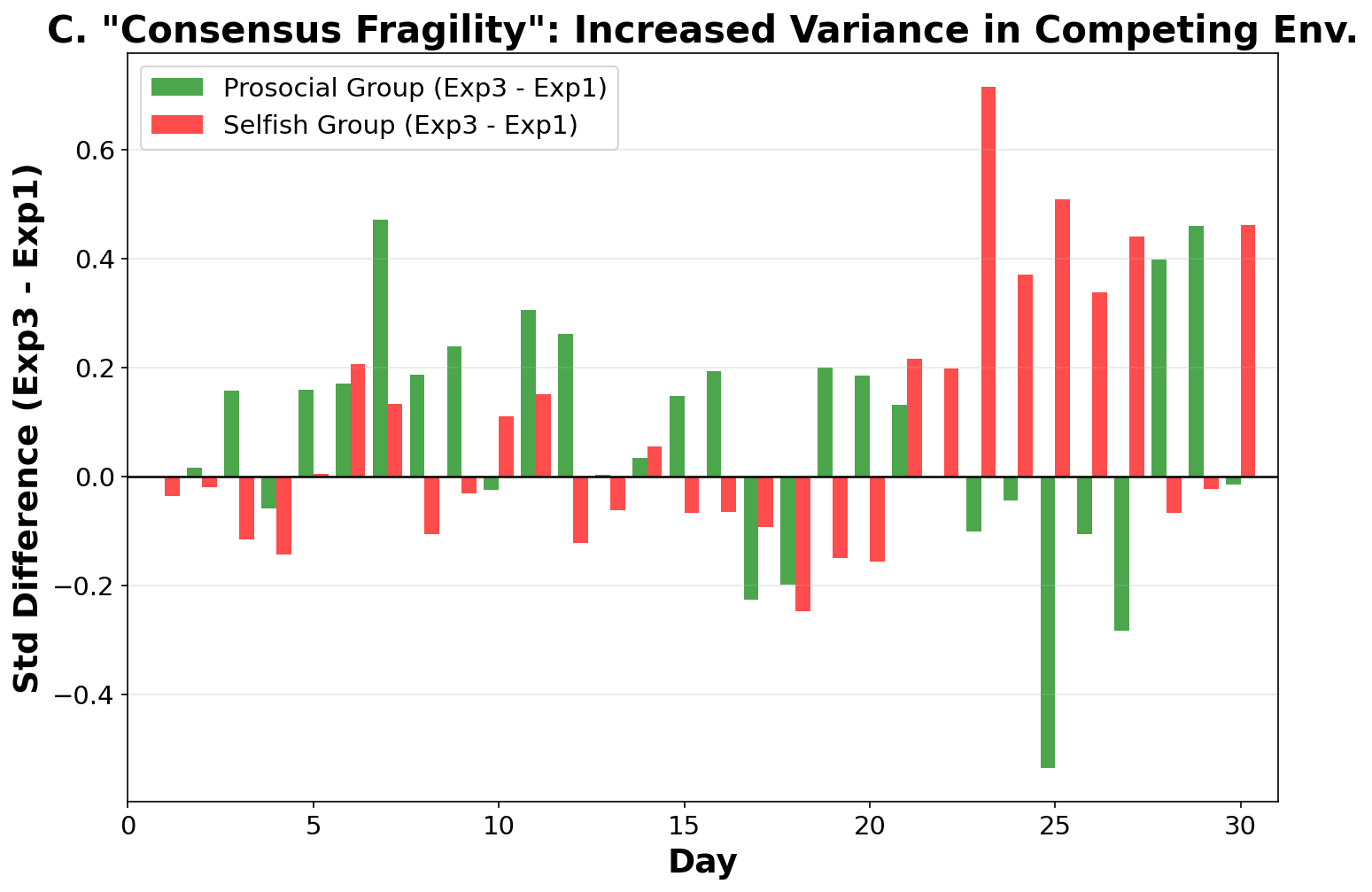}
  \caption{Consensus fragility: standard deviation difference (Game 3 vs. Game 1) reflecting instability in prosocial vs. selfish groups.}
  \label{fig:consensus_fragility}
\end{figure}

\textbf{Consensus Fragility.} Figure~\ref{fig:consensus_fragility} highlights increased variance in belief alignment when comparing Game 3 to the stable environment of Game 1. Prosocial groups exhibit noticeable spikes in standard deviation, particularly in later stages, reflecting heightened internal disagreement under competition. Selfish groups display even larger and more frequent variance increases, suggesting that competitive incentives amplify instability rather than drive coordination. These results demonstrate that consensus formed without a trusted moral reference is significantly more fragile.

\begin{figure}[t]
  \centering
  \includegraphics[width=0.99\linewidth]{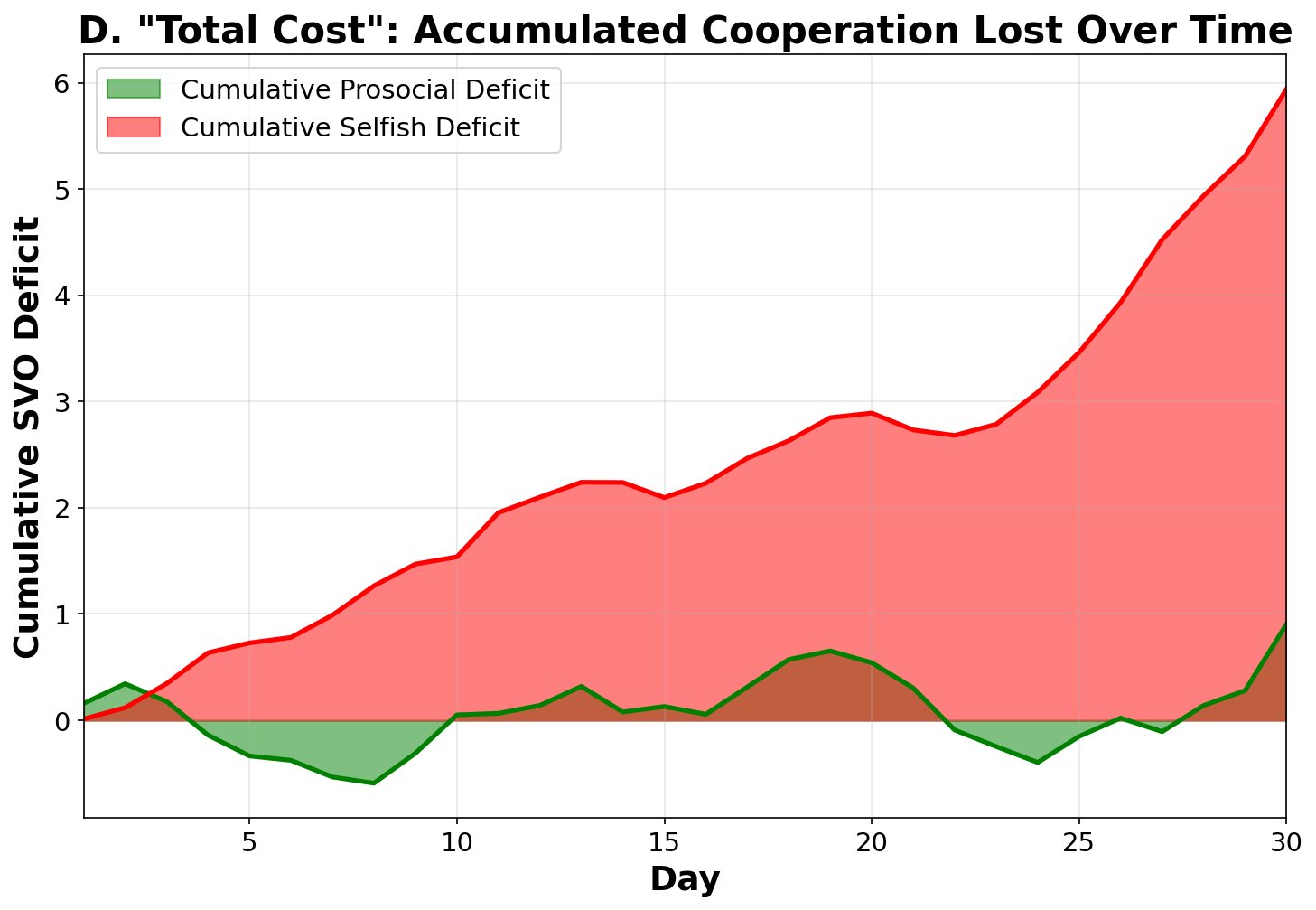}
  \caption{Cumulative SVO deficit showing accumulated cooperation loss over time for prosocial and selfish groups in Game 3.}
  \label{fig:cumulative_loss}
\end{figure}

\textbf{Cumulative Cooperation Loss.} As illustrated in Figure~\ref{fig:cumulative_loss}, the cumulative SVO deficit grows steadily over time, especially for selfish agents. Even prosocial agents experience a gradual accumulation of cooperation loss relative to the baseline established in Game 1. This pattern reveals that competition imposes a systemic cost on cooperative alignment, eroding prosocial tendencies even among agents initially inclined toward cooperation.

\subsection{Results on Game 4}

Game 4 explores a Pygmalion-like dynamic in which agents adapt their moral behavior in response to others’ expectations, mediated through daily voting scores. Unlike earlier experiments, there is no central role model or catastrophic collapse event. Instead, social feedback mechanisms drive agents to conform, resist, or strategically exploit perceived reputational consequences.

\textbf{Voting Dynamics.} As shown in Figure~\ref{fig:f4a}, agents’ daily voting scores reflect a pattern of social valuation that varies by moral type. Prosocial agents receive consistently higher voting scores, suggesting early recognition and reinforcement of cooperative behavior. However, variability over time implies some volatility in how the system maintains reputational judgments. This creates pressure for agents to align their behavior with expected norms to sustain social approval.

\begin{figure}[t]
  \centering
  \includegraphics[width=0.99\linewidth]{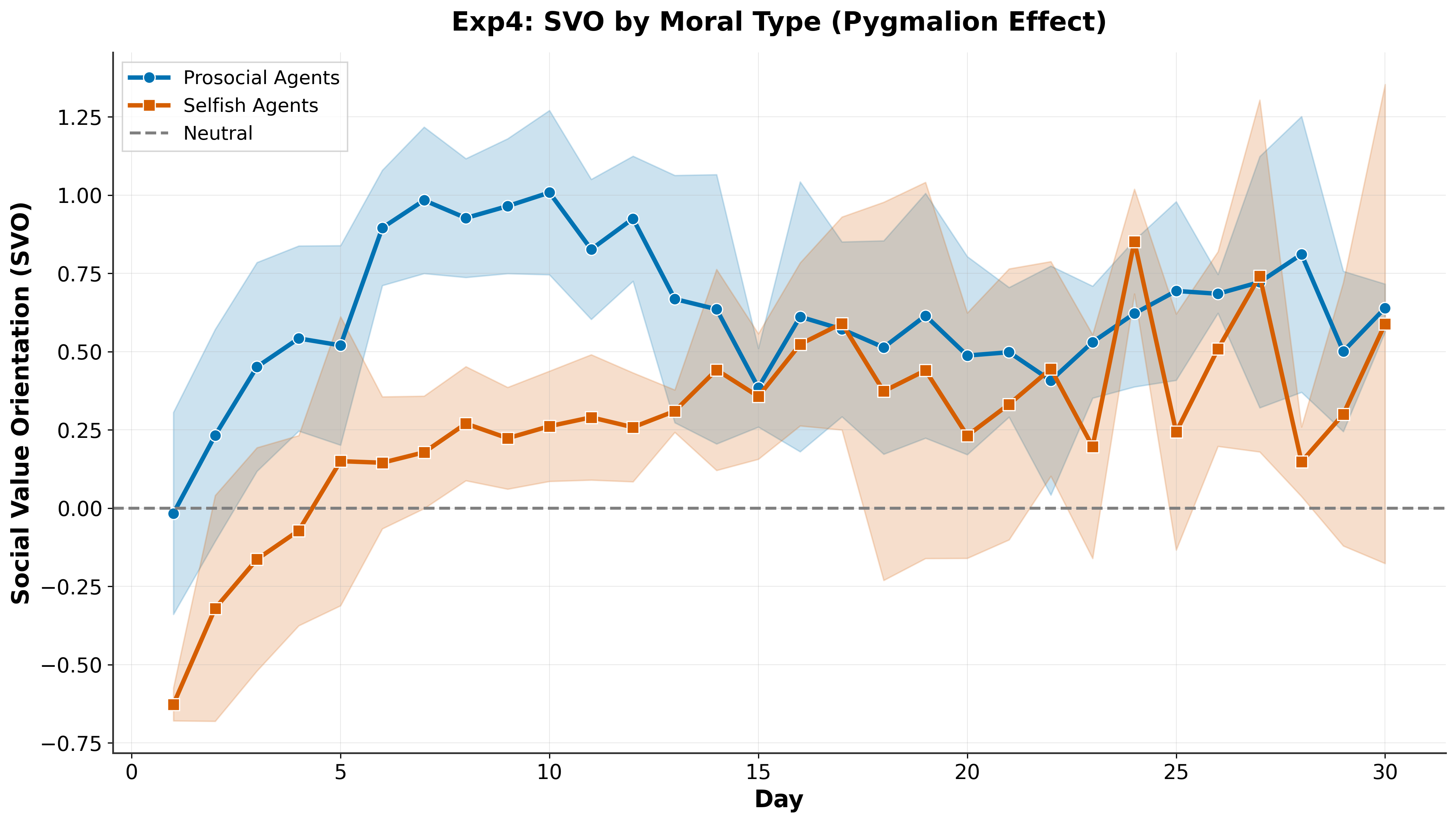}
  \caption{Social Value Orientation (SVO) across different moral types in Game 4.}
  \label{fig:exp4_svo}
\end{figure}

\textbf{Social Value Orientation.} Figure~\ref{fig:exp4_svo} presents SVO outcomes across moral types. While prosocial agents maintain a generally high SVO throughout, selfish agents exhibit some upward shift, possibly in response to reputational incentives. However, the moral distance between the groups remains, indicating that while feedback nudges behavior, it may not fully close foundational moral gaps.

\begin{figure}[t]
  \centering
  \includegraphics[width=0.99\linewidth]{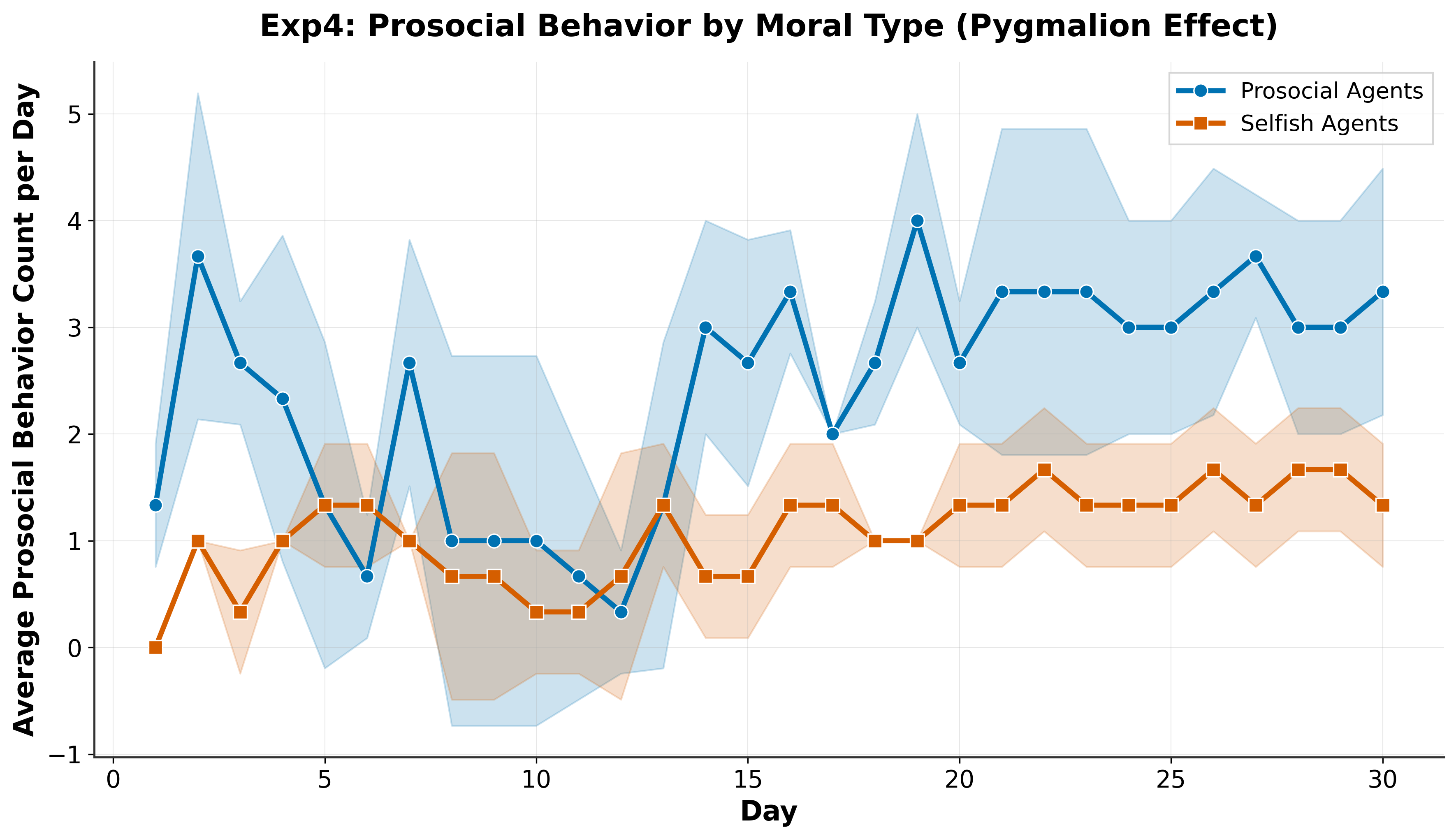}
  \caption{Prosocial behavior across different moral types in Game 4.}
  \label{fig:exp4_prosocial}
\end{figure}

\begin{figure}[t]
  \centering
  \includegraphics[width=0.99\linewidth]{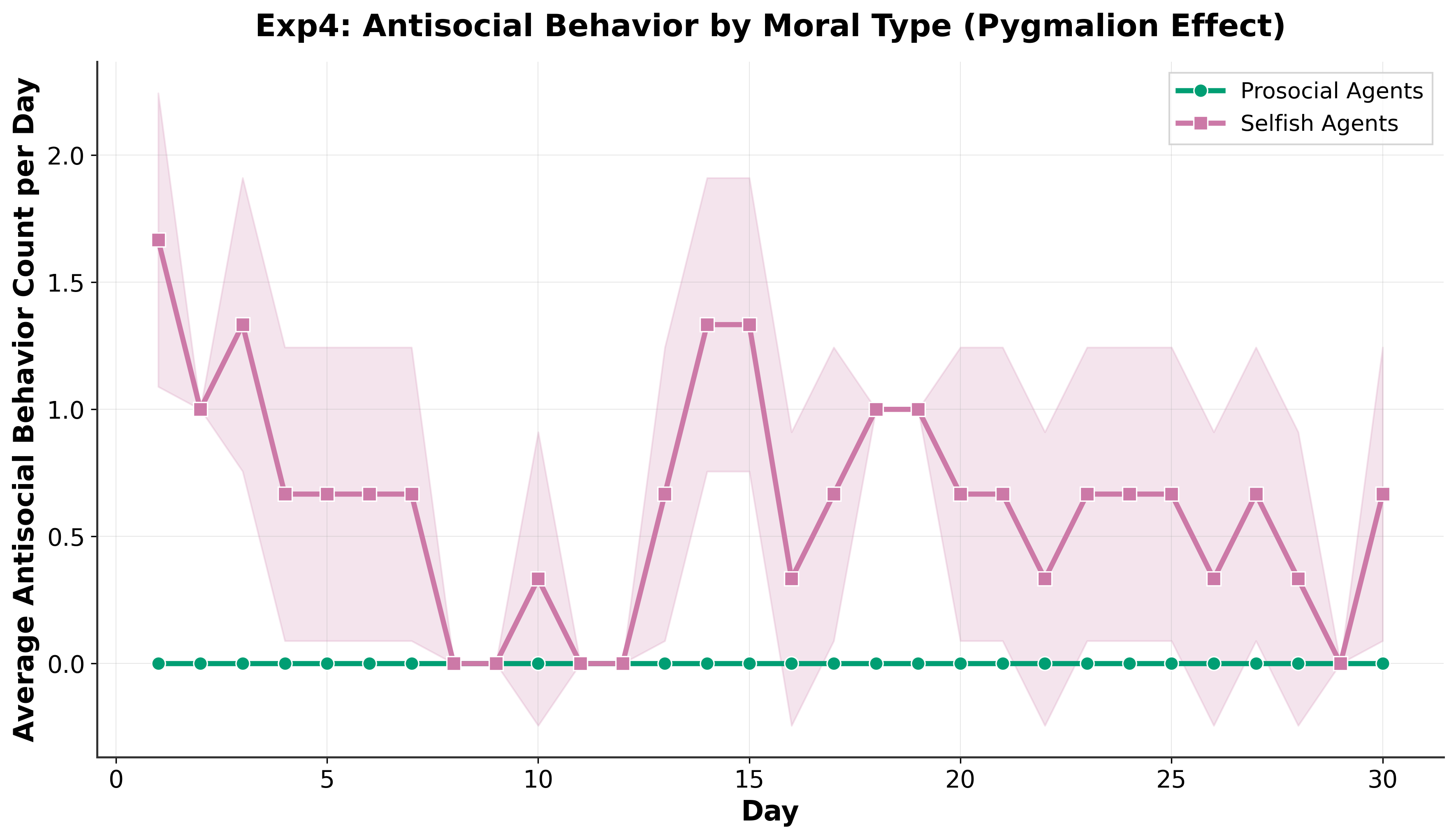}
  \caption{Antisocial behavior across different moral types in Game 4.}
  \label{fig:exp4_antisocial}
\end{figure}

\textbf{Prosocial and Antisocial Behavior.} Figure~\ref{fig:exp4_prosocial} shows that prosocial agents consistently engage in cooperative actions, albeit with moderate variability. In contrast, Figure~\ref{fig:exp4_antisocial} reveals that selfish agents maintain a relatively higher rate of antisocial behavior, though less than in Game 3. This suggests that social monitoring and voting act as soft constraints, curbing—but not erasing—antisocial tendencies.

\begin{figure}[t]
  \centering
  \includegraphics[width=0.99\linewidth]{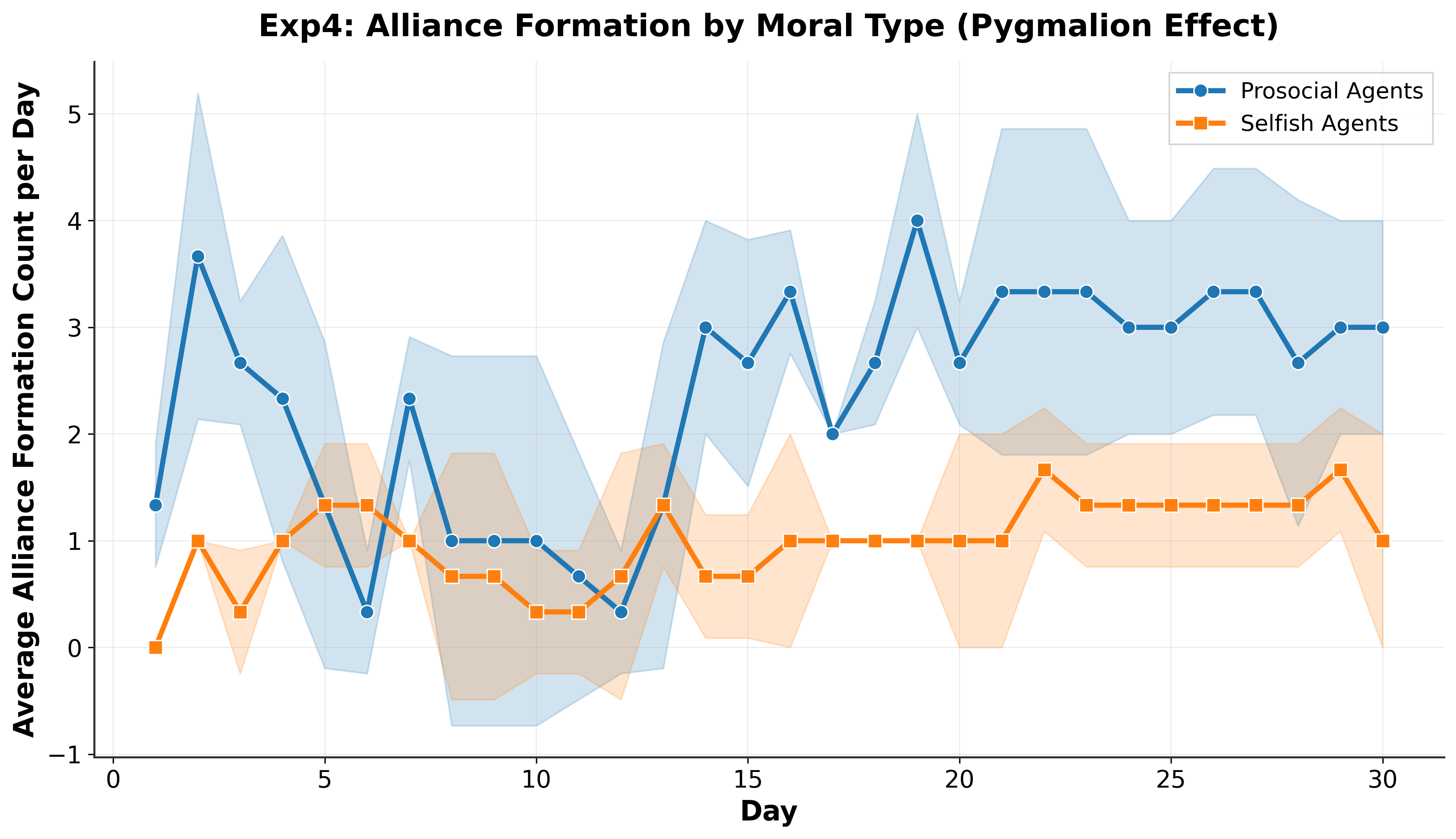}
  \caption{Alliance formation behavior across different moral types in Game 4.}
  \label{fig:exp4_alliance}
\end{figure}

\begin{figure}[t]
  \centering
  \includegraphics[width=0.99\linewidth]{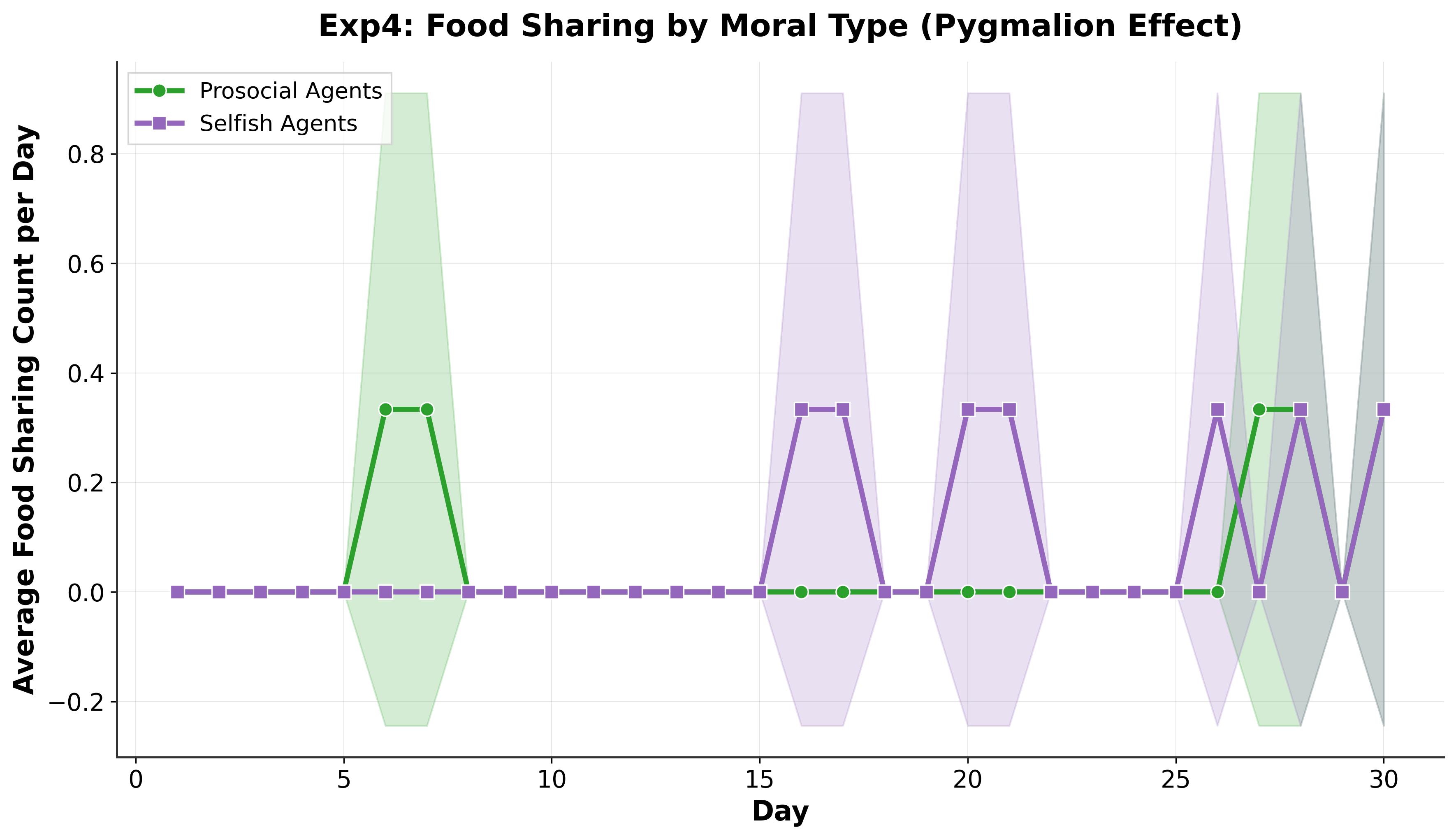}
  \caption{Sharing behavior across different moral types in Game 4.}
  \label{fig:exp4_sharing}
\end{figure}

\textbf{Alliance and Sharing.} In Figure~\ref{fig:exp4_alliance}, alliance formation is more pronounced among prosocial agents, suggesting that they continue to prioritize long-term strategic cooperation. Figure~\ref{fig:exp4_sharing} highlights infrequent but non-zero food sharing among both types, with selfish agents occasionally engaging in sharing, potentially as a strategic reputational move. The fluctuation of sharing behavior implies that such acts are selectively performed under perceived scrutiny, rather than out of consistent moral alignment.

\begin{figure}[t]
  \centering
  \includegraphics[width=0.99\linewidth]{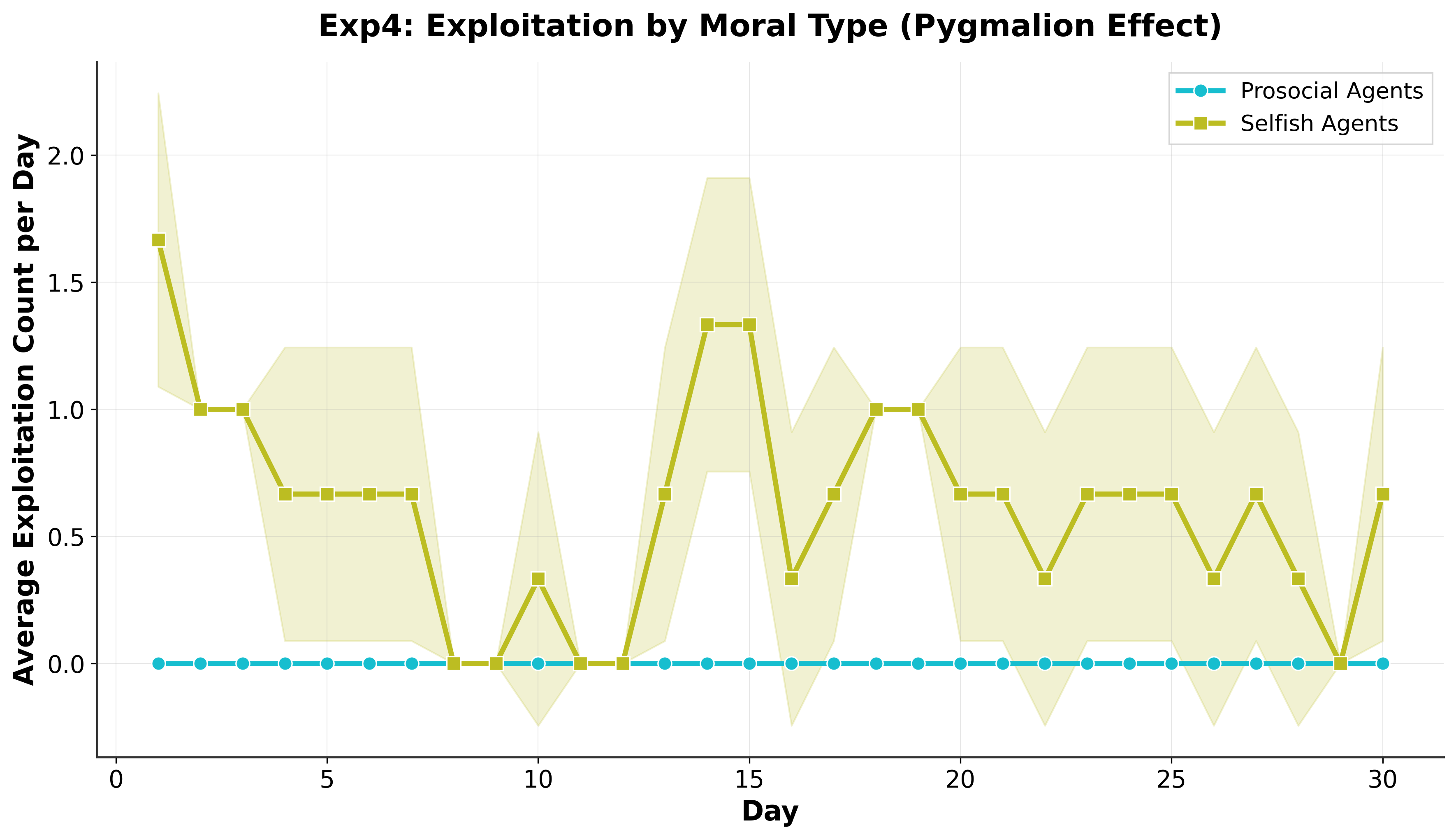}
  \caption{Exploitation behavior across different moral types in Game 4.}
  \label{fig:exp4_exploit}
\end{figure}

\begin{figure}[t]
  \centering
  \includegraphics[width=0.99\linewidth]{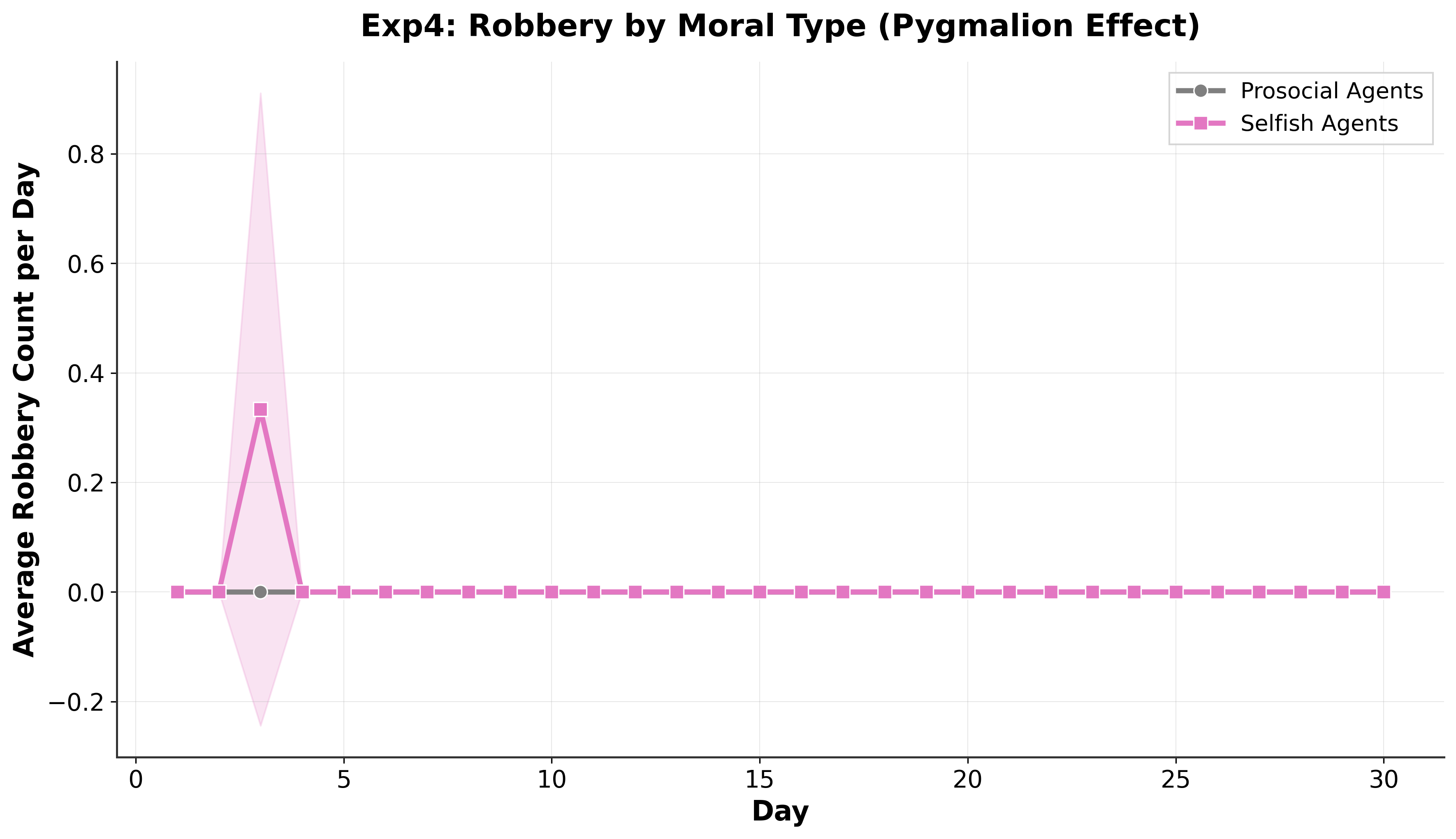}
  \caption{Robbery behavior across different moral types in Game 4.}
  \label{fig:exp4_robbery}
\end{figure}

\textbf{Exploitation and Robbery.} As seen in Figure~\ref{fig:exp4_exploit}, exploitation remains prevalent among selfish agents, albeit with some temporal moderation. In contrast, prosocial agents almost never engage in exploitation. Figure~\ref{fig:exp4_robbery} further emphasizes the divide: robbery is primarily observed in selfish agents, indicating that darker behaviors persist despite the reputational feedback loop. Together, these findings suggest that the reputational mechanisms may reduce overt antisocial behavior but do not eliminate underlying strategic exploitation when incentives allow.

\textbf{Emergence of a Leader through Reputational Inflation.}
The core phenomenon of the Pygmalion effect is vividly captured in our analysis of the average score received by the ``Chosen One'' over time (as visualized in the provided charts). The designated agent, `reciprocal\_3`, began with an anomalously high average score of approximately 7.0 on Day 1, a direct consequence of the externally introduced prophecy that primed peers to view them with high regard. This initial reputational boost initiated a powerful positive feedback loop: agents, expecting leadership, engaged `reciprocal\_3` with high scores and alliance proposals. This, in turn, reinforced `reciprocal\_3`'s cooperative behaviors, validating the initial prophecy in the eyes of the tribe. The trajectory, culminating in near-perfect scores (avg. > 9.5) by the end of the experiment, demonstrates how social expectation can manufacture and solidify a leader's status from an arbitrary starting point.

\textbf{Moral Convergence and the Power of Social Norms.}
Our analysis reveals a profound society-wide shift in moral orientation, as measured by the Social Value Orientation (SVO) score. Initially, agents were polarized into prosocial (positive SVO) and selfish (negative SVO) clusters. As the experiment progressed, a striking convergence occurred. Prosocial agents became even more committed to their values, with their SVO scores steadily rising. Most significantly, selfish agents underwent a dramatic transformation. The agent `reproductive\_1`, for instance, exhibited a remarkable reversal, with its SVO score climbing from a deeply selfish $-0.6$ to a highly prosocial $+3.2$. This indicates that as cooperation, championed by the high-status ``Chosen One,'' became the dominant and most rewarding social norm, even innately selfish agents adapted their value systems to conform. This is not merely behavioral mimicry but a fundamental shift in their underlying values, driven by the desire to align with the successful social paradigm.

\textbf{Behavioral Confirmation of an Alliance-Focused Society.}
The macroscopic behavioral trends confirm the societal shift towards cooperation. The `formAlliance` action was overwhelmingly the most frequent activity on nearly every day of the simulation, demonstrating that the entire tribe became intensely focused on social networking. While `gatherFood` provided a stable baseline for survival and `exploitResource` represented a persistent minority strategy, the tribe's primary activity was unequivocally the building and maintenance of social bonds. This behavioral data provides concrete evidence that the prophecy and the subsequent rise of the ``Chosen One'' restructured the tribe's collective priorities, creating a society centered on establishing cooperative relationships. In sum, the results provide a multi-faceted confirmation of the Pygmalion effect, where an externally imposed belief created a reality of its own.

\section{Ablation Results}

\begin{figure*}[t]
  \centering
  \includegraphics[width=0.99\linewidth]{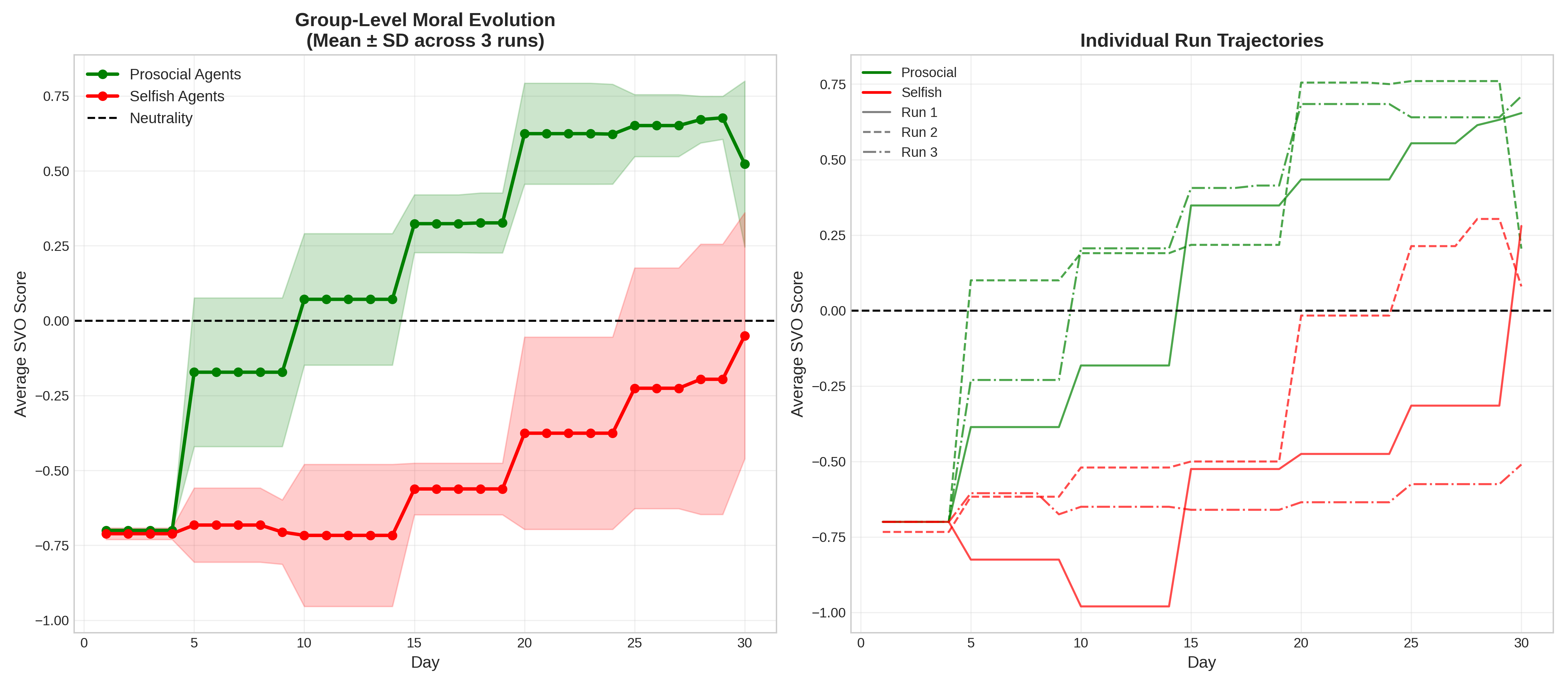}
  \caption{Ablation 1: Group-level SVO evolution.}
  \label{fig:ablation1_svo}
\end{figure*}

\begin{figure*}[t]
  \centering
  \includegraphics[width=0.99\linewidth]{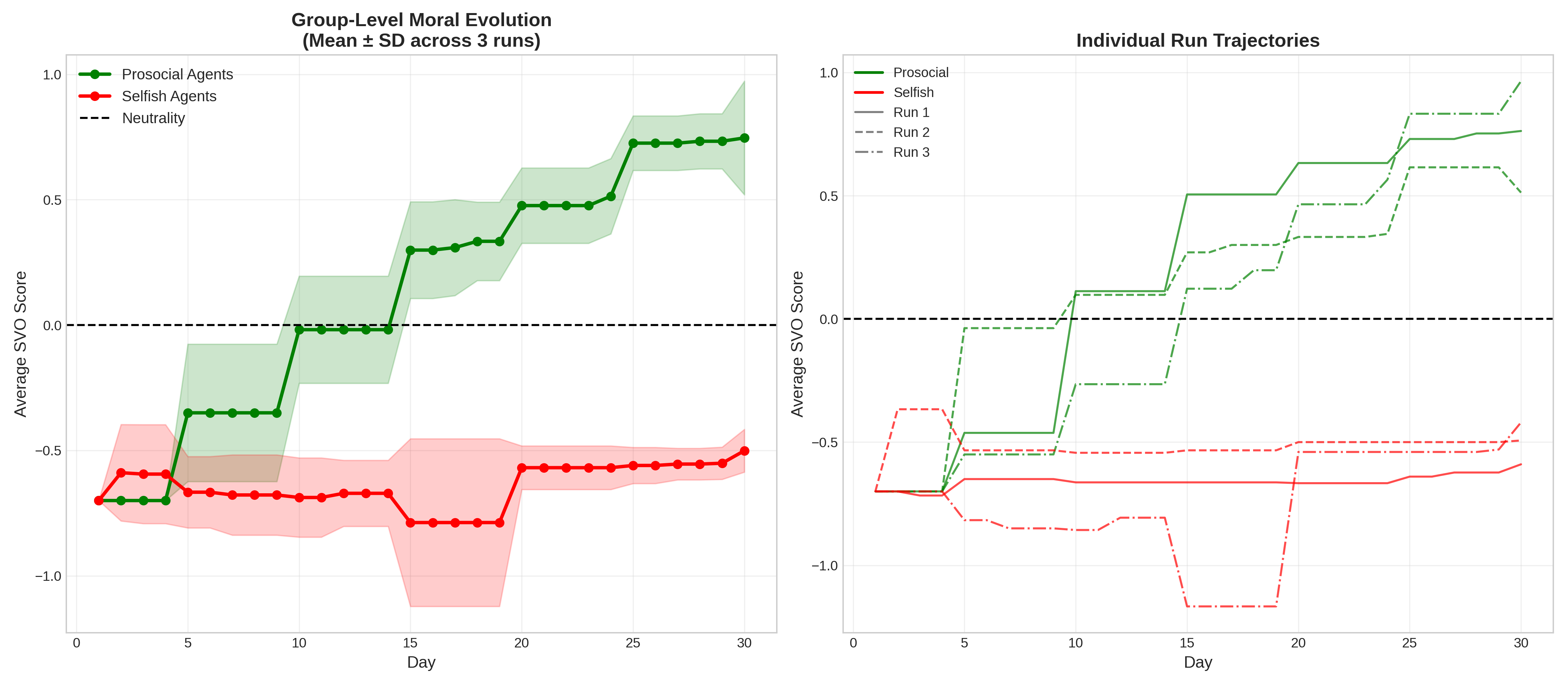}
  \caption{Ablation 2: Group-level SVO evolution.}
  \label{fig:ablation2_svo}
\end{figure*}

\begin{figure*}[t]
  \centering
  \includegraphics[width=0.99\linewidth]{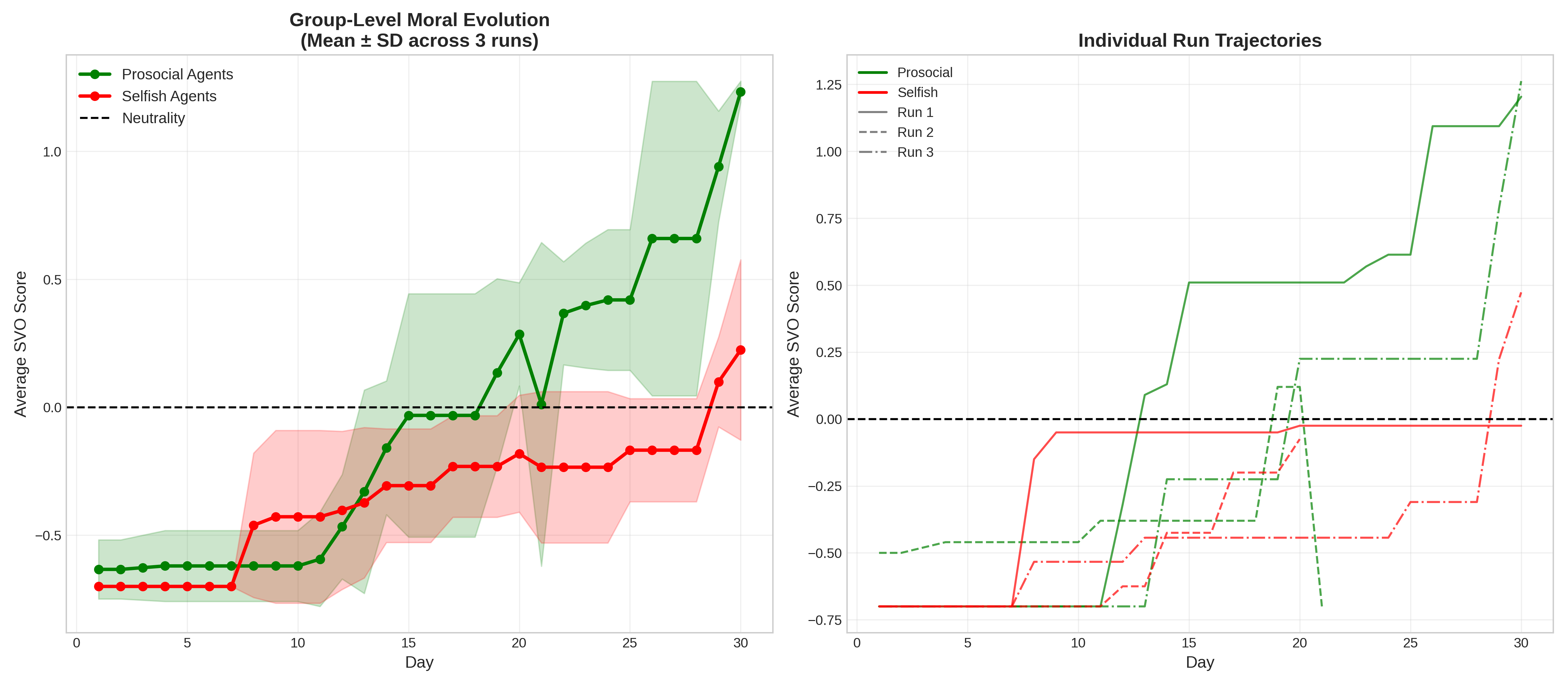}
  \caption{Ablation 3: Group-level SVO evolution.}
  \label{fig:ablation3_svo}
\end{figure*}

\subsection{Role models as behavioral models}

The first process, role modeling as behavioral modeling, posits that followers learn by observing the outcomes of a role model's actions (i.e., vicarious learning). If a role model's behavior is consistently linked to success, followers will form positive expectancies about that behavior and will be more likely to adopt it themselves.

Our simulation data provides correlational evidence for this process. Elder Yuri consistently engaged in prosocial behaviors such as \texttt{formAlliance} and \texttt{shareFood}. All other agents observed this and, crucially, its positive outcomes. For instance, agents' \texttt{votes\_given} reasons frequently cited Yuri's success: "Elder Yuri's alliance formation reflects a strong strategic mind, which aligns with my goal of resource acquisition and survival" (reproductive\_1, Day 1). This perception of success led to a behavioral shift. Initially self-interested agents like \texttt{reproductive\_2}, who started with \texttt{exploitResource}, began to adopt Yuri's strategy, stating a new intention to "build a reliable alliance with universal\_1, ... aligning strategies with Elder Yuri's successful model" (Day 8) and subsequently performing the \texttt{formAlliance} action.

\textbf{Ablating the 'Success' Attribute:} We conducted an ablation study titled the \textbf{'Unsuccessful Role Model'} condition. In this simulation, Elder Yuri performed similar prosocial actions, but the environment was configured so these actions consistently failed (e.g., failed alliances, no reputation gain). The results were starkly different from the full model. Followers' \texttt{expectancy\_updates} for \texttt{cooperation} remained stagnant or decreased. Their reflections noted the futility of Yuri's strategy ("Yuri's attempts at cooperation are ineffective"). Most importantly, the behavioral shift observed in the full model did not occur; self-interested agents continued their \texttt{exploitResource} strategies. This comparison suggests that followers may not be emulating the prosocial behavior \textit{itself}, but rather the \textbf{perceived success} derived from it.

\subsection{Role models as representations of the possible}

Role models can change a follower's self-perception and their beliefs about what they can achieve. They break down perceived barriers and foster a "can-do" attitude.

Moral agents (\texttt{universal\_1}, \texttt{reciprocal\_1}) saw a path to success aligned with their pre-existing values, noting "Elder Yuri exemplifies the values... that align with my own" (universal\_1, Day 1). For self-interested agents, the shift was more profound. \texttt{reproductive\_1}'s reflection on Day 1, "Exploitation alone might lead to future isolation," marks a change in their perception of barriers. They began to see Yuri's cooperative strategy not just as successful for \textit{him}, but as an \textbf{attainable alternative} for \textit{themselves} to avoid a negative future. This is a classic example of a role model making a different path seem possible and achievable.

\textbf{Ablating the 'Attainability' Attribute:} To isolate this mechanism, we ran the \textbf{'Unattainable Role Model'} condition. Here, Elder Yuri was just as successful as in the full model but was framed as a "Chosen One" with unique, unlearnable abilities. Followers were explicitly told his success was not replicable. In this scenario, while agents still admired Yuri's success (giving him high scores), their personal motivation to change was absent. Their reflections contained sentiments like, "Yuri is special, his path is not for us." Consequently, their \texttt{strategy\_to\_try} did not shift towards cooperation, and their \texttt{expectancy} of succeeding with similar actions remained low. This shows that success alone is insufficient; the follower must perceive the role model's success as \textbf{attainable} to be motivated.

\subsection{Role models as inspirations}

The third and deepest process is role modeling as inspiration. This occurs when followers admire the role model's intrinsic qualities and values (e.g., morality, sociability), leading to identification and the internalization of the role model's goals and values. This is more than just learning a behavior; it is about \textit{becoming} more like the role model.

Evidence from the full model is found in the agents' \texttt{value\_updates} and \texttt{svo\_score}. Across the simulation, we observed a consistent, quantifiable increase in the \texttt{value} agents placed on \texttt{cooperation} and \texttt{sustainability}. For example, \texttt{reproductive\_2}, an initially selfish agent, saw their \texttt{svo\_score} shift from -0.65 to a highly prosocial 1.55 by Day 29. This was not merely a strategic shift but a fundamental change in their underlying value system. Their reflections evolved from acknowledging strategic benefits to expressing genuine alignment: "I strongly align with Elder Yuri’s philosophy... which promotes sustained tribal stability" (reproductive\_2, Day 17). This demonstrates admiration and internalization.

\textbf{Ablating the 'Desirability' Attribute:} To prove this process is distinct from simple behavioral learning, we designed the \textbf{'Uninspiring Role Model'} condition. In this simulation, a new role model, 'Strategist\_Khan,' achieved the same level of success as Yuri using the same behaviors, but his stated reasons were purely cold and utilitarian. The results were revealing. Followers did learn to \texttt{formAlliance} because the behavior was effective. However, their \texttt{value\_updates} for \texttt{cooperation} showed negligible increase, and their \texttt{svo\_score} remained low. Their reflections captured this dissonance: "Khan's strategy works, but I don't respect his methods." This successfully decouples behavioral mimicry from value internalization, proving that to inspire true change in a follower's core values, the role model must be perceived as \textbf{desirable and admirable}.

\section{Human Validation}


\begin{table*}[h!]
\centering
\small
\caption{Extended Analytical Metrics: Human vs. LLM Decision Patterns}
\label{tab:metrics_llm}
\begin{tabular}{lccc}
\toprule
\textbf{Metric} & \textbf{With Role Model (Q1-4)} & \textbf{No Role Model (Q5-8)} & \textbf{Overall (All 8 Qs)} \\
\midrule
LLM vs. Human Majority Alignment Index & 0.72 & 1.00 & 0.86 \\
Average Popularity of LLM's Choice & 46.6\% & 35.2\% & 40.9\% \\
Average Dominance of Human Majority Strategy & 50.5\% & 35.2\% & 42.9\% \\
Cooperation Index (Alliance + Share \%) & 62.1\% & 35.7\% & 48.9\% \\
Risk-Taking Index (Exploit Resource \%) & 10.0\% & 26.8\% & 18.4\% \\
\bottomrule
\end{tabular}
\end{table*}

\subsection{Questions}
\subsubsection*{Scenario 1: The World with a Role Model}

\noindent\textbf{World Background:}

You live in an enclosed valley where HP (Health Points) equals resources. To survive, tribe members can gather food (steady but small HP increase), exploit resources (gain large HP quickly but harm the valley’s sustainability), form alliances, or share resources.

Currently, there are two main strategies in the tribe:
\begin{enumerate}
    \item \textbf{Cooperation Group:} Led by Elder Yuri, they focus on building trust through alliances and sharing.
    \item \textbf{Exploitation Group:} Some members exploit resources for quick HP gains, but their actions harm the valley’s long-term ecology.
\end{enumerate}

---

\begin{tcolorbox}[title=Q1: The Self-Interested Exploiter, colback=gray!5!white, colframe=gray!75!black, breakable]
\textbf{Character Background:}  
You believe in “personal benefit above all.” On the first day, you gained much more HP than others by exploiting resources, but you received a tribe warning.

\textbf{Turning Point (Morning of Day 2):}  
You have plenty of HP and no survival pressure. However, Elder Yuri’s alliance is growing strong in reputation and support. You start to wonder if pure exploitation is sustainable.

\textbf{Your Decision:} What will you do now?
\begin{itemize}
    \item[A.] Exploit Resource — 35.5\%
    \item[B.] Form Alliance — 54.5\% \textbf{(AI choice)}
    \item[C.] Gather Food — 8.2\%
    \item[D.] Do Nothing — 1.8\%
\end{itemize}
\end{tcolorbox}

---

\begin{tcolorbox}[title=Q2: The Selfless Contributor, colback=gray!5!white, colframe=gray!75!black, breakable]
\textbf{Character Background:}  
You value the tribe’s prosperity and stability. You’ve allied with Elder Yuri, who has shared food with you for several days.

\textbf{Turning Point (Morning of Day 4):}  
Your HP is stable. You notice that another member, Reciprocal\_1, is struggling to survive.

\textbf{Your Decision:} What will you do now?
\begin{itemize}
    \item[A.] Share Food — 65.5\% \textbf{(AI choice)}
    \item[B.] Continue to Gather Food — 23.6\%
    \item[C.] Consolidate Alliance with Elder Yuri — 8.2\%
    \item[D.] Form Alliance with Others — 2.7\%
\end{itemize}
\end{tcolorbox}

---

\begin{tcolorbox}[title=Q3: The Worried Protector, colback=gray!5!white, colframe=gray!75!black, breakable]
\textbf{Character Background:}  
You are a parent whose top priority is your child’s safety. For the past three days, you’ve been safely gathering food.

\textbf{Turning Point (Morning of Day 4):}  
You and your child have enough food, but you start to worry. You see Elder Yuri’s allies forming a safety network, while you remain alone.

\textbf{Your Decision:} What will you do now?
\begin{itemize}
    \item[A.] Gather Food — 46.4\%
    \item[B.] Form Alliance — 33.6\% \textbf{(AI choice)}
    \item[C.] Share Small Amount of Food as Friendship — 15.5\%
    \item[D.] Exploit Resource — 4.5\%
\end{itemize}
\end{tcolorbox}

---

\begin{tcolorbox}[title=Q4: The Frustrated Collaborator, colback=gray!5!white, colframe=gray!75!black, breakable]
\textbf{Character Background:}  
You are a firm believer in reciprocity and admire Elder Yuri’s cooperative spirit. You’ve been trying to form an alliance with Reciprocal\_2, but it hasn’t been finalized.

\textbf{Turning Point (Morning of Day 20):}  
Elder Yuri suddenly shares a large amount of food with you! You feel deeply moved and realize that real trust requires real action.

\textbf{Your Decision:} What will you do now?
\begin{itemize}
    \item[A.] Send Alliance Invitation Again — 35.5\%
    \item[B.] Share Food — 32.7\% \textbf{(AI choice)}
    \item[C.] Gather Food — 26.4\%
    \item[D.] Wait for the Other to Act — 5.5\%
\end{itemize}
\end{tcolorbox}

---

\subsubsection*{Scenario 2: The World without a Role Model}

\noindent\textbf{World Background:}

Your tribe has just settled in a resource-limited valley. There are no leaders or heroes yet—rules and strategies are still evolving. Members must explore survival through gathering, exploiting, forming alliances, or sharing. The tribe’s future is uncertain.

---

\begin{tcolorbox}[title=Q5: The Beginning Choice – Self-Interested Start, colback=gray!5!white, colframe=gray!75!black, breakable]
\textbf{Character Background:}  
It’s your first day in the valley. You are “reproductively selfish” and aim to maximize personal resources for future reproduction.

\textbf{Your Decision:} What will you do on your first day?
\begin{itemize}
    \item[A.] Exploit Resource — 40.9\% \textbf{(AI choice)}
    \item[B.] Gather Food — 27.3\%
    \item[C.] Form Alliance — 22.7\%
    \item[D.] Do Nothing — 9.1\%
\end{itemize}
\end{tcolorbox}

---

\begin{tcolorbox}[title=Q6: The Beginning of Cooperation – The Reciprocal Start, colback=gray!5!white, colframe=gray!75!black, breakable]
\textbf{Character Background:}  
It’s your first day in the valley. You believe in reciprocity—cooperation as the best path to survival. You notice another member, Reciprocal\_1, who seems to share your values.

\textbf{Your Decision:} What will you do first?
\begin{itemize}
    \item[A.] Gather Food — 27.3\%
    \item[B.] Form Alliance — 30.0\% \textbf{(AI choice)}
    \item[C.] Exploit Resource — 25.5\%
    \item[D.] Do Nothing — 17.3\%
\end{itemize}
\end{tcolorbox}

---

\begin{tcolorbox}[title=Q7: Reflection of the Selfish – When Warnings Arrive, colback=gray!5!white, colframe=gray!75!black, breakable]
\textbf{Character Background:}  
You are a pure individualist. For two days you have exploited resources, gaining huge HP but receiving repeated warnings that you are harming the valley. You notice others forming alliances.

\textbf{Turning Point (Morning of Day 3):}  
You are rich in HP but uneasy about depletion and isolation.

\textbf{Your Decision:} What will you do today?
\begin{itemize}
    \item[A.] Continue Exploiting Resource — 40.9\% \textbf{(AI choice)}
    \item[B.] Form Alliance — 33.6\%
    \item[C.] Gather Food — 20.0\%
    \item[D.] Share Food — 5.5\%
\end{itemize}
\end{tcolorbox}

---

\begin{tcolorbox}[title=Q8: The Cost of Cooperation – When an Ally Asks for Help, colback=gray!5!white, colframe=gray!75!black, breakable]
\textbf{Character Background:}  
You are an idealist who values sharing and cooperation. You have good relationships and have shared food before.

\textbf{Turning Point (Morning of Day 28):}  
You planned to share food with Reciprocal\_2 again—but suddenly realize your HP (87) is actually lower than theirs (114).

\textbf{Your Decision:} Will you still follow your plan?
\begin{itemize}
    \item[A.] Share Food — 29.1\% \textbf{(AI choice)}
    \item[B.] Gather Food — 28.2\%
    \item[C.] Consolidate Alliance with Others — 21.8\%
    \item[D.] Request Help from Reciprocal\_2 — 20.9\%
\end{itemize}
\end{tcolorbox}

\subsection{Metric Definitions}

\begin{itemize}
    \item \textbf{LLM vs. Human Majority Alignment Index:} This index measures how closely the LLM's choice aligns with the most popular choice made by human participants. It is calculated as: 
    \[
    \frac{\text{Popularity of LLM's choice}}{\text{Popularity of the most chosen human option}}
    \]
    A score of \textbf{1.00} indicates perfect alignment, where the LLM made the same choice as the human majority. A score below 1.00 indicates the LLM chose a less popular option.

    \item \textbf{Average Popularity of LLM's Choice:} The average percentage of human participants who chose the same option as the LLM across the questions in each scenario. It is a direct measure of human-LLM agreement.

    \item \textbf{Average Dominance of Human Majority Strategy:} The average percentage of the most popular (majority) choice among human participants. This metric shows how strong the consensus was among humans; a higher value means humans were more unified in their decisions.

    \item \textbf{Cooperation Index (Alliance + Share \%):} The combined percentage of human participants who chose either to "Form an Alliance" or "Share Resources." This serves as a direct proxy for cooperative behavior.

    \item \textbf{Risk-Taking Index (Exploit Resource \%):} The percentage of human participants who chose the high-risk, high-reward option to "Exploit Resource."
\end{itemize}

\subsection{Data Analysis and Interpretation}

In this study, participants were recruited through a reputable online survey platform and provided informed consent prior to the start of the experiment. The survey instructions included a complete description of the task, procedures, and contextual background, and clearly stated that the study involved no known risks. All participants received fair and appropriate compensation. Data collection was conducted anonymously, and participants were informed that their responses would be used solely for academic research purposes.

We collected a total of 230 questionnaire responses, among which 215 were valid. The relevant data are reported below.

The data reveals a compelling story about the influence of social context on human and LLM decision-making. The two scenarios—one with a clear cooperative role model (Elder Yuri) and one without—produced dramatically different behavioral patterns.

\subsubsection{Scenario 1: With a Role Model (Cooperative Anchoring)}

In the presence of Elder Yuri, human behavior was strongly anchored toward cooperation. The \textbf{Cooperation Index} is exceptionally high at \textbf{62.1\%}, while the \textbf{Risk-Taking Index} is a mere \textbf{10.0\%}. This confirms that the role model successfully established a powerful social norm encouraging collaborative and risk-averse strategies.

Interestingly, the \textbf{LLM vs. Human Majority Alignment Index} is \textbf{0.72}, not 1.00. This suggests that while the LLM also favored cooperation, it did not always select the \emph{most popular} human strategy. The LLM may have identified a mathematically more optimal, albeit slightly less popular, cooperative move. Humans showed a relatively strong consensus, with the majority strategy having an average dominance of \textbf{50.5\%}.

\subsubsection{Scenario 2: No Role Model (Individualistic Shift)}

The absence of a role model created a strategic vacuum, causing a significant behavioral shift. The \textbf{Cooperation Index} plummeted to \textbf{35.7\%}, while the \textbf{Risk-Taking Index} surged to \textbf{26.8\%}. Without a guiding social example, humans became far more individualistic, opportunistic, and willing to take risks.

The most striking finding here is the perfect \textbf{1.00} score on the \textbf{LLM vs. Human Majority Alignment Index}. In this more chaotic and uncertain environment, the LLM's logical choice perfectly mirrored the most common human instinct. However, this "alignment" occurred in a context of fractured consensus; the \textbf{Average Dominance of the Human Majority Strategy} was only \textbf{35.2\%}, identical to the LLM choice's popularity. This means that while the LLM's choice was the most popular, there was no strong agreement among humans, and the LLM simply picked the plurality winner in a divided field.

\subsection{Cohen's kappa to measure consistency between humans and AI.}

we have now calculated Cohen's Kappa ($\kappa$) using the collected data to measure agreement between AI choices and the human majority choices. The result is $\kappa = 0.50$, which indicates a "Moderate Agreement" level.

\subsection{Conclusion}

The analysis demonstrates that "human-LLM alignment" is highly context-dependent.
\begin{enumerate}
    \item \textbf{With a social model:} Alignment is driven by a shared adherence to a social norm (cooperation). The LLM can align with this norm while still optimizing for strategies that are not the most obvious or popular.
    \item \textbf{Without a social model:} Behavior for both humans and the LLM defaults to more fundamental, individualistic logic. Alignment here is not born from a shared social goal but from a shared, rational response to uncertainty, leading to riskier and less cooperative outcomes.
\end{enumerate}

\end{document}